\documentclass[
    aps,
    prb,
    twocolumn,
    groupedaddress,superscriptaddress,
    footinbib,
    floatfix,
    longbibliography,
    ]{revtex4-2} 

\usepackage[utf8]{inputenc}
\usepackage{amssymb,amsmath,mathtools,graphicx,xcolor}
\usepackage{booktabs} %
\usepackage{bm} %

\usepackage[bookmarksnumbered,              %
colorlinks = true,
linkcolor = black,
urlcolor  = blue,
citecolor = blue,
anchorcolor = blue]{hyperref}

\colorlet{EMlinkcolor}{blue!50!black}
\colorlet{SMlinkcolor}{blue!50!black}

\usepackage{xstring} %
\usepackage{bbm} %
\usepackage[capitalise]{cleveref} %

\usepackage{orcidlink} %
\graphicspath{{figures}}

\DeclarePairedDelimiterX\abs[1]{\lvert}{\rvert}{#1}
\DeclarePairedDelimiterX\ket[1]{\lvert}{\rangle}{#1}
\DeclarePairedDelimiterX\bra[1]{\langle}{\rvert}{#1}
\DeclarePairedDelimiterX\braket[2]{\langle}{\rangle}{#1\,\vert\,#2}
\DeclarePairedDelimiterX\ketbra[2]{\lvert}{\rvert}{#1\rangle\!\langle#2}
\DeclarePairedDelimiterX\projector[1]{\lvert}{\rvert}{#1\rangle\!\langle#1}
\DeclarePairedDelimiterX\expval[2]{\langle}{\rangle}{#1\,\vert\,#2\,\vert\,#1}
\DeclarePairedDelimiterX\matel[3]{\langle}{\rangle}{#1\,\vert\,#2\,\vert\,#3}
\DeclarePairedDelimiterX\ep[1]{\langle}{\rangle}{#1}

\newcommand{\del}{\partial}

\newcommand{\ccite}[1]{%
\IfSubStr{#1}{,}{refs.}{ref.~}\cite{#1}%
}
\newcommand{\Ccite}[1]{%
\IfSubStr{#1}{,}{Refs.~}{Ref.~}\cite{#1}%
}

\DeclareRobustCommand{\SuppMat}{\hyperref[sec:SuppMat]{\textcolor{SMlinkcolor}{Supplemental Material}}}
\DeclareRobustCommand{\SMref}[1]{Sec.~\ref{#1} of~\cite{Note1}}
\DeclareRobustCommand{\SM}{\cite{Note1}}
\newcommand{\beginsupplement}{%
    \setcounter{secnumdepth}{3}%
    \setcounter{section}{0}%
    \setcounter{equation}{0}%
    \setcounter{figure}{0}%
    \setcounter{table}{0}%
    \renewcommand{\thesection}{S\arabic{section}}%
    \renewcommand{\theequation}{S\arabic{equation}}%
    \renewcommand{\thefigure}{S\arabic{figure}}%
    \renewcommand{\thetable}{S\arabic{table}}%
    \renewcommand{\theHsection}{supp.\arabic{section}}%
    \renewcommand{\theHequation}{supp.\arabic{equation}}%
    \renewcommand{\theHfigure}{supp.\arabic{figure}}%
    \renewcommand{\theHtable}{supp.\arabic{table}}%
}

\begin{document}

\makeatletter
\def\maketitle{%
    \@author@finish
    \title@column\titleblock@produce
    \suppressfloats[t]%
}
\makeatother

\newcommand{\titlecontent}{Exact quantum geometry from sublattice symmetry:~\\Closed-form solution of the quarter-flux Harper-Hofstadter model}
\title{\titlecontent
}

\newcommand{\ITPTUB}{Institut für Physik und Astronomie, Technische Universit\"{a}t Berlin,
Hardenbergstr.~36, D-10623 Berlin, Germany}
\author{Isaac Tesfaye\,\orcidlink{0009-0001-4194-3916}}
\email{i.tesfaye@tu-berlin.de}
\affiliation{\ITPTUB}

\author{Andr\'{e} Eckardt\,\orcidlink{0000-0002-5542-3516}}
\email{eckardt@tu-berlin.de}
\affiliation{\ITPTUB}

\begin{abstract}
Quantum geometry has emerged as a guiding principle across atomic and
condensed-matter physics, shaping the topological responses of Bloch bands and the
stability of the correlated phases they host.
Sublattice symmetry, though common among bipartite lattice models, has not yet been
exploited to obtain closed-form quantum geometry in multiband systems.
For this purpose, we derive a general expression for the QGT of sublattice-symmetric
systems in terms of contributions from the individual sublattice sectors, and show
that this symmetry renders the Bloch Hamiltonian of a paradigmatic four-band model,
the quarter-flux Harper-Hofstadter model, anti-block-diagonal, analytically yielding
the spectrum, eigenstates, and full quantum geometric tensor (QGT), including the
Berry curvature and quantum metric, for all four bands. 
The model, describing charged particles on a two-dimensional square lattice subjected to a uniform
magnetic field, has recently been realized experimentally with ultracold atoms, photons,
and superconducting circuits. 
Finally, we evaluate fractional-Chern-insulator stability criteria analytically and quantify the lowest band of the quarter-flux Harper-Hofstadter model to be a nearly ideal Chern band.
Our approach opens a route for studying also the quantum geometry of other
sublattice-symmetric multiband systems.
\end{abstract}

\maketitle

\makeatletter
\let\origaddcontentsline\addcontentsline
\renewcommand{\addcontentsline}[3]{}
\makeatother

\hypersetup{linkcolor=blue} %
\emph{Introduction~--}
\phantomsection\label{sec:introduction}
The quantum geometry of Bloch bands has emerged as a unifying language across atomic and condensed matter physics~\cite{Provost1980,Resta2011,Torma2023,Yu2025a,Liu2025a,Bzdusek2026,Verma2026,Gao2026}.
The quantum geometric tensor (QGT), whose imaginary part is the Berry curvature and whose real part is the quantum metric~\cite{Fubini1904,Provost1980,Berry1984,Aharonov1987,Anandan1990,Resta2011,Ma2010,Kolodrubetz2017}, governs topological responses such as the quantized Hall conductance~\cite{Thouless1982,Niu1985,Avron1985,Xiao2010}, the superfluid weight of flat-band superconductors~\cite{Peotta2015,Julku2016,Liang2017,Xie2020,Herzog-Arbeitman2022,Torma2022,Tian2023}, orbital magnetism~\cite{Gao2019,Piechon2016}, (nonlinear) optical responses~\cite{Souza2008,Gianfrate2020,Ozawa2018a,Ozawa2019a,Topp2021,Ahn2022,Shinada2025,Du2021,Smith2022,Liu2025d,Gao2023a,Wang2023e}, and the localization of Wannier functions~\cite{Marzari1997,Resta1999,Souza2000,Marzari2012,Kruchkov2022}. 
Importantly, the QGT (which itself has also been generalized to other settings~\cite{Zhang2019,Zhu2021,ChenYe2024,Tesfaye2025,Oancea2026,Salerno2020a,Weinberg2017a,Zhou2024c,He2026,Cuerda2024,Matraszek2025,Montag2026,Solnyshkov2021,Liao2021}) also governs a growing family of quantum-geometric bounds on observables~\cite{Roy2014,Mera2021,Ozawa2021,Onishi2024a,Onishi2024b,Onishi2025,Yu2024b,Yu2022,Yu2025b,Jankowski2024b,Jankowski2025a,Shinada2025,Kruchkov2025,Lim2026,Azam2026,Bouhon2023a,Zhao2026a}. 
A prominent example of such a bound is related to the stabilization of fractional Chern insulators (FCIs)~\cite{Neupert2011,Sheng2011,Regnault2011,Sun2011,Hafezi2007,Parameswaran2013,Bergholtz2013}, which are lattice analogs of fractional quantum Hall states~\cite{Laughlin1983,Moore1991,Read1999,DasSarma2005}. 
Such states have recently been realized in small systems of ultracold atoms~\cite{Leonard2023a,Lunt2024a,Kwan2026} and photons~\cite{Clark2020,Wang2024e}. 
When a Chern band saturates this quantum-geometric bound, it is said to be ideal (or vortexable)~\cite{Parameswaran2012,Roy2014,Mera2021,Ozawa2021,Ledwith2023,Fujimoto2025a,Liu2025b,Wang2021,Zhao2026,Li2026a}, as its quantum geometry then mimics that of the lowest Landau level (LLL) and becomes an especially favorable FCI host~\cite{Roy2014,Jackson2015,Ledwith2023,Wang2021}. 

Assessing these criteria typically requires the full quantum geometry of the Bloch band, yet closed-form expressions are scarce~\cite{Barnett2012,Bauer2016}. 
For two-band models, the QGT follows from the Bloch vector, but Chern bands relevant for FCIs generically live in topological multiband settings where analytic solutions for both spectrum and quantum geometry are rare.
A paradigmatic multiband system is the Harper-Hofstadter (HH) model~\cite{Harper1955,Azbel1964,Hofstadter1976}, describing charged particles tunneling on a square lattice that is threaded by a uniform magnetic flux $\Phi=2\pi\alpha$ per plaquette. 
This textbook example of a topological band insulator with nonzero Chern numbers underlying the quantized Hall conductances~\cite{Thouless1982,Xiao2010} has become a workhorse of quantum simulation, realized with ultracold atoms, both in two-dimensional optical lattices \cite{Aidelsburger2013,Miyake2013,Aidelsburger2015,Tai2017} and by combining a one-dimensional optical lattice with a synthetic dimension~\cite{Stuhl2015,Mancini2015}, in photonic systems~\cite{Hafezi2011}, and with superconducting qubits~\cite{Roushan2017}.
Despite its prominence, and although a rational flux $\alpha=p/q$ per plaquette (with $p,q$ coprime) reduces the problem to a $q\times q$ Bloch Hamiltonian, a closed-form solution for the model has not been reported beyond the $q=3$ three-band case~\cite{Barnett2012,Bauer2016}, for $q>3$ both the spectrum and the quantum geometry have so far been accessible only numerically or perturbatively~\cite{Harper2014,Bauer2016}. 

In this work, we analytically solve the quarter-flux ($\alpha\!=\!1/4$) Harper-Hofstadter model, obtaining closed-form expressions for the spectrum, the eigenstates, and the quantum geometry of all four bands as a function of quasimomentum $\bm{k}$. From the latter we also evaluate the momentum-based FCI-stability criteria analytically and show that the lowest Hofstadter band nearly approaches an ideal Chern band.
The key observation of our work, allowing us to find closed-form expressions, is that among all possible choices of magnetic unit cell (MUC) for $\alpha=1/4$, the symmetric $2\times2$ MUC uniquely preserves the sublattice symmetry of the model in quasimomentum space. 
This renders the $4\times4$ Bloch Hamiltonian anti-block-diagonal, reducing the problem to a singular-value decomposition of a remaining $2\times2$ block. 
Moreover, we show and use that for \emph{any} sublattice-symmetric system, the Abelian and non-Abelian QGT decompose into contributions from the two sublattice sectors.

\emph{Sublattice symmetry in the quarter-flux Harper-Hofstadter model~--}
\phantomsection\label{sec:Sublattice-Symmetry-HH-Model} 
We study the Harper-Hofstadter Hamiltonian~\cite{Harper1955,Azbel1964,Hofstadter1976} in the Landau gauge
\begin{align}
    \hat{H}
    =-J\sum_{m,n}(e^{-i\Phi n}\,\ketbra{m\!+\!1,n}{m,n}
    \!+\!\ketbra{m,n\!+\!1}{m,n}\!+\!\text{h.c.})
    \label{eq:HH-Ham-Landau}
\end{align}
with a quarter flux per plaquette of $\Phi=2\pi \alpha=\pi/2$ ($\alpha=1/4$) and a hopping amplitude $J$. 
Here, $\ket{m,n}$ denotes the single-particle state localized at lattice site $(m,n)$, where $m$ ($n$) denotes the discrete position along the $x$ ($y$) direction of the square lattice of size $M \times N$, with $M$ ($N$) the number of lattice sites along $x$ ($y$). 

Under periodic boundary conditions, the Hamiltonian~\eqref{eq:HH-Ham-Landau} at rational flux $\alpha=p/q=1/4$ admits a magnetic translation symmetry with respect to a \emph{magnetic unit cell} (MUC) enclosing $q=4$ plaquettes of total flux $2\pi$~\cite{Zak1964,Zak1964a,Bernevig2013,Aidelsburger2016}. 
Its area $A_{\text{MU}}=4a^2$ is fixed (lattice constant $a=1$ in the following), 
but its shape is not, giving three choices for $\alpha=1/4$: (i) $1\times4$, (ii) $2\times2$, and (iii) $4\times1$, each with four sublattice sites $A,B,C,D$, as shown in~\cref{fig:Hofstadter-Energy-Spectrum}(a)-(b). 
A generalized Bloch theorem based on the associated magnetic translation operators (MTOs) yields, for each choice, a $4\times4$ Bloch Hamiltonian $H(\bm{k})$ acting on the sublattice basis states $\ket{\bm{k},\tau}$, $\tau=A,B,C,D$, with bands $E_n(\bm{k})$, $n\in\{1,2,3,4\}$, and quasimomentum $\bm{k}=(k_x,k_y)$.
Different from the commonly studied case (i), we focus on case (ii) with the symmetric square MUC, for which the Bloch Hamiltonian reads~\cite{Aidelsburger2015} 
\begin{align}
    &H(\bm{k})=\nonumber \\
    &-2J\begin{pmatrix}
        0 & \cos(k_x a) & \cos(k_y a) & 0 \\
        \cos(k_x a) & 0 & 0 & -\sin(k_y a) \\
        \cos(k_y a) & 0 & 0 & -i \sin(k_x a) \\
        0 & -\sin(k_y a) & i \sin(k_x a) & 0
    \end{pmatrix}. \label{eq:Bloch-Ham-Case-II}
\end{align}
The Bloch Hamiltonians for the other two choices (i) and (iii) of the MUC are given in Sec.~\ref{app:Magnetic-Translation-Symmetry-HH-Model} of the \SuppMat~\footnote{See \SuppMat~at [URL] for detailed derivations of the results, which includes Refs.~\cite{Zak1964a,Zak1964,Harper1955,Hofstadter1976,Bernevig2013,Aidelsburger2016,Aidelsburger2015,Sutherland1986,Lieb1989,Ludwig2015,Ramachandran2017,Yu2017,Xiao2024,Theel2025,Nicolau2026,Altland1997,Horn2012,Zirnbauer1996,Ryu2010,Manjunath2021,Manjunath2020,Zhang2022c,Zhang2025b,Chiu2016,Nakahara2003,Aidelsburger2013,Miyake2013,Resta2011,Mera2022,Graf2021,Pozo2020,Harper2014,Ma2010,Rezakhani2010,Wilczek1984,Peotta2015,Xiao2010,Ledwith2023,Fujimoto2025a,Liu2025b,Wang2025a,Girvin1986,Roy2014,Mera2021,Ozawa2021,Parameswaran2012,Varjas2022,Onishi2024a,Onishi2024b,Onishi2025a,Marzari1997,Marzari2012,Brouder2007,Hastings2010}.}.
We stress that all three choices of the MUC, and hence the Bloch Hamiltonians $H(\bm{k})$, originate from the same parent Hamiltonian~\eqref{eq:HH-Ham-Landau} in a \emph{fixed} gauge~\footnote{We stress that the different choices of the MUCs are \emph{not} related to different gauge choices of~\cref{eq:HH-Ham-Landau}. All of our results apply equally to other gauge choices, such as the symmetric gauge or the Landau gauge with Peierls phases increasing along the horizontal direction, which will, however, admit different MTOs~\cite{Bernevig2013,Aidelsburger2016}.}, are unitarily equivalent, and share the same energy spectrum as a function of $\bm{k}$~\SM.
The first goal of this work is to find an exact analytical solution for the single-particle eigenstates and energies as a function of quasimomentum $\bm{k}$ by diagonalizing \eqref{eq:Bloch-Ham-Case-II}.

To this end, we utilize the chiral symmetry that is present for the Harper-Hofstadter model~\eqref{eq:HH-Ham-Landau}.
A chiral symmetry is defined as a unitary involution $\hat{S}$ ($\hat{S}^2=\mathbbm{1}$) that anti-commutes with the Hamiltonian, $\{\hat{H},\hat{S}\}=0$,~\cite{Sutherland1986,Lieb1989,Ludwig2015,Ramachandran2017,Xiao2024} and splits the Hilbert space $\mathcal{H}=\mathcal{H}_+\oplus\mathcal{H}_-$ into two orthogonal subspaces~$\mathcal{H}_\pm$ of dimension $d_\pm =\dim\mathcal{H}_{\pm}$ corresponding to the eigenvalues $\pm 1$ of $\hat{S}$.
In the eigenbasis of $\hat{S}$, $\hat{H}$ is anti-block-diagonal and its spectrum is symmetric about zero, with non-zero energies $E_{\pm,j}=\pm\sigma_j$ set by the singular values $\sigma_j$ of the off-diagonal block $D$. 
The eigenstates of each chiral pair always take the form~\cite{Ryu2010}
\begin{align}
    \ket{\Psi_{\epsilon, j}}=\frac{1}{\sqrt{2}}\begin{pmatrix}\, \ \ket{u_j} \\ \epsilon \ket{v_j} \end{pmatrix},
    \label{eq:Sublattice-Symmetric-Eigenstates}
\end{align}
where $\ket{u_j}\in \mathcal{H}_+$ ($\ket{v_j}\in \mathcal{H}_-$) are the left (right) singular vectors of $D$~\cite{Horn2012}. 
Here, $\epsilon=\pm 1$ labels the energy pair $E_{\pm,j}$ and $j=1,\ldots,r$ with $r=\mathrm{rank}(D)$.
\begin{figure}[t]
    \begin{center}
        \includegraphics[width=\columnwidth]{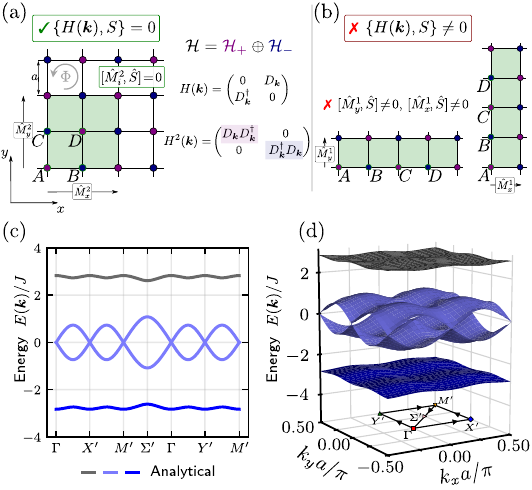}
        \caption{
        Sublattice symmetry and analytical spectrum of the quarter-flux Harper-Hofstadter model.
        (a) Real space lattice of HH-model with a quarter flux $\Phi=\pi/2$ per plaquette, the symmetric $2\times 2$ MUC (green shaded area) with its four sublattice sites $A,B,C,D$. 
        The sublattice symmetry $\hat S=(-1)^{m+n}\projector{m,n}$ corresponding to the bipartite checkerboard pattern (purple/blue colored sites for the even/odd sublattice), 
        commutes with the associated magnetic translation operators (MTOs) $\hat M_i^2$ of the MUC, $[\hat S,\hat M_i^2]=0$, and is preserved in momentum space via $S=\mathrm{diag}(1,-1,-1,1)$, and
        anti-commutes with the Bloch Hamiltonian~\eqref{eq:Bloch-Ham-Case-II}, $\{H(\bm{k}),S\}=0$.
        (b): Same as (a) for the other two choices of the MUC, (i) and (iii), of size $1\times 4$ and $4\times 1$, respectively, showing that the sublattice symmetry is not preserved in momentum space as $\{H(\bm{k}),S\}\neq 0$ due to $[\hat S,\hat M_i^1]\neq 0$~\SM. 
        (c) Energy spectrum $E_n(\bm{k})$~\eqref{eq:Energy-Spectrum-Hofstadter-Exact-quarter-Flux-momentum} along the high-symmetry path [defined in (d)] in the FBZ with solid lines indicate the analytical results. 
        (d)$E_n(\bm{k})$ over the full FBZ.
        }
        \label{fig:Hofstadter-Energy-Spectrum}
    \end{center}
\end{figure}

The chiral symmetry results from the bipartite checkerboard pattern $s_{m,n}=(-1)^{m+n}$ of the real-space square lattice, which allows us to divide the lattice sites into even (odd) sites with $s_{m,n}=+1$ ($s_{m,n}=-1$), as illustrated in~\cref{fig:Hofstadter-Energy-Spectrum}(a)-(b) via the purple (blue-colored) lattice sites, respectively.
Thus, for $\hat H$~\eqref{eq:HH-Ham-Landau} we have $\{\hat{H},\hat{S}\}=0$,
with 
$\hat{S}=\sum_{m,n} s_{m,n} \projector{m,n}$,
because there are only inter-sublattice couplings (between even and odd sites), but no intra-sublattice couplings (between even and even or odd and odd sites). 

Now, the key idea of our work is realizing that, when going to momentum space, the distinction between a chiral and a sublattice symmetry becomes crucial. 
Namely, a sublattice symmetry is a chiral symmetry whose operator is diagonal in the site basis, assigning a fixed sign $s=\pm1$ to every site and thereby partitioning the lattice into two sign-definite sublattices and providing a \emph{grading}~\SM.
In real space the checkerboard operator $\hat{S}$ is both, since its even-odd pattern $s_{m,n}$ is itself such a grading, with the ``even'' $s=1$ sublattice containing states with $\tau = A $ and $D$ and the ``odd''  $s=-1$ sublattice containing states with $\tau = B$ and $C$~[\cref{fig:Hofstadter-Energy-Spectrum}(a)]. 
In momentum space, however, it is only the choice (ii) of the MUC with the Bloch Hamiltonian $H(\bm{k})$~\eqref{eq:Bloch-Ham-Case-II} for which the chiral operator remains diagonal in the sublattice basis $\ket{\bm{k},\tau}$ of~\eqref{eq:Bloch-Ham-Case-II}, and hence preserves the sublattice symmetry. 
Indeed, in~\eqref{eq:Bloch-Ham-Case-II} we see that there are no coupling terms between even sublattice sites ($A$ and $D$) or odd sublattice sites ($B$ and $C$), but only coupling terms connecting even with odd sublattice sites and vice versa~[\cref{fig:Hofstadter-Energy-Spectrum}(a)].
We stress that this is \emph{not} the case for the other two choices of the MUC (i) and (iii), whose associated magnetic translation operators (MTOs) include a single-site translations that do not commute with the real-space sublattice symmetry operator $\hat{S}$~[\cref{fig:Hofstadter-Energy-Spectrum}(b)], so that $\hat{S}$ is not preserved as a grading in momentum space.
In these cases $H(\bm{k})$ contains diagonal terms, such that no even-odd grading of the four sublattices $\tau$ can anti-commute with it~\SM. 

With time-reversal and particle-hole symmetry absent and a $\bm{k}$-independent chiral symmetry operator present, the model~\eqref{eq:Bloch-Ham-Case-II} belongs to the chiral class AIII of the Altland-Zirnbauer classification~\cite{Altland1997,Zirnbauer1996}, for which no winding-number invariant should be present in two dimensions~\cite{Ryu2010}. 
This is consistent with the absence of an energy gap at zero energy~[\cref{fig:Hofstadter-Energy-Spectrum}(c)-(d)], making the ground state of non-interacting fermions topologically trivial at Fermi energy of $E_F=0$. 
However, when only the lowest band is occupied, the Fermi energy $E_F$ takes a negative value in gap above the lowest band, which breaks the chiral symmetry. 
This places the model into symmetry class A, which allows for a topological Chern invariant in even spatial dimensions, consistent with non-zero Chern number of the lowest band~\SM. 

\emph{Analytical spectrum of the quarter-flux Harper-Hofstadter model~--}
\phantomsection\label{sec:Analytical-Solution} 
From the illustration in~\cref{fig:Hofstadter-Energy-Spectrum}(a), we see
that the sublattice sites $A$ and $D$ ($B$ and $C$) within the $2\times2$ MUC form the even (odd) sublattice sector, such that sublattice symmetry operator in its eigenbasis reads 
$S = \mathrm{diag}(+1,+1,-1,-1)$. 
A site permutation of~\eqref{eq:Bloch-Ham-Case-II} brings it to the anti-block-diagonal form 
$H_S(\bm{k})=\big(\begin{smallmatrix}0 & D_{\bm{k}}\\ D^{\dagger}_{\bm{k}} & 0\end{smallmatrix}\big)$, with the $2\times2$ block
    \begin{align}
        D_{\bm{k}}=-2J\begin{pmatrix}
            \cos(k_x a) & \cos(k_y a) \\
            -\sin(k_y a) & i \sin(k_x a)
        \end{pmatrix},
        \label{eq:D-matrix}
    \end{align}
mapping the odd to the even sublattice sector. 
Squaring gives $H_S^2=\mathrm{diag}(D_{\bm{k}}D^{\dagger}_{\bm{k}},D^{\dagger}_{\bm{k}}D_{\bm{k}})$, so the energies are the square roots of the eigenvalues of the two Hermitian blocks, $M_{\bm{k}}\equiv D^{\dagger}_{\bm{k}}D_{\bm{k}}$ and $\tilde{M}_{\bm{k}}\equiv D_{\bm{k}}D^{\dagger}_{\bm{k}}$. 
Their non-zero eigenvalues, with eigenvectors $|v_j\rangle$ and $|u_j\rangle$, respectively, are the same and coincide with the squared singular values $\sigma_j^2$ of $ D_{\bm{k}}=U_{\bm{k}} \Sigma_{\bm{k}} V^{\dagger}_{\bm{k}}=\sum^{r=2}_{j=1}\sigma_j \ket{u_j}\bra{v_j}$~\cite{Horn2012}.
We write $M_{\bm{k}}= D^{\dagger}_{\bm{k}}D_{\bm{k}}=h_0\sigma_0+\bm{h}\cdot\bm{\sigma}$ in Pauli form with $h_0=4J^2$ and, in the shorthand notation $c_{i\nu}\equiv\cos(ik_\nu a)$, $s_{i\nu}\equiv\sin(ik_\nu a)$, the Bloch vector $\bm{h}=(h_x,h_y,h_z)=4J^2(c_xc_y,\,s_xs_y,\,\tfrac12(c_{2x}-c_{2y}))$ (omitting the $\bm{k}$-dependence for brevity). 
Its magnitude is $h\equiv\abs{\bm{h}}=J^2\sqrt{2Z}$ with $Z=6+c_{4x}+c_{4y}$, so the eigenvalues are $\lambda_j=h_0+\eta_j\abs{\bm{h}}$ with $\eta_j=(-1)^{j+1}$ ($j\in\{1,2\}$)
Taking square roots, the four energy bands $E_{\pm,j}=\pm\sqrt{\lambda_j}$ of model follow as
    \begin{align}
        E_{\pm,j}(\bm{k})/J=\pm\sqrt{4+\eta_j\sqrt{2}\sqrt{6+\cos(4 k_x a)+\cos(4 k_y a)}},
        \label{eq:Energy-Spectrum-Hofstadter-Exact-quarter-Flux-momentum}
    \end{align}
which is the first main result of this work~\SM.
In~\cref{fig:Hofstadter-Energy-Spectrum}(c)-(d) we show the exact analytical energy spectrum~\eqref{eq:Energy-Spectrum-Hofstadter-Exact-quarter-Flux-momentum} along a high-symmetry path in the first magnetic Brillouin zone (FBZ)~[\cref{fig:Hofstadter-Energy-Spectrum}(c)] as well as over the entire FBZ (d).
Ordering the spectrum from lowest ($n=1$) to highest ($n=4$) energy, we have the four bands $E_n(\bm{k})\in\{E_{-,1},E_{-,2},E_{+,2},E_{+,1}\}$. 
The middle two bands $E_{-,2}$ and $E_{+,2}$ touch at certain high-symmetry points in the FBZ, forming Dirac cones, and hence should actually be thought of as a single \emph{super-band}~\cite{Aidelsburger2015}. 
From the sublattice symmetry it follows that the spectrum is symmetric around zero, $E_{-,j}(\bm{k})=-E_{+,j}(\bm{k})$~\eqref{eq:Energy-Spectrum-Hofstadter-Exact-quarter-Flux-momentum} with $j \in \{1,2\}$. 

Having obtained the exact energy spectrum, let us now compute the eigenstates.
To this end, note that the right singular vectors $\ket{v_j}$ of $D_{\bm{k}}$ are also the eigenvectors of $M_{\bm{k}}$, hence fixed by the Bloch vector $\bm{h}$,
\begin{align}
    \ket{v_{j}}=\frac{1}{\sqrt{2h(h+\eta_j h_z)}}
    \begin{pmatrix} h+\eta_j h_z \\ \eta_j(h_x+i h_y) \end{pmatrix},
    \label{eq:em:right-singular-vectors}
\end{align}
with $\eta_j=(-1)^{j+1}$, satisfying $M_{\bm{k}}\ket{v_j}=\sigma_j^2\ket{v_j}$, while the left singular vectors follow from the SVD relation $\ket{u_j}=D_{\bm{k}}\ket{v_j}/\sigma_j$ with $\sigma_j=E_{+,j}>0$~\eqref{eq:Energy-Spectrum-Hofstadter-Exact-quarter-Flux-momentum}. 
Note that this SVD relation is crucial as it fixes the relative phase between $\ket{v_j}$ and $\ket{u_j}$, and is what makes $\ket{\Psi_{\epsilon,j}}$~\eqref{eq:Sublattice-Symmetric-Eigenstates} an eigenstate of $H_S(\bm{k})$.
This gives the four energy eigenstates $\ket{\Psi_{\epsilon,j}}=(\ket{u_j},\epsilon\ket{v_j})^T/\sqrt{2}$~\eqref{eq:Sublattice-Symmetric-Eigenstates} for the four bands $E_{\epsilon,j}=\epsilon E_{+,j}$ ($\epsilon=\pm1$, $j\in\{1,2\}$), explicitly given by
\begin{align}
    \ket{\Psi_{\epsilon,j}}
    &=
    \begin{pmatrix}
        -\dfrac{2J}{E_{+,j}}\left[\,c_x(h+\eta_j h_z)+\eta_j c_y(h_x+i h_y)\,\right]\\[.28cm]
        -\dfrac{2J}{E_{+,j}}\left[\,-s_y(h+\eta_j h_z)+i\eta_j s_x(h_x+i h_y)\,\right]\\[.28cm]
        \epsilon\,(h+\eta_j h_z)\\[.18cm]
        \epsilon\eta_j\,(h_x+i h_y)
    \end{pmatrix} 
    \nonumber \\
    \times & \frac{1}{\sqrt{2}\,\sqrt{2h(h+\eta_j h_z)}}
    \label{eq:em:All-Eigenstates-explicit}
\end{align}
In particular, the ground state of the lowest band $E_{-,1}$, the state typically populated in cold-atom realizations of the quarter-flux Harper-Hofstadter model~\cite{Aidelsburger2013,Aidelsburger2015,Miyake2013}, is $\ket{\Psi_{-,1}}=(\ket{u_1},-\ket{v_1})^T/\sqrt2$.
An equivalent Bloch-sphere representation in the polar and azimuthal angles $(\theta_{\bm{k}},\varphi_{\bm{k}})$ of $\hat{\bm{h}}=\bm{h}/h$ and the full derivation are given in~\SMref{app:all-eigenstates-Hofstadter}. 

\emph{QGT for sublattice-symmetric systems~--}
\phantomsection\label{sec:Exact-Quantum-Geometry-FCI-Stability}
The sublattice structure that solves the spectrum also allows us to obtain the full quantum-geometrical characterization for the model in closed form.
We find that for any sublattice-symmetric system, whose eigenstates take the form of~\eqref{eq:Sublattice-Symmetric-Eigenstates}, the Abelian and the non-Abelian quantum geometric tensor (QGT)~\cite{Resta2011,Ma2010} decompose into contributions from the two sublattice sectors. 
In particular, the Abelian QGT $Q^n_{\mu\nu}= \bra{\partial_\mu \Psi_n} (1 - \projector{\Psi_n}) \ket{\partial_\nu \Psi_n}=g^n_{\mu\nu}-\tfrac{i}{2}\Omega^n_{\mu\nu}$ of an isolated band $n\equiv (\epsilon,j)$, always takes the form
\begin{align}
    Q^{j}_{\mu\nu} = \tfrac{1}{2}\big(Q^{(u_j)}_{\mu\nu} + Q^{(v_j)}_{\mu\nu}\big) + \tfrac{1}{4}\Delta A^{j}_{\mu}\Delta A^{j}_{\nu},
    \label{eq:Abelian-QGT-Sublattice-Symmetric}
\end{align}
where $Q^{(u_j)}$ ($Q^{(v_j)}$) is the QGT of the singular vector $\ket{u_j}$ ($\ket{v_j}$), and $\Delta A^{j}_{\mu}\equiv A^{(u_j)}_{\mu}-A^{(v_j)}_{\mu}$ is the difference of their Berry connections $A^{(w)}_{\mu}=i\bra{w}\partial_\mu w\rangle$ with $w\in \{u_j,v_j\}$~\footnote{Note that $\Delta A^{j}_{\mu}$ is gauge invariant under a local $U(1)$ gauge transformation $\ket{v_j}\to e^{i\chi}\ket{v_j}$, since the SVD relation $\ket{u_j}=D\ket{v_j}/\sigma_j$ also leads to $\ket{u_j}\to e^{i\chi}\ket{u_j}$ such that both $A^{(w)}_{\mu}\to A^{(w)}_{\mu}-\partial_\mu \chi$, leaving $\Delta A^{j}_{\mu}$ unchanged.}.
Note that Eq.~\eqref{eq:Abelian-QGT-Sublattice-Symmetric} is independent of $\epsilon$, any two chiral partner bands $E_{\pm,j}$ (that are gapped) always share the same quantum geometry.
Splitting it into its real and symmetric part, the quantum metric $g^j_{\mu\nu}=\mathfrak{Re}[Q^j_{\mu\nu}]$, and its imaginary and antisymmetric part, the Berry curvature $\Omega^j_{\mu\nu}=-2\mathfrak{Im}[Q^j_{\mu\nu}]$, while using that $\Delta A^{j}_{\mu}\Delta A^{j}_{\nu}$ is purely real and symmetric, we obtain
\begin{align}
    g^j_{\mu\nu} &= \tfrac{1}{2}\big(g^{(u_j)}_{\mu\nu} + g^{(v_j)}_{\mu\nu}\big) + \tfrac{1}{4}\Delta A^{j}_{\mu}\Delta A^{j}_{\nu}, \nonumber \\
    \Omega^j_{\mu\nu} &= \tfrac{1}{2}\big(\Omega^{(u_j)}_{\mu\nu} + \Omega^{(v_j)}_{\mu\nu}\big).
    \label{eq:Quantum-Metric-Berry-Curvature-Sublattice-Symmetric}
\end{align}
We find that the connection differences enter the metric only, while the Berry curvature is simply given by the sector average.
Thus, both members of a chiral pair carry identical Berry curvature, $\Omega^{j}_{\mu\nu} \equiv \Omega^{(+,j)}_{\mu\nu}=\Omega^{(-,j)}_{\mu\nu}$, and therefore identical Chern numbers, $C_{+,j}=C_{-,j}$.

For degenerate bands the non-Abelian QGT $\mathcal{Q}_{\mu\nu}^{ab}=\braket{\del_\mu \Psi_a}{\del_\nu \Psi_b}- \sum_c A_{\mu}^{ac}A_{\nu}^{cb} =G^{ab}_{\mu\nu}-\tfrac{i}{2}F^{ab}_{\mu\nu}$~\cite{Ma2010,Rezakhani2010,Ding2024} becomes relevant, with the sum running over the states $c$ of the degenerate group. 
For a group containing both chiral partners $(\pm,j)$ of each of its bands, with $a\equiv (\epsilon',j')$ and $b\equiv (\epsilon,j)$, it admits an analogous sublattice decomposition
\begin{align}
    \mathcal{Q}_{\mu\nu}^{(\epsilon',j'),(\epsilon,j)} =\tfrac{1}{2}\big[
            \mathcal{Q}^{(u_{j'},u_j)}_{\mu\nu} + \epsilon' \epsilon\, \mathcal{Q}^{(v_{j'},v_j)}_{\mu\nu} \big].
    \label{eq:non-Abelian-QGT-Sublattice-Symmetric}
\end{align}
where $\mathcal{Q}^{(u_{j'},u_j)}=\braket{\del_\mu u_{j'}}{\del_\nu u_j}-\sum_{j''} A^{(u_{j'},u_{j''})}_\mu A^{(u_{j''},u_j)}_\nu$ is the non-Abelian QGT of the left singular vectors $\ket{u_j}$, with the non-Abelian (interband) Berry connection~\cite{Wilczek1984} $A^{(u_{j'},u_j)}_\mu=i\braket{u_{j'}}{\del_\mu u_j}$, and similarly for the right singular vectors $\ket{v_j}$.
For a group of a single sign $\epsilon$ (bands touching on one side of zero energy), the unoccupied chiral partner bands instead restore the matrix analogue $\tfrac{1}{4}(\Delta A_\mu\Delta A_\nu)_{j'j}$ of the Abelian $\Delta A$ term in~\eqref{eq:Abelian-QGT-Sublattice-Symmetric}, which leaves $\mathrm{tr}\,F_{\mu\nu}$ unchanged but increases $\mathrm{tr}\,G_{\mu\mu}$. 
The derivations of~\eqref{eq:Abelian-QGT-Sublattice-Symmetric} and~\eqref{eq:non-Abelian-QGT-Sublattice-Symmetric}, including general degenerate groups, are given in~\SMref{app:quantum-geometry-all-states}.

\emph{Application to the HH-model~--}
Here, each sector is a two-level Bloch problem set by $M=D^\dagger D$ and $\tilde M=DD^\dagger$ and their Bloch vectors $\bm{h},\tilde{\bm{h}}$, so the sector QGTs, and hence the full $Q^j_{\mu\nu}$~\eqref{eq:Abelian-QGT-Sublattice-Symmetric} can be obtained analytically. 
For the quarter-flux ground band this yields the analytical expression for the Berry curvature
\begin{align}
    \Omega^1_{xy}(\bm{k}) = -\sqrt{2}\,a^2\,\frac{6-\cos(4 k_x a)-\cos(4 k_y a)}{[\,6+\cos(4 k_x a)+\cos(4 k_y a)\,]^{3/2}},
    \label{eq:Berry-Curvature-Exact-Expression}
\end{align}
together with the exact quantum metric components
\begin{align}
    g^1_{xx}/a^2 &=\frac{Z(2-c_{4x}) -s_{4x}^2}{Z^2} + \frac{s^2_{2y}c^2_{2x}}{\mathcal{G}_1^2}, \nonumber\\
    g^1_{yy}/a^2 &=\frac{Z(2-c_{4y}) -s_{4y}^2}{Z^2} + \frac{s^2_{2x}c^2_{2y}}{\mathcal{G}_1^2}, \nonumber\\
    g^1_{xy}/a^2 &= -\frac{s_{4x} s_{4y}}{Z^2} - \frac{s_{2x}s_{2y}c_{2x}c_{2y}}{\mathcal{G}_1^2},
    \label{eq:em:quantum-metric-components}
\end{align}
with $\mathcal{G}_1=2\sqrt{2Z}+Z$, which by the $\epsilon$-independence of~\eqref{eq:Abelian-QGT-Sublattice-Symmetric} are simultaneously those of the highest band $E_{+,1}$. 
Since $\Omega^1_{xy}<0$ throughout the FBZ, the curvature never changes sign, consistent with the Chern number $C_1\equiv C_{-,1} = \int_{\text{BZ}} (\Omega^1_{xy}/2\pi)\,d^2k = -1$ of the lowest Hofstadter band, which, due to~\eqref{eq:Quantum-Metric-Berry-Curvature-Sublattice-Symmetric} coincides with that of its chiral partner (highest band), $C_4\equiv C_{+,1}=C_1=-1$.

Applying the non-Abelian decomposition~\eqref{eq:non-Abelian-QGT-Sublattice-Symmetric} to the degenerate middle super-band, $(\epsilon',j')=(+,2)$ and $(\epsilon,j)=(-,2)$, the non-Abelian QGT becomes a $2\times2$ matrix in the degenerate subspace with the trace of its non-Abelian Berry curvature
    \begin{align}
        \mathrm{tr}\,F_{xy}/\,a^2 &= 2\sqrt{2}\,\frac{6-\cos(4 k_x a)-\cos(4 k_y a)}{[\,6+\cos(4 k_x a)+\cos(4 k_y a)\,]^{3/2}}\nonumber  \\
        &= -2\,\Omega^1_{xy},
        \label{eq:non-Abelian-Berry-Curvature-trace}
    \end{align}
consistent with the conservation law that the total Berry curvature summed over all bands vanishes~\cite{Xiao2010}, and fixing the super-band's total Chern number to $C_{2+3}\equiv C_{(-,2),(+,2)}=(2\pi)^{-1}\int_{\mathrm{BZ}}\mathrm{tr}\,F_{xy}\,d^2k=-2\,C_1=2$.
The trace of the non-Abelian quantum metric follows analogously,
    \begin{align}
        \mathrm{tr}(G_{xx})/a^2&= \frac{2[Z(2-c_{4x}) -s_{4x}^2]}{Z^2}, \nonumber \\
        \mathrm{tr}(G_{yy})/a^2&= \frac{2[Z(2-c_{4y}) -s_{4y}^2]}{Z^2}, \nonumber \\
        \mathrm{tr}(G_{xy})/a^2&=\frac{-2s_{4x} s_{4y}}{Z^2}.
        \label{eq:non-Abelian-quantum-metric-trace}
    \end{align}
with all derivations given in~\SMref{app:quantum-geometry-all-states}. 

\emph{A nearly ideal Chern band~--}
\phantomsection\label{sec:Nearly-Ideal-Chern-Band}
Several band-geometric criteria quantify a band's stability to host fractional Chern insulators (FCIs), most of them based on lowest-Landau-level (LLL) ``mimicry''.
These are momentum-based conditions reproducing the QGT of the LLL~\cite{Parameswaran2012,Parameswaran2013,Roy2014,Jackson2015,Claassen2015,Bauer2016,Wang2021,Mera2021,Ozawa2021,Mera2021a}, or, more recently, real-space conditions such as vortexability, aimed at platforms without a simple momentum-space description~\cite{Ledwith2023,Fujimoto2025a,Liu2025b}. 
We test these criteria on the lowest Chern band using our exact QGT. 
Positive semi-definiteness of the QGT yields a hierarchy of local inequalities
$\mathrm{tr}\,g(\bm{k}) \ge 2\sqrt{\det g(\bm{k})} \ge \abs{\Omega_{xy}(\bm{k})}$ in 2D~\cite{Roy2014,Peotta2015,Mera2021,Ozawa2021}. 
A band saturating the local trace defect $D^1(\bm{k})=\mathrm{tr}\,g^1(\bm{k})-\abs{\Omega^1_{xy}(\bm{k})}\ge 0$ everywhere is a so-called ideal (vortexable) Chern band on the lattice~\cite{Jackson2015,Ozawa2021,Ledwith2023}. 
The integrated version of this trace defect, the (global) trace-condition~\cite{Ledwith2023},
\begin{align}
    \mathcal{T}_{\mathrm{HH}} = \int_{\mathrm{BZ}}D^1(\bm{k})\,d^2k \geq 0,
    \label{eq:trace-defect}
\end{align}
is a single number that summarizes the deviation from ideality.  
Saturating it with uniform curvature further leads to the famous Girvin-MacDonald-Platzman algebra~\cite{Girvin1986} with Landau-level form factors~\cite{Parameswaran2012,Roy2014,Varjas2022,Wang2025a}. 
The lowest band of the quarter-flux Harper-Hofstadter model realizes this approximately~\SM: 
using the exact curvature~\eqref{eq:Berry-Curvature-Exact-Expression} and exact metric~\eqref{eq:Quantum-Metric-Berry-Curvature-Sublattice-Symmetric} gives us $D^1(\bm{k})$ in closed form, which we depict in~\cref{fig:trace-defect}. 
\begin{figure}[tb]
    \begin{center}
        \includegraphics[width=\columnwidth]{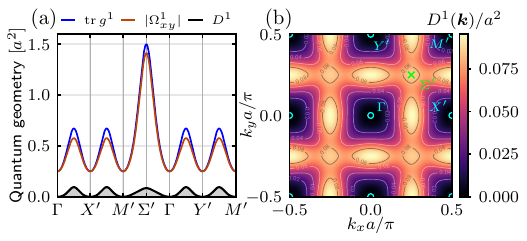}
        \caption{Local trace-condition (ideal-band) defect $D^1(\bm{k})=\mathrm{tr}\,g^1(\bm{k})-\abs{\Omega^1_{xy}(\bm{k})}\ge 0$~\eqref{eq:trace-defect} of the quarter-flux Harper-Hofstadter ground band $\ket{\Psi_{-,1}}$ 
        obtained from the analytical Berry curvature~\eqref{eq:Berry-Curvature-Exact-Expression} and the quantum metric~\eqref{eq:Quantum-Metric-Berry-Curvature-Sublattice-Symmetric}. 
        (a)~$D^1(\bm{k})$ along the high-symmetry path $\Gamma$--$X'$--$M'$--$\Sigma'$--$\Gamma$--$Y'$--$M'$ [as in~\cref{fig:Hofstadter-Energy-Spectrum}], together with $\mathrm{tr}\,g^1$ (blue) and $\abs{\Omega^1_{xy}}$ (red) overlaid, which fluctuate across the FBZ, but coincide at the high-symmetry points ($\Gamma$, $X'$, $Y'$, $M'$ cyan circles in (b)), so the shaded gap $D^1$ (black) stays small. 
        (b)~$D^1(\bm{k})$ depicted over the 2D FBZ.
        The defect is pointwise non-negative and vanishes exactly at the cyan-colored high-symmetry points where the band is locally an ideal LLL while it peaks near the mid-point of $\Gamma$--$X'$ (and its $C_4$ equivalent). 
        Its integral over the entire Brillouin zone equals the global trace-condition $\mathcal{T}_{\mathrm{HH}}$\eqref{eq:trace-defect}.}
        \label{fig:trace-defect}
    \end{center}
\end{figure}
Although the ground band is not geometrically flat, with its Berry curvature and metric fluctuating across the Brillouin zone~[\cref{fig:trace-defect}(b)], it falls short of saturating the global trace condition~\eqref{eq:trace-defect} by only $\approx 8.3\%$ of the topological bound $2\pi\abs{C_1}$~\cite{Onishi2024a,Onishi2024b}, so that the band is ``nearly ideal'' in the integrated sense~\SM.
Note that this defect is also directly measurable: $\mathcal{T}_{\mathrm{HH}}$ controls the small-$\bm q$ static structure factor, $S(\bm{q})\sim K_{\mu\nu}q^\mu q^\nu+\mathcal{O}(q^4)$ through the so-called quantum weight $K_{\mu\nu}=(2\pi)^{-1}\int_{\mathrm{BZ}}g_{\mu\nu}\,d^2k$~\cite{Onishi2024a,Onishi2024b}, which is isotropic here with $K\equiv K_{xx}+K_{yy}=\abs{C_1}+\mathcal{T}_{\mathrm{HH}}/2\pi\approx1.083 > \abs{C_1}=1$, exceeding the topological bound $K\ge\abs{C_1}=1$~\cite{Onishi2024a,Onishi2024b}~(see~\SMref{app:quantum-weight-sf} for details).

This near-ideality is complemented by the flatness of the band itself, a typical requirement for FCI stability, since it appears favorable when interactions dominate the kinetic energy scale~\cite{Bergholtz2013,Tang2011,Neupert2011}
\footnote{Beyond the case where $W \ll U \ll \Delta$, FCIs can also be stabilized in scenarios where the interaction strength $U$ exceeds the band gap $\Delta$~\cite{Wang2011,Sheng2011,Kourtis2014,Motruk2017}.}. 
From the spectrum~\eqref{eq:Energy-Spectrum-Hofstadter-Exact-quarter-Flux-momentum} the exact bandwidth and gap are $W/J=2\sqrt{2}-\sqrt{4+2\sqrt{2}}$ and $\Delta/J=2\sqrt{2-\sqrt{2}}$, giving the exact flatness ratio $f=\Delta/W\approx 7.1097$~\cite{Goldman2015}.
Without fine tuning this is flatter than the Haldane model~\cite{Haldane1988}, which at its optimal fine-tuned point has a flatness ratio of $f\approx 6.0136$~\cite{Leite2021,Neupert2011,Grushin2015}.
Models with even flatter (or exactly flat) bands exist~\cite{Tang2011,Sun2011,Wang2011,Wang2011a,Wang2012a,Hu2011,Neupert2011,Liu2013,Kapit2010,Trescher2012,Yang2012,Mera2021a}, but they rely on (fine-tuned) longer-range hoppings, such as the exponentially decaying couplings of the Kapit-Mueller model~\cite{Kapit2010}, 
or on fine-tuned system sizes at which the Hofstadter bands themselves become exactly flat~\cite{Scaffidi2014}, all of which remain experimentally challenging to realize. 
The quarter-flux Harper-Hofstadter model, by contrast, needs only nearest-neighbor tunneling and has already been realized on several platforms~\cite{Aidelsburger2013,Miyake2013,Aidelsburger2015,Stuhl2015,Hafezi2011,Mancini2015,Tai2017,Roushan2017}. 

\emph{Summary~--}
\phantomsection\label{sec:summary}
We have analytically solved the quarter-flux Harper-Hofstadter model by exploiting its sublattice symmetry that survives in quasimomentum space only for a certain symmetric magnetic unit cell. 
The resulting anti-block-diagonal Bloch Hamiltonian reduces the four-band problem to a single $2\times2$ block, yielding the spectrum and all four eigenstates analytically. 
We then showed that for any such sublattice-symmetric system the quantum geometric tensor decomposes into contributions from the two sublattice sectors, giving us analytical expressions for the Berry curvature and quantum metric of the gapped bands, and their non-Abelian counterparts for the degenerate middle pair.
For the ground band this exposes an energetically flat Chern band that is nearly FCI-ideal, which we quantified via the quantum-geometric trace condition. 
As our QGT decomposition is completely general and holds for any sublattice-symmetric Bloch Hamiltonian, we envision that it can be applied to other multi-band systems, which, provided the off-diagonal block $D$ is small enough, may even be obtained in closed form.

Within the Hofstadter family with $\alpha=p/q$ flux per plaquette this is restrictive as the sublattice symmetry survives in momentum space only for $q$ divisible by four 
(and $q=8$ already requires solving a $4\times4$ block).
Thus, non-trivial $N$-band candidates ($N=d_++d_->2$) are more natural elsewhere. 
For instance, with balanced sublattices ($d_+=d_-$) the BBH quadrupole insulator~\cite{Benalcazar2017,Benalcazar2017a} and the square-octagon lattice~\cite{Pal2018,Nunes2020,Mukherjee2026} are four-site bipartite models, where $D$ is a $2\times2$ block. 
For a majority sublattice ($d_+\neq d_-$), the dice and Lieb lattices~\cite{Sutherland1986,Julku2016} as well as other chiral flat-band networks~\cite{Ramachandran2017} are interesting candidates to study. 
Here, $D$ is rectangular with $\abs{d_+-d_-}$ zero modes, so the QGT decomposition covers the dispersive bands, while the unpaired zero-mode flat sector requires separate treatment~\cite{Mera2022}.
When larger blocks resist closed-form diagonalization, it might also still be interesting to combine the decomposition with eigenprojector methods for $N$-band quantum geometry~\cite{Graf2021}.

Finally, our analytical results might help to understand, why the Harper-Hofstadter model at quarter flux hosts a bosonic Laughlin-like FCI ground state at half filling of the lowest band, even in the limit of infinitely strong on-site interactions~\cite{Motruk2017}.

\emph{Acknowledgments~--}
We thank Luis Steinfadt for helpful discussions and Bruno Mera for the suggestion to look into FCI stability conditions. 
We also acknowledge funding by the Deutsche Forschungsgemeinschaft (DFG, German Research Foundation) via the Research Unit FOR 5688 (Project No. 521530974).  
I.T.~is grateful for support support by the Studienstiftung des deutschen Volkes. 

\emph{Data Availability~--}
No data were created or analyzed in this study. The data presented in the figures can be reproduced from the presented equations in the manuscript. 

\bibliography{main}

\cleardoublepage
\makeatletter
\let\addcontentsline\origaddcontentsline
\makeatother

\beginsupplement
\setcounter{page}{1}

\title{Supplemental Material for ``\titlecontent''}
\maketitle
\phantomsection\label{sec:SuppMat}
\hypersetup{linkcolor=blue} %
\tableofcontents

\bigskip

This Supplemental Material is organized as follows. 
In~\cref{app:Magnetic-Translation-Symmetry-HH-Model} we review the magnetic translation symmetry and the generalized Bloch theorem for the three magnetic-unit-cell choices, specialized to the quarter-flux case.
\cref{app:Chiral-Sublattice-Symmetry} provides more details regarding the sublattice symmetry, establishing the class AIII assignment, and briefly discusses the $C_4$ symmetry of the quarter-flux model.
In~\cref{app:Analytical-Solution} we provide more details on the SVD-based derivation of the exact spectrum, after which in~\cref{app:all-eigenstates-Hofstadter} the four closed-form eigenstates of the symmetric $2a\times 2a$ MUC are derived. 
Then in~\cref{app:quantum-geometry-all-states} we derive the quantum geometry of sublattice-symmetric systems, including the non-Abelian quantum geometric tensor of the degenerate middle bands. \cref{app:fci-stability-conditions} establishes the hierarchy of FCI stability conditions and applies it to the lowest Chern band. Finally, in~\cref{app:quantum-weight-sf} we briefly relate the static structure factor to the quantum weight and its lower bound. 
\medskip

\section{Magnetic translation symmetry in the Harper-Hofstadter model}
\label{app:Magnetic-Translation-Symmetry-HH-Model}
In this section, we briefly review the magnetic translation symmetry~\cite{Zak1964a,Zak1964} of the Harper-Hofstadter model for the specific case of a quarter flux per plaquette. For a more extensive discussion on magnetic translation symmetries, we refer to~\Ccite{Bernevig2013,Aidelsburger2016}.
We consider the Harper-Hofstadter Hamiltonian~\cite{Harper1955,Hofstadter1976} 
in the Landau gauge 
$(\phi^x_{m,n},\phi^y_{m,n})=(-\Phi n,0)$ given by
    \begin{widetext}
        \begin{align}
            \hat{H}&=-J\sum_{m,n} (e^{i\phi^x_{m,n}}\ket{m+1,n}\!\bra{m,n}
            +e^{i\phi^y_{m,n}}\ket{m,n+1}\!\bra{m,n}+\text{h.c.})
            \nonumber \\
            &
            =-J\sum_{m,n} (e^{-i\Phi n}\ket{m+1,n}\!\bra{m,n}
            +\ket{m,n+1}\!\bra{m,n}+\text{h.c.}),
            \label{eq:app:HH-Ham-gen}
        \end{align}
    \end{widetext}
    with a flux per plaquette of $\Phi=2\pi\alpha$ and a hopping amplitude $J$.
    Here $\ket{m,n}$ denotes the single-particle state localized at lattice site $(m,n)$, where $m$ ($n$) denotes the discrete position along the $x$ ($y$) direction of the square lattice of size $M \times N$, with $M$ ($N$) the number of lattice sites along $x$ ($y$). 
    We work in the single-particle sector throughout, so the results are independent of particle statistics.

Then, for periodic boundary conditions (PBC), the Hamiltonian~\eqref{eq:app:HH-Ham-gen} for rational values of flux per plaquette, $\alpha=p/q$ with $p,q \in \mathbb{Z}$ being coprime integers, admits a magnetic translation symmetry with respect to a \emph{magnetic unit cell} (MUC) of size $k\times \ell$ with $k\ell=q\times \mathbb{Z}$~\cite{Zak1964,Zak1964a}.
In other words, one can construct a full set of commuting operators if one can find a region of dimension $k\times \ell$, called \emph{supercell}, which encloses multiple plaquettes whose total flux is an integer multiple of $2\pi$ (magnetic flux quantum).
The magnetic unit cell is then defined as the supercell with the smallest dimension which satisfies this condition,
    \begin{align}
        k\ell\Phi=k\ell 2\pi \alpha=2\pi p \frac{k\ell}{q}\overset{!}{=}2\pi \lambda,\, \lambda\in \mathbb{Z},
        \label{eq:app:SuperCell-IntegerFlux-Cond}
    \end{align}
where the solution is given by $k\ell=q$~\cite{Bernevig2013,Aidelsburger2016}.
Note that only the total area of the magnetic unit cell, $A_{\text{MU}}=(k\times\ell)a^2$ is fixed, where the specific dimension is not. Due to the enlarged magnetic unit cell of size $k \times \ell$ with $k \ell=q$, 
we obtain a reduced first (magnetic) Brillouin zone of size, 
$0 \leq k_x < 2\pi/(ka)$ (or $-\pi/(ka) \leq k_x < \pi/(ka)$) and $0 \leq k_y < 2\pi/(\ell a)$ (or $-\pi/(\ell a) \leq k_y < \pi/(\ell a)$), with $\bm{k}=(k_x,k_y)$ being the quasimomenta of the system and $a$ the lattice spacing constant.
Because of the enlarged magnetic unit cell the allowed quasimomenta for the first (magnetic) Brillouin zone, for a system of size $M\times N$ with PBC, are given by $k_x=2\pi \alpha_x/(M a)$ and $k_y=2\pi \alpha_y/(N a)$, where $\alpha_x \in \{0,1,\ldots,M/k-1\}$ and $\alpha_y \in \{0,1,\ldots,N/\ell-1\}$ are integers and where $k$ ($\ell$) is the dimension of the MUC along $x$ ($y$).
For simplicity, we will assume in the following, that $M$ ($N$) are always integer multiple of $k$ ($\ell$). 

In the case of the Landau gauge specified above, one can find the full set of commuting operators 
$\{\hat{H},\hat{M}^ k_x,\hat{M}^{\ell}_y\}$ where the so-called commuting magnetic translation operators (MTOs) $\hat{M}^{k}_x$ and $\hat{M}^{\ell}_y$ are given by~\cite{Aidelsburger2016,Bernevig2013}
    \begin{align}
          \hat{M}^k_x&=\sum_{m,n} \ket{m+k,n}\!\bra{m,n}, \nonumber  \\
           \hat{M}^{\ell}_y&=\sum_{m,n} e^{-i 2\pi \alpha \ell m}\ket{m,n+ \ell}\!\bra{m,n}.
        \label{eq:app:HH-MTOs}
    \end{align}
Then the simultaneous single-particle eigenstates $\ket{\psi}=\sum_{m,n}\psi_{m,n}\ket{m,n}$
of the set of commuting operators $\{\hat{H},\hat{M}^ k_x,\hat{M}^{\ell}_y\}$ are obtained 
by using the \emph{generalized Bloch theorem} based on the MTOs $\hat{M}^ k_x$ and $\hat{M}^{\ell}_y$ 
\begin{align}
    &\bra{m,n}\hat{M}_x^k\ket{\psi}=e^{i\mu^x_{m,n}(k)}\psi_{m-k,n}=e^{-ik_xka}\psi_{m,n}, \nonumber \\
    &\bra{m,n}\hat{M}_y^{\ell} \ket{\psi}=e^{i\mu^y_{m,n}(\ell)}\psi_{m,n-\ell}=e^{-ik_y\ell a}\psi_{m,n},
    \label{eq:app:Generalized-XY-Bloch-Theorem}
\end{align}
where $\mu^x_{m,n}(k)$ ($\mu^y_{m,n}(\ell)$) denote the phases arising from the application of operators~$\hat{M}_x^k$ ($\hat{M}_y^{\ell}$)~\cite{Aidelsburger2016,Bernevig2013}.
\begin{figure}[tb]
    \begin{center}
        \includegraphics[width=\columnwidth]{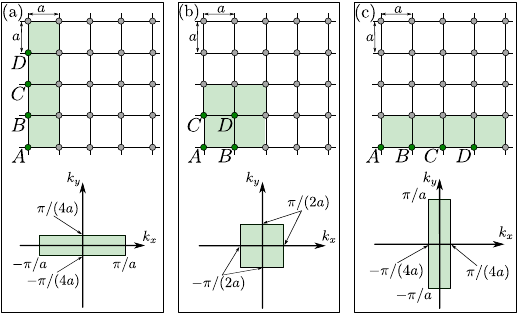} 
        \caption{Admissible magnetic unit cells (MUC) and first magnetic Brillouin zones (FBZ) 
        for a square lattice with $\alpha=1/4$ flux per plaquette.
        Three different possibilities for the choice of the MUC are available. Each MUC contains necessarily $q=4$ lattices sites (dark green circles), labeled by the letters $A,B,C,D$.
        (a) [Choice (i)] Rectangular choice for the magnetic unit cell along $y$ enclosing an area of 
        $A_{\text{MU}}=1a\times 4a$ (green shaded area) with an associated FBZ of size $-\pi/a \leq k_x < \pi/a$ and $-\pi/(4 a) \leq k_y < \pi/(4 a)$ in quasimomentum space.
        (b) [Choice (ii)] Symmetric square-shaped choice for the MUC, $A_{\text{MU}}=2a\times 2a$ leading to a FBZ of size $-\pi/(2a) \leq k_i < \pi/(2a)$ for $i \in \{x,y\}$.
        (c) [Choice (iii)] Rectangular choice for the magnetic unit cell along $x$ covering an area of 
        $A_{\text{MU}}=4a\times 1a$ with a corresponding FBZ of extent $-\pi/(4a) \leq k_x < \pi/(4a)$ and $-\pi/a \leq k_y < \pi/a$.}
        \label{fig:App:MUC}
    \end{center}
\end{figure}

\subsection{Quarter-flux case}
Hence, for the specific case of a quarter flux per plaquette, $\alpha=1/4$, there are three different choices for the magnetic unit cell (MUC), namely (i) $A_{\text{MU}}=ka \times \ell a =1a\times 4a$, (ii) $A_{\text{MU}}=2a\times 2a$ and (iii) $A_{\text{MU}}=4a\times 1a$, as illustrated in~\cref{fig:App:MUC}.
Each MUC consists of four sublattice sites $A,B,C,D$ and four plaquettes enclosing a total area of $A_{\text{MU}}=4a^2$, which is pierced by a total flux of $4\times 2\pi \alpha=2\pi$, as required by the condition above~\eqref{eq:app:SuperCell-IntegerFlux-Cond}.

Using a generalized Bloch theorem based on the magnetic translation operators associated with the MUC~\cite{Bernevig2013,Aidelsburger2016}, we can construct the single-particle quasimomentum Bloch Hamiltonian $H(\bm{k})$ associated to the eigenstates $\ket{\bm{k},\mu} =\sum_{m,n}\phi_{m,n}^{\bm{k},\mu}\ket{m,n}$ for the three different choices of the MUC, which we label as cases (i), (ii) and (iii).
Here, $\bm{k}=(k_x,k_y)$ is the quasimomentum of the system and $\mu \in \{1,2,3,4\}$ denotes band index for the system.
Since each MUC contains four sublattice sites, the Bloch Hamiltonian $H(\bm{k})$ is a $4\times 4$ matrix acting on the so-called Bloch function $\bm{u}_{\mu}(\bm{k})=(u^A_{\mu}(\bm{k}),u^B_{\mu}(\bm{k}),u^C_{\mu}(\bm{k}),u^D_{\mu}(\bm{k}))^T$, with the eigenvalue equation $H(\bm{k})\bm{u}_{\mu}(\bm{k})=E_{\mu}(\bm{k})\bm{u}_{\mu}(\bm{k})$.

For our purposes the case of (ii) is the relevant one because it is the only case for which the sublattice symmetry of the quarter-flux Harper-Hofstadter model manifests itself, which we discuss in more detail now.
In the case (ii) of the symmetric MUC, we have $k=\ell=2$ such that MTOs $\hat{M}^{2}_x$ and $\hat{M}^{2}_y$ of~\eqref{eq:app:HH-MTOs} take the form
    \begin{align}
        \hat{M}^2_x&=\sum_{m,n} \ket{m+2,n}\!\bra{m,n}, \nonumber \\
        \hat{M}^{2}_y&=\sum_{m,n} e^{-i\pi m}\ket{m,n+2}\!\bra{m,n}.
        \label{eq:app:QuarterFlux-Case-II-MTOs}
    \end{align}
The phase factor in $\hat{M}^{2}_y$ somewhat complicates the conditions imposed by the generalized Bloch theorem~\eqref{eq:app:Generalized-XY-Bloch-Theorem}.
We find the single-particle eigenstates of the system in this case, 
$\ket{\psi}=\sum_{m,n}\psi_{m,n} \ket{m,n}$, by proposing an ansatz for the coefficients 
$\psi_{m,n}$, which must satisfy the generalized Bloch theorem~\eqref{eq:app:Generalized-XY-Bloch-Theorem}
    \begin{align}
        &\matel{m,n}{\hat{M}_x^2}{\psi}=\psi_{m-2,n}=e^{-ik_x 2a}\psi_{m,n}, \nonumber \\
        &\matel{m,n}{\hat{M}_y^2}{\psi}=e^{-i\pi m} \psi_{m,n-2}=e^{-ik_y 2a}\psi_{m,n}.
        \label{eq:app:QuarterFlux-Case-II-Bloch-Theorem}
    \end{align}
An Ansatz which satisfies these conditions~\eqref{eq:app:QuarterFlux-Case-II-Bloch-Theorem} is given by~\cite{Aidelsburger2015}
    \begin{align}
        \psi_{m,n}=e^{ik_x ma}e^{ik_y na}  \times
                        \begin{cases}
                            u^A, \, &\text{for $m,n$ even,}\\ 
                            u^B e^{in\pi/2}, \, &\text{for $m$ odd, $n$ even,}\\ 
                            u^C, \, &\text{for $m$ even, $n$ odd,} \\
                            u^D e^{in\pi/2}, \, &\text{for $m,n$ odd,} 
                        \end{cases}
        \label{eq:app:QuarterFlux-Case-II-Ansatz}
    \end{align}
where $u^{i}$ denote the complex amplitudes of the Bloch functions on the sublattice sites 
$i\in\{A,B,C,D\}$ in each MUC, as shown in \cref{fig:App:MUC}(b).
Inserting this ansatz into the eigenvalue equation $H\ket{\psi}=E\ket{\psi}$, we can obtain the Bloch Hamiltonian $H(\bm{k})$ for case (ii) given by
\begin{align}
    &H^{\text{(ii)}}(\bm{k})=\nonumber \\
    &-2J\begin{pmatrix}
        0 & \cos(k_x a) & \cos(k_y a) & 0 \\
        \cos(k_x a) & 0 & 0 & -\sin(k_y a) \\
        \cos(k_y a) & 0 & 0 & -i \sin(k_x a) \\
        0 & -\sin(k_y a) & i \sin(k_x a) & 0
    \end{pmatrix}, \label{eq:app:Bloch-Ham-Case-II}
\end{align}

For completeness, let us also briefly discuss the Bloch Hamiltonians $H(\bm{k})$ for case (i) and case (iii).
In the commonly studied case of vertically aligned rectangular MUC (i) with $A_{\text{MU}}=1a\times 4a$ [\cref{fig:App:MUC}(a)] there are no phase factors in either of the MTOs $\hat{M}_x^1$ and $\hat{M}_y^{4}$. 
This simplifies the conditions imposed by the generalized Bloch theorem~\eqref{eq:app:Generalized-XY-Bloch-Theorem} such that a simple plane-wave ansatz, $\psi_{m,n}=e^{ik_x ma}e^{ik_y na} u^n$ with $u^{n-4}=u^n$, suffices for the coefficients of the single-particle eigenstate in this case~\cite{Aidelsburger2016,Bernevig2013}. 
It follows that the Bloch Hamiltonian for case (i) is then given by
\begin{align}
    H^{\text{(i)}}(\bm{k})&=-J\begin{pmatrix}
        h_1 & e^{i k_y a} & 0 & e^{-i k_y a} \\
        e^{-i k_y a} & h_2 & e^{i k_y a} & 0 \\
        0 & e^{-i k_y a} & h_3 & e^{i k_y a} \\
        e^{i k_y a} & 0 & e^{-i k_y a} & h_4
    \end{pmatrix}, \label{eq:app:Bloch-Ham-Case-I}
\end{align}
where $h_j=2\cos(k_x a + (\pi/2)j)$ for $j \in \{1,2,3,4\}$.

For case (iii), we have that $\hat{M}_y^1$ carries a phase factor of $e^{-i\pi m/2}$, so that the ansatz, 
$\psi_{m,n}=e^{ik_x ma}e^{ik_y na}e^{-i \pi mn/2} u^m$ with $u^{m-4}=u^m$, leads to the following Bloch Hamiltonian
    \begin{align}
        H^{\text{(iii)}}(\bm{k})&=-J\begin{pmatrix}
            g_1 & e^{i k_x a} & 0 & e^{-i k_x a} \\
            e^{-i k_x a} & g_2 &  e^{i k_x a} & 0 \\
            0 & e^{-i k_x a} & g_3 & e^{i k_x a} \\
            e^{i k_x a} & 0 & e^{-i k_x a} & g_4
        \end{pmatrix} \label{eq:app:Bloch-Ham-Case-III}
    \end{align}
where $g_j=2\cos(k_y a - (\pi/2)j)$ for $j \in \{1,2,3,4\}$.

We stress that the three MUC choices describe the same physics and do \emph{not} correspond to different gauges (e.g. Landau vs symmetric gauge). 
Choosing a MUC is choosing how to group the lattice into supercells for the generalized Bloch theorem~\eqref{eq:app:Generalized-XY-Bloch-Theorem}, a labeling choice made independently of the gauge, since the admissibility condition of integer flux through the supercell~\eqref{eq:app:SuperCell-IntegerFlux-Cond} is gauge invariant. 
Fixing the single Landau-gauge Hamiltonian~\eqref{eq:app:HH-Ham-gen}, all three MUCs supply commuting MTOs for that \emph{one} $\hat{H}$~\eqref{eq:app:HH-Ham-gen}, so the three $4\times4$ Bloch Hamiltonians are the same operator written in three magnetic-Bloch bases, with eigenstates that are the same physical states relabeled. 

\section{Chiral and sublattice symmetry}
\label{app:Chiral-Sublattice-Symmetry}
Here, we collect some general properties of finite-dimensional chiral-symmetric systems that are used in the main text, after which we discuss their application to the three MUC choices of the quarter-flux Harper-Hofstadter model. 
Since the two terms, \emph{chiral} and \emph{sublattice} symmetry, are often used interchangeably, we clarify the distinction in the following. 

A \emph{chiral symmetry} is a unitary involution $\hat{S}$ (with $\hat{S}^2=\mathbbm{1}$, and hence $\hat{S}=\hat{S}^{-1}=\hat{S}^\dagger$) that anti-commutes with the Hamiltonian, $\{\hat{H},\hat{S}\}=0$~\cite{Sutherland1986,Lieb1989,Ludwig2015,Ramachandran2017,Yu2017,Xiao2024,Theel2025,Nicolau2026},
\begin{align}
    \hat{S}^2=\mathbbm{1}, \quad \{\hat{H},\hat{S}\}=0.
    \label{eq:app:chiral-symmetry-def}
\end{align}
In the sense of the tenfold way~\cite{Altland1997} only these two properties~\eqref{eq:app:chiral-symmetry-def}, if present, place a Hamiltonian in class AIII for instance. 

Now, a \emph{sublattice symmetry} is the stronger requirement that the involution is also a \emph{grading} of the lattice, i.e., that $\hat{S}$ be diagonal in the local site basis $\ket{\tau}$ with entries $\pm1$,
\begin{align}
    \hat{S}=\sum_{\tau}s_\tau\projector{\tau}, \qquad s_\tau=\pm1,
    \label{eq:app:grading-definition}
\end{align}
so that each site carries a definite sign and the eigenspaces $\mathcal{H}_\pm$ are spanned by sets of sites rather by arbitrary superpositions of them. 
Every grading is a chiral operator, while the converse is not true. 
We stress that the grading, and not chirality alone, is what the following sections use. It gives the off-diagonal block $D$ introduced below the meaning of an inter-sublattice hopping matrix, and it makes a decomposition into the $\mathcal{H}_\pm$ sectors a decomposition by sublattice sites.

For completeness, below we list some general properties of \emph{chirally}-symmetric systems, which are used throughout this work, which all follow the two properties~\eqref{eq:app:chiral-symmetry-def} alone.

    \emph{Eigenspace splitting.} As a unitary involution $\hat{S}$ has eigenvalues $\pm1$, so the Hilbert space splits into its eigenspaces, $\mathcal{H}=\mathcal{H}_+\oplus\mathcal{H}_-$ with $d_\pm=\dim\mathcal{H}_\pm$ so that in its eigenbasis $\hat{S}=\mathrm{diag}(\mathbbm{1}_{d_+},-\mathbbm{1}_{d_-})$.

    \emph{Anti-block-diagonal form.} In the $\hat{S}$-eigenbasis the Hamiltonian is purely off-diagonal, $\hat{H}=\begin{psmallmatrix}0 & D\\ D^\dagger & 0\end{psmallmatrix}$ with $D:\mathcal{H}_-\to\mathcal{H}_+$ and vice versa $D^\dagger:\mathcal{H}_+\to\mathcal{H}_-$. 
    The diagonal blocks are forced to vanish by $\{\hat{H},\hat{S}\}=0$ together with $\hat{S}=\mathrm{diag}(\mathbbm{1},-\mathbbm{1})$.

    \emph{Spectral symmetry.} The spectrum is symmetric about zero. That means if $\hat{H}\ket{\psi_E}=E\ket{\psi_E}$ then $\hat{H}(\hat{S}\ket{\psi_E})=-E(\hat{S}\ket{\psi_E})$, such that $\hat{S}\ket{\psi_E}$ is an eigenstate with energy $-E$. 

    \emph{Singular-value spectrum.} Squaring the Hamiltonian gives $\hat{H}^2=\mathrm{diag}(DD^\dagger, D^\dagger D)$, whose non-zero eigenvalues are the squared singular values $\sigma_j^2$ of $D$. 
    Hence, the non-zero energies of $\hat H$ come in chiral pairs $E_{\pm,j}=\pm\sigma_j$ with $j=1,\dots,r$ and $r=\mathrm{rank}(D)$. This follows from the SVD, $D=U\Sigma V^\dagger=\sum_{j=1}^r\sigma_j \ketbra{u_j}{v_j}$ and the fact that $DD^\dagger$ and $D^\dagger D$ share their non-zero spectrum~\cite{Horn2012}.

    \emph{Paired eigenstates.} The eigenstates of a pair $\pm\sigma_j$ always take the form of~\eqref{eq:Sublattice-Symmetric-Eigenstates} 
    \begin{align}
        \ket{\Psi_{\epsilon,j}}=\frac{1}{\sqrt2} \begin{pmatrix}
             \ket{u_j} \\[.1cm]
            \epsilon \ket{v_j}
        \end{pmatrix}
        \label{eq:app:Sublattice-Symmetric-Eigenstates}
    \end{align}
    with $\epsilon=\pm1$, where $\ket{u_j}\in\mathcal{H}_+$ and $\ket{v_j}\in\mathcal{H}_-$ are the left and right singular vectors of $D$, respectively. It follows from inserting the SVD relations $D\ket{v_j}=\sigma_j\ket{u_j}$ and $D^\dagger\ket{u_j}=\sigma_j\ket{v_j}$ into the energy eigenvalue equation $\hat {H}\ket{\Psi_{\epsilon,j}}=\epsilon\sigma_j\ket{\Psi_{\epsilon,j}}$ (in the chiral basis).

    \emph{Protected zero modes.} The dimension of the kernel of $\hat H$ (the zero-energy subspace) is $d_0=\dim\ker\hat{H}=|d_+-d_-|+2(\min(d_+,d_-)-r)\ge|d_+-d_-|$, and the zero modes $\ket{\psi_0}$ can be chosen within a single chiral sector, i.e., $\ket{\psi_0}\in\ker D^\dagger\subset\mathcal{H}_+$ or $\ker D\subset\mathcal{H}_-$, following from the rank-nullity theorem applied to $D$~\cite{Sutherland1986,Lieb1989,Ramachandran2017,Nicolau2026}. Whenever $\hat{S}$ is a (grading) sublattice symmetry~\eqref{eq:app:grading-definition}, these sectors are sublattices and the zero modes are supported on one sublattice only.

\subsection{Sublattice symmetry, class A/AIII, and $C_4$ in the quarter-flux Harper-Hofstadter model}
\label{app:subsec:Sublattice-Symmetry-QuarterFlux}
We now apply these properties to the quarter-flux Harper-Hofstadter model, where the bipartite checkerboard sign $s_{m,n}=(-1)^{m+n}$ defines the real-space chiral operator $\hat{S}=\sum_{m,n}s_{m,n}\projector{m,n}$.
The Hamiltonian~\eqref{eq:app:HH-Ham-gen} anti-commutes with $\hat{S}$, since the square lattice is bipartite with neither diagonal couplings nor on-site potentials, and in real space $\hat{S}$ is manifestly a grading~\eqref{eq:app:grading-definition}, being diagonal in the site basis with entries $s_{m,n}=\pm1$.
The two notions coincide there, so that the question is whether this coincidence survives in quasimomentum space.

When $\hat{S}$ commutes with the magnetic translations of the chosen cell, the two can be simultaneously diagonalized, so that $\hat{S}$ descends to a $\bm{k}$-independent diagonal matrix in the cell basis $\ket{\bm{k},\tau}$ and $\{H(\bm{k}),S\}=0$ at fixed $\bm{k}$.
This holds for the symmetric cell (ii), where $[\hat{S},\hat{M}^2_x]=[\hat{S},\hat{M}^2_y]=0$.
For the rectangular cells the single-site translations instead anti-commute, $\{\hat{S},\hat{M}^1_x\}=0$ for (i) and $\{\hat{S},\hat{M}^1_y\}=0$ for (iii), thereby exchanging the even-odd sublattice labels~[\cref{fig:Hofstadter-Energy-Spectrum}(a)-(b)].
The operators themselves are collected in~\cref{app:Magnetic-Translation-Symmetry-HH-Model}, and the two cases are the class-I versus class-II distinction mentioned in~\Ccite{Xiao2024}.

The consequence for the Bloch Hamiltonians follows from splitting the checkerboard sign into its two factors, $s_{m,n}=(-1)^m(-1)^n$.
In the symmetric cell both factors act internally, so that $\hat{S}$ descends to the $\bm{k}$-independent grading $S^{\text{(ii)}}=\mathrm{diag}(1,-1,-1,1)$ with
\begin{align}
    S^{\text{(ii)}} H^{\text{(ii)}}(\bm{k})\,S^{\text{(ii)}} = -H^{\text{(ii)}}(\bm{k}),
    \label{eq:app:chiral-caseII}
\end{align}
so the spectrum is reflected about zero energy at each \emph{fixed} $\bm{k}$. 
For the rectangular cells only $(-1)^n$ stays internal, becoming $-S_\pi$, while $(-1)^m=e^{i\pi m}$ acts, after the partial Fourier transform along $x$, as a momentum shift $k_x\to k_x+\pi/a$. The chiral relation therefore acquires that shift,
\begin{align}
    S_\pi  H^{\text{(i)}}(k_x,k_y)\,S_\pi &= -H^{\text{(i)}}(k_x+\pi/a,\;k_y),
    \label{eq:app:chiral-shift-caseI}
\end{align}
where $S_\pi=\mathrm{diag}(1,-1,1,-1)$ written in the sublattice site basis of the case-(i) ansatz~\eqref{eq:app:Bloch-Ham-Case-I}, and analogously $S_\pi H^{\text{(iii)}}(k_x,k_y)S_\pi=-H^{\text{(iii)}}(k_x,k_y+\pi/a)$ for case (iii).
The energies~\eqref{eq:app:Energy-Spectrum-Hofstadter-Exact-quarter-Flux-momentum} then satisfy $E(k_x,k_y)=-E(k_x+\pi/a,k_y)$ in case (i) and $E(k_x,k_y)=-E(k_x,k_y+\pi/a)$ in case (iii), so the chiral partner of a state at $\bm{k}$ is displaced by half a reciprocal lattice vector.

A diagonal $S=\mathrm{diag}(s_j)$ would give $\{H,S\}_{jj}=2s_jH_{jj}$, which vanishes only if every diagonal entry does, while $H^{\text{(i)}}(\bm{k})$~\eqref{eq:app:Bloch-Ham-Case-I} carries the on-site terms $-Jh_j$ with $h_j=2\cos(k_xa+(\pi/2)j)$. 
Note that these $h_j$ are $\bm{k}$-dependent, antiperiodic kinetic terms ($h_{j+2}=-h_j$), not a static on-site potential, and are fully compatible with chirality, as we will see below. 
Contrary to this, a global chemical potential $\mu\mathbbm{1}$ is a different matter, as required to ensure quarter-filling for instance, since $\mathbbm{1}$ is chiral-even, $\{H-\mu\mathbbm{1},S\}=-2\mu S\neq0$ for \emph{every} chiral involution $S$, so a nonzero $\mu$ breaks the chiral symmetry and moves the model from class AIII to class A, as discussed in the main text. 
The lowest-band Chern number is accordingly an ordinary class-A invariant, a property of the isolated band protected by its gap. 
Thus, for cases (i) and (iii) a $\bm{k}$-independent chiral operator nevertheless exists,
\begin{align}
    S^{\text{(i)}}=S^{\text{(iii)}}=\begin{pmatrix}
        0 & 0 & 1 & 0 \\
        0 & 0 & 0 & -1 \\
        1 & 0 & 0 & 0 \\
        0 & -1 & 0 & 0
    \end{pmatrix},
    \label{eq:app:chiral-op-caseI}
\end{align}
which is Hermitian and unitary with $(S^{\text{(i)}})^2=\mathbbm{1}$ and $\{H^{\text{(i)}}(\bm{k}),S^{\text{(i)}}\}=0$ at fixed $\bm{k}$, and likewise for case (iii) with~\eqref{eq:app:Bloch-Ham-Case-III}.
By being off-diagonal it is only a chiral operator mapping site $j$ to $j+2$. 
There is therefore no contradiction with $\{\hat{S},\hat{M}^1_x\}=0$, since $S^{\text{(i)}}$ is the Bloch representative not of $\hat{S}$ but of $\hat{S}'=\hat{S}\hat{M}^2_y$, which still anti-commutes with $\hat{H}$, i.e., $\{\hat{H},\hat{S}'\}=0$. 
However, $\hat{S}'$ now commutes with both magnetic translations of cell (i), $[\hat{S}',\hat{M}^1_x]=[\hat{S}',\hat{M}^4_y]=0$, because $\hat{M}^2_y$ anti-commutes with $\hat{M}^1_x$ just as $\hat{S}$ does. 

The three magnetic-Bloch bases are related by $\bm{k}$-independent unitaries, which are given by
\begin{align}
    W H^{\text{(i)}}(\bm{k})\,W^{\dagger}=H^{\text{(ii)}}(\bm{k}),
    \quad
    W=\frac{1}{\sqrt{2}}\begin{pmatrix}
        0 & 1 & 0 & 1 \\
        0 & -1 & 0 & 1 \\
        1 & 0 & 1 & 0 \\
        -i & 0 & i & 0
    \end{pmatrix},
    \label{eq:app:MUC-intertwiner-W}
\end{align}
with columns labeled by the sublattice sites $j=1,\ldots,4$ of cell (i) and rows by $\{A,B,C,D\}$ of cell (ii). 

Note that the $\pm1$ eigenvectors of $S^{\text{(i)}}$ are the superpositions $(\ket{1}\pm\ket{3})/\sqrt{2}$ and $(\ket{2}\mp\ket{4})/\sqrt{2}$ of the cell-(i) sites. By rotating $H^{\text{(i)}}$ and $H^{\text{(iii)}}$ into this $S^{\text{(i)}}$-eigenbases we can also bring them to the anti-block-diagonal form $\big(\begin{smallmatrix}0 & D\\ D^\dagger & 0\end{smallmatrix}\big)$ as well, with the off-diagonal blocks
\begin{align}
    D^{\text{(i)}}_{\bm{k}}=\begin{pmatrix}
        2J\sin(k_x a) & -2J\cos(k_y a) \\
        2iJ\sin(k_y a) & 2J\cos(k_x a)
    \end{pmatrix},
    \\
    D^{\text{(iii)}}_{\bm{k}}=\begin{pmatrix}
        -2J\sin(k_y a) & -2J\cos(k_x a) \\
        2iJ\sin(k_x a) & 2J\cos(k_y a)
    \end{pmatrix}.
    \label{eq:app:D-caseI-caseIII}
\end{align}
These are \emph{not} equal to $D^{\text{(ii)}}_{\bm{k}}$~\eqref{eq:app:D-matrix}, but they share its singular values: in all three cases $\mathrm{tr}(D^\dagger D)=8J^2$ and $\det(D^\dagger D)=2J^4\,[\,2-\cos(4k_x a)-\cos(4k_y a)\,]$, so that they reproduce the same spectrum~\eqref{eq:app:Energy-Spectrum-Hofstadter-Exact-quarter-Flux-momentum} as expected. 
What our exact solution relies on is the \emph{grading}, namely, only in case (ii) does the anti-block form live on the physical sublattice basis. 
There, $D^{\text{(ii)}}_{\bm{k}}$ is a genuine inter-sublattice hopping matrix, its singular vectors are site-basis Bloch functions~\eqref{eq:app:Sublattice-Symmetric-Eigenstates}, and the sublattice decomposition of the quantum geometry becomes a decomposition into the two \emph{physical} sublattices. 
The abstract superposition sectors of cases (i) and (iii) provide none of this, even though their quantum geometry itself is identical in all three MUC descriptions, as it must be for a gauge-invariant observable.
To summarize, the grading of case (ii) provides two things instead.
First, it gives the closed-form eigenstates~\eqref{eq:app:Sublattice-Symmetric-Eigenstates}, from which the quantum geometry follows analytically. 
Second, it decomposes that geometry into the two sublattice sectors $\mathcal{H}_{\pm}$, which is physically meaningful because those sectors are spanned by lattice sites~(see~\cref{app:quantum-geometry-all-states}).

With time-reversal and particle-hole symmetry absent and a $\bm{k}$-independent chiral operator present, all three descriptions of $H(\bm{k})$ belong to the chiral class AIII of the tenfold Altland-Zirnbauer classification~\cite{Zirnbauer1996,Altland1997}, as they must, being three magnetic-Bloch bases of one and the same Hamiltonian which dictates that in $d=2$ dimension, at half-filling at zero energy, there is no topological invariant associated to the band below that gap~\cite{Ryu2010}.
This is consistent with the fact that the HH-model at quarter-flux does not feature any isolated band at half-filling as it is gapless at zero energy there. 
Introducing a chemical potential $\mu\neq0$ to ensure quarter-filled lowest band breaks the chiral symmetry and moves the model to class A, where the lowest band is isolated and carries a nonzero Chern number $C_1=-1$ consistent with AZ-class A in $d=2$. 

\phantomsection\label{app:subsec:PointGroup-Degeneracy}
Let us briefly discuss the point-group symmetry of the square lattice in the presence of the flux.
The bare rotation $\mathcal{P}=\sum_{m,n}\ket{-n,m}\bra{m,n}$ maps the Landau gauge along $x$ into the Landau gauge along $y$, and is therefore not a symmetry by itself.
The symmetry is the composite $\mathcal{C}_4=\mathcal{G}\mathcal{P}$ with the compensating gauge phase $\chi_{m,n}=-\Phi mn$ (assuming for simplicity a linear system size of a multiple of $q=4$). 
Restricting $e^{-i\Phi mn}$ to the four sites of the symmetric cell gives $\mathrm{diag}(1,1,1,i)$, since $mn$ is odd only at $D=(a,a)$, while $\mathcal{P}$ acts within the cell as the transposition $\Pi_{BC}$, so that we have 
\begin{align}
   R^{\text{(ii)}}&=
   \begin{pmatrix}
        1 & 0 & 0 & 0 \\
        0 & 0 & 1 & 0 \\
        0 & 1 & 0 & 0 \\
        0 & 0 & 0 & i
    \end{pmatrix},
    \nonumber \\
    R^{\text{(ii)}} H^{\text{(ii)}}(k_x,k_y) R^{\text{(ii)}\dagger}&=H^{\text{(ii)}}(-k_y,k_x),
    \label{eq:app:C4-operator-caseII}
\end{align}
with $(R^{\text{(ii)}})^4=\mathbbm{1}$, so that $C_4$ symmetry is represented linearly. 
The rotation commutes with the sublattice operator, $[R^{\text{(ii)}},S^{\text{(ii)}}]=0$, so that $C_4$ acts as an ordinary crystalline symmetry on top of the chiral class AIII or class A (if a chemical potential is added). 
We caution that, in analogy with the chiral operator discussed above, the existence of such a constant $R$ does \emph{not} single out the symmetric cell. 
Namely, the intertwining condition~\eqref{eq:app:MUC-intertwiner-W} allows for a similar construction also  for the MUCs associated to cases (i) and (iii). 
Since $W H^{\text{(iii)}}(k_x,k_y)W^{\dagger}=H^{\text{(ii)}}(-k_y,k_x)$, the unitary $W^{\text{(iii)}}=R^{\text{(ii)}\dagger}W$ relates case (iii) to case (ii) at the same quasimomentum. 

We mention in passing that the $C_4$ symmetry endows the Chern bands with a concrete, quantized secondary invariant, the \emph{discrete shift} $\mathcal{S}$ as detailed in~\cite{Manjunath2021,Manjunath2020}.
The discrete shift is the lattice analog of the Wen-Zee shift, defined for $U(1)$-charge-conserving fermions with $C_M$ rotational symmetry and valued in $\mathbb{Z}_M$.
For $C_4$ ($M=4$) and spinless fermions the crystalline response theory fixes $\mathcal{S}\equiv C/2 \pmod 1$, so that a band with \emph{odd} Chern number necessarily carries a \emph{half-integer} shift~\cite{Zhang2022c,Zhang2025b}, so $\mathcal{S}=-1/2 \pmod 1$ for the lowest band of the quarter-flux Harper-Hofstadter model here. 
Physically, $\mathcal{S}$ governs a curvature/defect response distinct from the chiral edge mode: a $\pi/2$ disclination centered on a $C_4$-invariant site binds a fractional charge $\bar{Q}=\mathcal{S}/4=1/8 \pmod 1$ on top of the background filling $\nu_0=1/4$~\cite{Zhang2022c,Zhang2025b}.

\section{Details on analytical solution of the quarter-flux Harper-Hofstadter model}
\label{app:Analytical-Solution}
We collect some of the algebraic details behind the exact solution summarized in the main text, which will also be used later for the derivation of the eigenstates~(Sec.~\ref{app:all-eigenstates-Hofstadter}) and the exact quantum geometry~(Sec.~\ref{app:quantum-geometry-all-states}).
For the symmetric $2a\times2a$ MUC (ii), the bipartite checkerboard sign $s_{m,n}=(-1)^{m+n}$~(see~\cref{app:Chiral-Sublattice-Symmetry}) assigns the sites $A,D$ to the even sector $\mathcal{H}_+=\text{span}\{\ket{A},\ket{D}\}$ and $B,C$ to the odd sector $\mathcal{H}_-=\text{span}\{\ket{B},\ket{C}\}$, so that the sublattice operator in its eigenbasis is $S=\mathrm{diag}(\mathbbm{1}_2,-\mathbbm{1}_2)$.
Rotating $H^{\text{(ii)}}(\bm{k})$~\eqref{eq:Bloch-Ham-Case-II} into the eigenbasis $\{\ket{A},\ket{D},\ket{B},\ket{C}\}$ by the permutation $Q$ (with $Q\ket{A}=\ket{A}$, $Q\ket{B}=\ket{D}$, $Q\ket{C}=\ket{B}$, $Q\ket{D}=\ket{C}$) brings it to the anti-block-diagonal form enforced by the sublattice symmetry,
    \begin{align}
        H^{\text{(ii)}}_S(\bm{k})=Q^{\dagger} H^{\text{(ii)}}(\bm{k}) Q
        =\begin{pmatrix}
            0 & D_{\bm{k}} \\
            D^{\dagger}_{\bm{k}} & 0
        \end{pmatrix},
        \label{eq:app:Bloch-Ham-Case-II-Sublattice-Basis}
    \end{align}
with the $2\times 2$ block $D_{\bm{k}}:\mathcal{H}_- \to \mathcal{H}_+$ given by
    \begin{align}
        D_{\bm{k}}=-2J\begin{pmatrix}
            \cos(k_x a) & \cos(k_y a) \\
            -\sin(k_y a) & i \sin(k_x a)
        \end{pmatrix},
        \label{eq:app:D-matrix}
    \end{align}
as~\eqref{eq:D-matrix} in the main text.
Squaring $H^{\text{(ii)}}_S$~\eqref{eq:app:Bloch-Ham-Case-II-Sublattice-Basis} block-diagonalizes it into the two Hermitian $2\times2$ blocks $\tilde{M}_{\bm{k}}=D_{\bm{k}}D^{\dagger}_{\bm{k}}$ (acting on $\mathcal{H}_+$) and $M_{\bm{k}}=D^{\dagger}_{\bm{k}}D_{\bm{k}}$ (acting on $\mathcal{H}_-$),
    \begin{align}
        (H^{\text{(ii)}}_S(\bm{k}))^2=\begin{pmatrix}
            \tilde{M}_{\bm{k}} & 0 \\
            0 & M_{\bm{k}}
        \end{pmatrix}.
        \label{eq:app:Bloch-Ham-Case-II-Squared}
    \end{align}
Both admit a Pauli decomposition $M_{\bm{k}}=h_0\sigma_0+\bm{h}\cdot\bm{\sigma}$ and $\tilde{M}_{\bm{k}}=h_0\sigma_0+\tilde{\bm{h}}\cdot\bm{\sigma}$, with $h_\mu=\text{Tr}[M_{\bm{k}}\sigma_\mu]/2$, $\tilde{h}_\mu=\text{Tr}[\tilde{M}_{\bm{k}}\sigma_\mu]/2$, $h_0=\tilde{h}_0=4J^2$, (dropping the explicit $\bm{k}$-dependence from now on) and the
Bloch vectors
    \begin{align}
        \bm{h}=4J^2\begin{pmatrix}
            c_{x} c_{y} \\
            s_{x} s_{y} \\
            \tfrac{1}{2}(c_{2x}-c_{2y})
        \end{pmatrix}, \quad
        \tilde{\bm{h}}=4J^2\begin{pmatrix}
            -c_{x} s_{y} \\
            s_{x} c_{y} \\
            \tfrac{1}{2}(c_{2x}+c_{2y})
        \end{pmatrix},
        \label{eq:app:Bloch-Vectors-M-Mtilde}
    \end{align}
with the same shorthand $c_{i\nu}=\cos(i k_\nu a)$, $s_{i\nu}=\sin(i k_\nu a)$ ($\nu\in\{x,y\}$, $i\in\mathbb{N}^+$) as in the main text. 
The two Bloch vectors have equal magnitude, $h\equiv\abs{\bm{h}}=\abs{\tilde{\bm{h}}}\equiv\tilde{h}=J^2\sqrt{2Z}$ with $Z=6+c_{4x}+c_{4y}$, reflecting that $M$ and $\tilde{M}$ share their non-zero eigenvalues~\cite{Horn2012},
    \begin{align}
        \lambda_{j}&=h_0+\eta_j h,
        \nonumber \\ \lambda_{1,2}/J^2&= 4\pm \sqrt{2}\sqrt{6+c_{4x}+c_{4y}},
        \label{eq:app:Eigenvalues-DdaggerD-matrix}
    \end{align}
with the eigenvalue selector $\eta_j=(-1)^{j+1}\in\{+1,-1\}$ (not to be confused with the energy sign $\epsilon=\pm1$ of the chirally symmetric pairs).
These are the squared singular values $\lambda_j=\sigma_j^2$ of the singular value decomposition of $D_{\bm{k}}$,
    \begin{align}
        D_{\bm{k}}=U_{\bm{k}} \Sigma_{\bm{k}} V^{\dagger}_{\bm{k}}=\sum^{r}_{j=1}\sigma_j \ket{u_j}\bra{v_j},
        \label{eq:app:SVD-D-matrix}
    \end{align}
with rank $r=\mathrm{rank}(D_{\bm{k}})=2$ and $\sigma_j\equiv E_{+,j}>0$, from which
    \begin{align}
        M&=D^\dagger D=V\Sigma^\dagger \Sigma V^{\dagger} = \sum^{r}_{j=1}\sigma^2_j \ket{v_j}\bra{v_j}
        , \nonumber \\
         \tilde{M}&=DD^\dagger=U \Sigma \Sigma^\dagger U^\dagger = \sum^{r}_{j=1}\sigma^2_j \ket{u_j}\bra{u_j}. 
        \label{eq:app:SVD-D-matrix-M-Mtilde}
    \end{align}
Hence the right and left singular vectors $\ket{v_j}$ and $\ket{u_j}$ are simultaneously the eigenvectors of $M$ and $\tilde{M}$ with eigenvalue $\lambda_j=\sigma_j^2$.
Taking square roots, the four bands $E_{\pm,j}=\pm\sqrt{\lambda_j}$ read
    \begin{align}
        E_{\pm,1}(\bm{k})/J=\pm \sqrt{4+\sqrt{2}\sqrt{6+\cos(4 k_x a)+\cos(4 k_y a)}}, \nonumber \\
        E_{\pm,2}(\bm{k})/J=\pm \sqrt{4-\sqrt{2}\sqrt{6+\cos(4 k_x a)+\cos(4 k_y a)}},
        \label{eq:app:Energy-Spectrum-Hofstadter-Exact-quarter-Flux-momentum}
    \end{align}
reproducing the main-text spectrum~\eqref{eq:Energy-Spectrum-Hofstadter-Exact-quarter-Flux-momentum}.
\section{Details on the eigenstates of the quarter-flux Harper-Hofstadter model for the symmetric $2a\times 2a$ magnetic unit cell (MUC)}
\label{app:all-eigenstates-Hofstadter}
Having obtained the exact energy spectrum~\eqref{eq:app:Energy-Spectrum-Hofstadter-Exact-quarter-Flux-momentum}, we now construct all four eigenstates $\ket{\Psi_{\epsilon,j}}$ of $H_S(\bm{k})$~\eqref{eq:app:Bloch-Ham-Case-II-Sublattice-Basis} in closed form.
Recall from~Sec.~\ref{app:Analytical-Solution} and~\eqref{eq:app:SVD-D-matrix-M-Mtilde} that the eigenvectors $\ket{v_j}$ of $M$ corresponding to the eigenvalues $\lambda_j$ are simultaneously the right singular vectors of $D$~\eqref{eq:app:SVD-D-matrix}.
Expressed directly through the Bloch-vector components in the (north-pole) gauge, $\ket{v_j}$
reads~\cite{Chiu2016,Nakahara2003}
 \begin{align}
    \ket{v_{j}}=\frac{1}{\sqrt{2h(h+\eta_j h_z)}}
    \begin{pmatrix}
        h+\eta_j h_z \\[.1cm]
        \eta_j(h_x+i h_y)
    \end{pmatrix},
    \label{eq:app:Right-Singular-Vectors-D-matrix-Pauli}
 \end{align}
which is the gauge used throughout the quantum-geometry analysis as well.
Equivalently, parametrizing the unit vector $\hat{\bm{h}}=\bm{h}/h=(\cos\varphi_{\bm{k}}\sin\theta_{\bm{k}},\sin\varphi_{\bm{k}}\sin\theta_{\bm{k}},\cos\theta_{\bm{k}})^T$ by its polar and azimuthal angles $(\theta_{\bm{k}},\varphi_{\bm{k}})$, the same states take the Bloch-sphere form $\ket{v_1}=\cos(\theta_{\bm{k}}/2)\ket{A}+\sin(\theta_{\bm{k}}/2)e^{i \varphi_{\bm{k}}}\ket{B}$ and $\ket{v_2}=\sin(\theta_{\bm{k}}/2)\ket{A}-\cos(\theta_{\bm{k}}/2)e^{i \varphi_{\bm{k}}}\ket{B}$, located at $\pm \hat{\bm{h}}$ on the Bloch sphere, where $\ket{A}$ ($\ket{B}$) denotes the north (south) pole.

The left singular vectors then follow from the SVD relation $\ket{u_{j}}=D_{\bm{k}}\ket{v_j}/\sigma_j$~\eqref{eq:app:SVD-D-matrix}. Using $D_{\bm{k}}=-2J\left(\begin{smallmatrix} c_x & c_y\\ -s_y & i s_x\end{smallmatrix}\right)$~\eqref{eq:app:D-matrix}, this yields
\begin{widetext}
    \begin{align}
        \ket{u_{j}}=\frac{-2J}{\sigma_j\sqrt{2h(h+\eta_j h_z)}}\begin{pmatrix}
            c_x(h+\eta_j h_z)+\eta_j c_y(h_x+i h_y) \\[.1cm]
            -s_y(h+\eta_j h_z)+i\eta_j s_x(h_x+i h_y)
        \end{pmatrix}.
        \label{eq:app:Left-Singular-Vectors-D-matrix-Pauli}
    \end{align}

We emphasize that $\ket{u_j}$ must be taken from this SVD relation, and \emph{not} by independently gauge-fixing an eigenvector $\ket{u_j}$ of $\tilde{M}$~\eqref{eq:app:SVD-D-matrix-M-Mtilde}. Although $\ket{u_j}$ is such an eigenvector (with Bloch vector $\tilde{\bm{h}}$, see~Sec.~\ref{app:Analytical-Solution}), only the relative $u$/$v$ phase,  correctly fixed by $D\ket{v_j}=\sigma_j\ket{u_j}$, makes $(\ket{u_j},\epsilon\ket{v_j})^T$ an eigenstate of $H_S$. 
An independent gauge choice would break this fixed phase relationship and differ by a $\bm{k}$-dependent phase. 
This is the same phase-locking mechanism that renders the Berry-connection difference $\Delta A^j_\mu$~\eqref{eq:app:Connection-Differences-Exact-Same-Gauge-Expression} gauge invariant in the following quantum-geometry analysis of~\cref{app:quantum-geometry-all-states}. 

With $\ket{v_j}$~\eqref{eq:app:Right-Singular-Vectors-D-matrix-Pauli} and $\ket{u_j}$~\eqref{eq:app:Left-Singular-Vectors-D-matrix-Pauli} at hand, the four eigenstates $\ket{\Psi_{\epsilon,j}}$ of $H_S$~\eqref{eq:app:Bloch-Ham-Case-II-Sublattice-Basis} follow from the general sublattice-symmetric form of~\eqref{eq:app:Sublattice-Symmetric-Eigenstates}, with $\epsilon=\pm1$ and $j\in\{1,2\}$, so that $(\epsilon,j)$ enumerates the four bands $E_{\epsilon,j}=\epsilon E_{+,j}$~\eqref{eq:app:Energy-Spectrum-Hofstadter-Exact-quarter-Flux-momentum}. 
Inserting~\eqref{eq:app:Right-Singular-Vectors-D-matrix-Pauli} and~\eqref{eq:app:Left-Singular-Vectors-D-matrix-Pauli} gives the explicit closed form
\begin{align}
    \ket{\Psi_{\epsilon,j}}=\frac{1}{\sqrt{2}\,\sqrt{2h(h+\eta_j h_z)}}
    \begin{pmatrix}
        -\dfrac{2J}{E_{+,j}}\left[\,c_x(h+\eta_j h_z)+\eta_j c_y(h_x+i h_y)\,\right]\\[.28cm]
        -\dfrac{2J}{E_{+,j}}\left[\,-s_y(h+\eta_j h_z)+i\eta_j s_x(h_x+i h_y)\,\right]\\[.28cm]
        \epsilon\,(h+\eta_j h_z)\\[.18cm]
        \epsilon\eta_j\,(h_x+i h_y)
    \end{pmatrix}
    \label{eq:app:All-Eigenstates-explicit}
\end{align}
in the sublattice eigenbasis, with $\eta_j=(-1)^{j+1}$ and $E_{+,j}=\sqrt{h_0+\eta_j h}$~\eqref{eq:app:Energy-Spectrum-Hofstadter-Exact-quarter-Flux-momentum}.

Writing out the two branches explicitly, the outer pair of bands $E_{\pm,1}$ ($j=1$, $\eta_1=+1$) reads
    \begin{align}
        \ket{\Psi_{\epsilon,1}}=\frac{1}{\sqrt{2}\,\sqrt{2h(h+h_z)}}
        \begin{pmatrix}
            -\frac{2J}{E_{+,1}}\left[\,c_x(h+h_z)+c_y(h_x+i h_y)\,\right] \\[.25cm]
            -\frac{2J}{E_{+,1}}\left[\,-s_y(h+h_z)+i s_x(h_x+i h_y)\,\right] \\[.25cm]
            \epsilon\,(h+h_z) \\[.15cm]
            \epsilon\,(h_x+i h_y)
        \end{pmatrix},
        \label{eq:app:Eigenstates-Hofstadter-momentum}
    \end{align}
whose lower member $\ket{\Psi_{-,1}}$ ($\epsilon=-1$) is the ground state that is typically populated in cold-atom experiments realizing the quarter-flux Harper-Hofstadter model~\cite{Aidelsburger2013,Aidelsburger2015,Miyake2013}, and whose upper member $\ket{\Psi_{+,1}}$ ($\epsilon=+1$) is its chiral partner, obtained by flipping the sign of the two lower (sublattice-odd) components.
The inner pair of bands $E_{\pm,2}$ ($j=2$, $\eta_2=-1$) reads
    \begin{align}
        \ket{\Psi_{\epsilon,2}}=\frac{1}{\sqrt{2}\,\sqrt{2h(h-h_z)}}
        \begin{pmatrix}
            -\frac{2J}{E_{+,2}}\left[\,c_x(h-h_z)-c_y(h_x+i h_y)\,\right] \\[.25cm]
            -\frac{2J}{E_{+,2}}\left[\,-s_y(h-h_z)-i s_x(h_x+i h_y)\,\right] \\[.25cm]
            \epsilon\,(h-h_z) \\[.15cm]
            -\epsilon\,(h_x+i h_y)
        \end{pmatrix}.
        \label{eq:app:All-Eigenstates-middle}
    \end{align}
Together with the formula~\eqref{eq:app:All-Eigenstates-explicit}, \cref{eq:app:Eigenstates-Hofstadter-momentum,eq:app:All-Eigenstates-middle} provide the exact analytical expressions for all four eigenstates of the quarter-flux Harper-Hofstadter model in the sublattice eigenbasis. 
The corresponding states in the original site basis of $H(\bm{k})$~\eqref{eq:Bloch-Ham-Case-II} are simply recovered by the fixed permutation $\ket{\Psi}=Q\ket{\Psi_{\epsilon,j}}$~\eqref{eq:app:Bloch-Ham-Case-II-Sublattice-Basis}, which just swaps the components. 
We keep all expressions in the sublattice ordering in the following. 

For the ground state~$\ket{\Psi_{-,1}}$~\eqref{eq:app:Eigenstates-Hofstadter-momentum}, inserting the explicit Bloch-vector components $h_x=4J^2 c_x c_y$, $h_y=4J^2 s_x s_y$, $h_z=2J^2(c_{2x}-c_{2y})$ and $h=J^2\sqrt{2Z}$~\eqref{eq:app:Bloch-Vectors-M-Mtilde} and simplifying the trigonometric factors yields the fully closed form expression
    \begin{align}
        \ket{\Psi_{-,1}}=\frac{1}{\sqrt{2}\,\sqrt{2\sqrt{2Z}\,\big[\sqrt{2Z}+2(c_{2x}-c_{2y})\big]}}
        \begin{pmatrix}
            -\dfrac{2}{\sqrt{4+\sqrt{2Z}}}\big[\,c_x\big(\sqrt{2Z}+4c_x^2\big)+2i\,s_x s_{2y}\,\big] \\[.28cm]
            -\dfrac{2}{\sqrt{4+\sqrt{2Z}}}\big[\,-s_y\big(\sqrt{2Z}+4s_y^2\big)+2i\,c_y s_{2x}\,\big] \\[.28cm]
            -\big[\sqrt{2Z}+2(c_{2x}-c_{2y})\big] \\[.18cm]
            -4\,(c_x c_y+i\,s_x s_y)
        \end{pmatrix},
        \label{eq:app:GS-closed-form}
    \end{align}
\end{widetext}
with $Z=6+c_{4x}+c_{4y}$, where we used $h+h_z=J^2[\sqrt{2Z}+2(c_{2x}-c_{2y})]$ and $E_{+,1}=J\sqrt{4+\sqrt{2Z}}$. 
This is the exact analytical ground state of the quarter-flux Harper-Hofstadter model for the symmetric $2a\times2a$ MUC (ii)~[\cref{fig:App:MUC}(b)] and solves $H_S(\bm{k})\ket{\Psi_{-,1}}=E_{-,1}(\bm{k})\ket{\Psi_{-,1}}$~\eqref{eq:app:Bloch-Ham-Case-II-Sublattice-Basis}. 
The other three eigenstates follow analogously from~\cref{eq:app:Eigenstates-Hofstadter-momentum,eq:app:All-Eigenstates-middle}.

Finally, let us briefly comment on the degeneracy of the inner two bands $E_{\pm,2}$, which form a single super-band that touches at zero energy wherever $\sigma_2=E_{+,2}=\sqrt{h_0-h}$ vanishes, i.e. $h=h_0$, or equivalently $Z=8$ ($c_{4x}=c_{4y}=1$), at the high-symmetry points seen as band touchings in~\cref{fig:Hofstadter-Energy-Spectrum}. 
At these Dirac points the construction $\ket{u_2}=D_{\bm{k}}\ket{v_2}/\sigma_2$ becomes ill-defined and the two states $\ket{\Psi_{\pm,2}}$~\eqref{eq:app:All-Eigenstates-middle} become degenerate, so that only the two-dimensional subspace they span is well-defined and any $U(2)$ rotation within it is an equally valid eigenbasis. 
Away from these points all four eigenstates~\eqref{eq:app:All-Eigenstates-explicit} are smooth, up to the coordinate (gauge) singularity of the north-pole parametrization at $h+\eta_j h_z\to0$, which is present for all two-band systems anyway~\cite{Nakahara2003}.

\section{Quantum Geometry of sublattice symmetric systems.}
\label{app:quantum-geometry-all-states}
This section is organized in two parts, the quantum geometry of isolated (gapped) bands in~\cref{app:Abelian-Quantum-Geometry} and that of degenerate groups of bands in~\cref{app:Non-Abelian-Quantum-Geometry}. 
Each part first derives the general result for an arbitrary sublattice-symmetric system and then applies it to the quarter-flux Harper-Hofstadter model.

\subsection{Abelian quantum geometry of isolated bands}
\label{app:Abelian-Quantum-Geometry}

\subsubsection{General sublattice decomposition}
\label{app:subsubsec:Abelian-General}
The quantum geometric tensor of a state $\ket{\Psi}$ is defined as~\cite{Resta2011}
\begin{align}
    Q_{\mu\nu} &= \bra{\partial_\mu \Psi} (1 - \ket{\Psi}\bra{\Psi})\ket{\partial_\nu \Psi}= \braket{\del_\mu \Psi}{\del_\nu \Psi} - A_{\mu} A_{\nu}
    \nonumber \\
    &= g_{\mu\nu} - \frac{i}{2}\Omega_{\mu\nu},
    \label{eq:app:Quantum-Geometric-Tensor}
\end{align}
where $A_{\mu}=i\braket{\Psi}{\del_\mu \Psi}$ is the (Abelian) Berry connection, and $g_{\mu\nu}=\text{Re}[Q_{\mu\nu}]=g_{\nu \mu}$ and $\Omega_{\mu\nu}=-2\text{Im}[Q_{\mu\nu}] = -\Omega_{\nu \mu}$ are the quantum metric and the Berry curvature, respectively. It is of course implicitly assumed that the state $\ket{\Psi}\equiv \ket{\Psi(\bm{\lambda})}$ depends on a set of parameters $\bm{\lambda}=(\lambda^1,\lambda^2,\ldots, \lambda^N)$, and that $\del_{\mu} \equiv \del/\del \lambda^{\mu}$, with $\mu \in \{1,2,\ldots,N\}$.
For a system with an $N$-dimensional parameter space the quantum geometric tensor, being hermitian, has $N^2$ independent components. 

We now derive the sublattice decomposition of the QGT quoted in the main text~\eqref{eq:Abelian-QGT-Sublattice-Symmetric}. Recall from the main text that the eigenstates of any sublattice symmetric $H$ take the form~\eqref{eq:app:Sublattice-Symmetric-Eigenstates}, with $\ket{u_j}\in\mathcal{H}_+$ and $\ket{v_j}\in\mathcal{H}_-$ the left and right singular vectors of $D$ sharing the singular value $\sigma_j$, and $\epsilon=\pm1$ labeling the chiral energy pair $E_{\pm,j}=\pm\sigma_j$.
Inserting~\eqref{eq:app:Sublattice-Symmetric-Eigenstates} into the definition~\eqref{eq:app:Quantum-Geometric-Tensor} and using $A_{\mu} = \frac{1}{2}(A^{(u_j)}_{\mu} + A^{(v_j)}_{\mu})$ together with $\braket{\del_\mu \Psi_{\epsilon,j}}{\del_\nu \Psi_{\epsilon,j}} = \frac{1}{2}(\braket{\del_\mu u_j}{\del_\nu u_j} + \braket{\del_\mu v_j}{\del_\nu v_j})$ yields
    \begin{align}
        Q^{j}_{\mu\nu} = \frac{1}{2}\left(
        Q^{(u_j)}_{\mu\nu} + Q^{(v_j)}_{\mu\nu} \right) + \frac{1}{4}\Delta A^{j}_{\mu}\Delta A^{j}_{\nu},
        \label{eq:app:Quantum-Geometric-Tensor-Sublattice-Symmetric}
    \end{align}
where $Q^{(u_j)}_{\mu\nu}=\braket{\del_\mu u_j}{\del_\nu u_j} - A^{(u_j)}_{\mu} A^{(u_j)}_{\nu}$ is the QGT~\eqref{eq:app:Quantum-Geometric-Tensor} and $A^{(u_j)}_{\mu} = i\braket{u_j}{\del_\mu u_j}$ the Berry connection of $\ket{u_j}$ (similarly for $v_j$), and where
\begin{align}
    \Delta A^{j}_{\mu} \equiv A^{(u_j)}_{\mu} - A^{(v_j)}_{\mu}
    \label{eq:app:Berry-Connection-Difference}
\end{align}
is the sector connection difference. As stressed in the main text, $\Delta A^{j}_{\mu}$ is gauge invariant only because the SVD relation $\ket{u_j} = D \ket{v_j}/\sigma_j$ ties the two sectors. 
Under $\ket{v_j}\to e^{i\chi}\ket{v_j}$ it forces $\ket{u_j} \to e^{i\chi} \ket{u_j}$, so both connections shift by $\del_\mu \chi$ such that the difference is unchanged. 
Splitting~\eqref{eq:app:Quantum-Geometric-Tensor-Sublattice-Symmetric} into real and imaginary parts, and using that $\Delta A^{j}_{\mu}\Delta A^{j}_{\nu}$ is real and symmetric, gives the metric and curvature decomposition~\eqref{eq:Quantum-Metric-Berry-Curvature-Sublattice-Symmetric} of the main text, which we restate here for convenience,
    \begin{align}
        g^j_{\mu\nu} &= \frac{1}{2}\left(
        g^{(u_j)}_{\mu\nu} + g^{(v_j)}_{\mu\nu} \right) + \frac{1}{4}\Delta A^{j}_{\mu}\Delta A^{j}_{\nu}, \nonumber \\
        \Omega^j_{\mu\nu} &= \frac{1}{2}\left(
        \Omega^{(u_j)}_{\mu\nu} + \Omega^{(v_j)}_{\mu\nu} \right).
        \label{eq:app:Quantum-Metric-Berry-Curvature-Sublattice-Symmetric}
    \end{align}
Since $Q^{j}_{\mu\nu}$ is independent of $\epsilon$, the QGT is always the same for the two chiral partners $\ket{\Psi_{+,j}}$ and $\ket{\Psi_{-,j}}$ as long as the energy pair is non-degenerate. 
The sector form~\eqref{eq:app:Quantum-Geometric-Tensor-Sublattice-Symmetric}, in which the connection difference $\Delta A^j_\mu$ enters the metric holds for arbitrary sublattice dimensions $d_\pm$ and, as shown in~\cref{app:Non-Abelian-Quantum-Geometry}, extends to the non-Abelian QGT, where the $\Delta A$ term drops out for degenerate groups containing both chiral partners of every band and survives as a matrix otherwise.
This highlights the role of virtual-interband quantum metric effects~\cite{Mera2022} in sublattice-symmetric multiband systems. 

\subsubsection{Application to the quarter-flux Harper-Hofstadter model}
\label{app:subsubsec:Abelian-HH}
The sector contributions follow from the $M_{\bm{k}}=D^{\dagger}_{\bm{k}}D_{\bm{k}}$ and $\tilde{M}_{\bm{k}}=D_{\bm{k}}D^{\dagger}_{\bm{k}}$ machinery of~Sec.~\ref{app:Analytical-Solution}, that is, $\ket{v_j}$ and $\ket{u_j}$ are also the eigenvectors of the two-level blocks $M$ and $\tilde{M}$~\eqref{eq:app:SVD-D-matrix-M-Mtilde}, with Bloch vectors $\bm{h}$ and $\tilde{\bm{h}}$~\eqref{eq:app:Bloch-Vectors-M-Mtilde}.
This allows us to utilize the established machinery of two-band quantum geometry~\cite{Graf2021,Pozo2020}. To this end, note that the quantum metric $g_{\mu\nu}$ and Berry curvature $\Omega_{\mu \nu}$~\eqref{eq:app:Quantum-Geometric-Tensor} of a generic two-level system are fully determined by a (normalized) Bloch vector $\hat{\bm{h}}=\bm{h}/h$ as those in~\eqref{eq:app:Bloch-Vectors-M-Mtilde} and given by
\begin{align}
    g_{\mu\nu} &= \frac{1}{4}\left( \del_\mu \hat{\bm{h}} \cdot \del_\nu \hat{\bm{h}} \right)
    \nonumber \\
    &= \frac{1}{4 h^4}\left[ h^2 \del_\mu \bm{h} \cdot \del_\nu \bm{h} - (\bm{h}\cdot \del_\mu \bm{h})(\bm{h}\cdot \del_\nu \bm{h}) \right], 
    \label{eq:app:Quantum-Metric-Two-Level-System} \\
    \Omega_{\mu\nu} &=\mp \frac{1}{2}\hat{\bm{h}} \cdot (\del_\mu \hat{\bm{h}} \times \del_\nu \hat{\bm{h}}) = \mp\frac{1}{2 h^3}\bm{h} \cdot (\del_\mu \bm{h} \times \del_\nu \bm{h}).
    \label{eq:app:Berry-Curvature-Two-Level-System}
\end{align}
Note that for the two eigenstates of a two-level system pointing to $\pm \hat{\bm{h}}$ on the Bloch sphere, 
the quantum metric~\eqref{eq:app:Quantum-Metric-Two-Level-System} is the same, while the Berry curvature switches sign, as indicated by the $\mp$ in~\eqref{eq:app:Berry-Curvature-Two-Level-System}. 
In our two-dimensional parameter space $\bm{k}=(k_x,k_y)$ the indices $\mu$ and $\nu$ take values in $\{x,y\}$, e.g., $\del_{x} \equiv \del/\del k_{x}$, such that we have single independent component of the Berry curvature $\Omega_{xy} \equiv \Omega$ and three independent components of the quantum metric $g_{xx}$, $g_{yy}$, and $g_{xy}$. 
As we focus here first on the ground state $\ket{\Psi_{-,1}}$~\eqref{eq:app:Eigenstates-Hofstadter-momentum} (and its chirally symmetric partner, the state $\ket{\Psi_{+,1}}$ corresponding to the highest energy band), the relevant eigenvectors of $M$ and $\tilde{M}$ corresponding to the eigenvalue $\lambda_1 = (E_{+,1})^2=h_0 + h$ are $\ket{v_1}$ and $\ket{u_1}$ pointing to $+\hat{\bm{h}}$ and $+\hat{\tilde{\bm{h}}}$ respectively. 

\begin{figure*}[tb]
    \begin{center}
        \includegraphics[width=2\columnwidth]{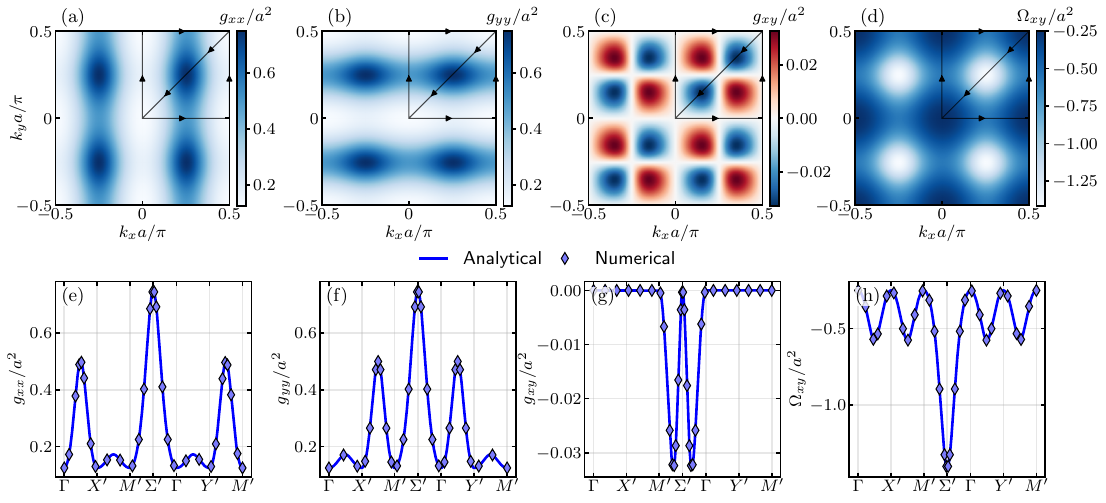}
        \caption{
            Quantum metric components and Berry curvature of the lowest (and highest) energy band of the quarter-flux Harper-Hofstadter model.
            (a)-(d) Analytically obtained quantum metric components $g^1_{xx}$~\eqref{eq:app:Quantum-Metric-xx-Exact-Expressions-Simplified}, $g^1_{yy}$~\eqref{eq:app:Quantum-Metric-yy-Exact-Expressions-Simplified}, $g^1_{xy}$~\eqref{eq:app:Quantum-Metric-xy-Exact-Expressions-Simplified}, and Berry curvature $\Omega^1_{xy}$~\eqref{eq:app:Berry-Curvature-Exact-Expression} as function of $\bm{k}$ over the FBZ,
            for the lowest and highest energy band $E_{\pm,1}$, corresponding to the eigenstate $\ket{\Psi_{\pm,1}}$~\eqref{eq:app:Eigenstates-Hofstadter-momentum}.
            (e)-(h) Same as (a)-(d) but along the high-symmetry path $\Gamma$--$X'$--$M'$--$\Sigma'$--$\Gamma$--$Y'$--$M'$ (as indicated via solid lines with arrows in (a)-(d), see also~\cref{fig:Hofstadter-Energy-Spectrum} in the main text) where solid lines (diamonds) depict the analytical (numerical) results.
        }
        \label{fig:App:QGT-Band1and4}
    \end{center}
\end{figure*}

We proceed by first computing the derivatives $\del_{\mu}\bm{h}$ and $\del_{\mu}\tilde{\bm{h}}$ of the Bloch vectors~\eqref{eq:app:Bloch-Vectors-M-Mtilde} which are given by,
    \begin{align}
        \del_{x} \bm{h} &= 4J^2 a \begin{pmatrix}
            -s_{x} c_{y} \\
            c_{x} s_{y} \\
            -s_{2x}        \end{pmatrix}, &&\mkern-14mu
        \del_{y} \bm{h} = 4J^2 a \begin{pmatrix}
            -c_{x} s_{y} \\
            s_{x} c_{y} \\
            s_{2y}        \end{pmatrix}, \nonumber \\
        \del_{x} \tilde{\bm{h}} &= 4J^2 a \begin{pmatrix}
            s_{x} s_{y} \\
            c_{x} c_{y} \\
            -s_{2x}        \end{pmatrix}, &&\mkern-14mu
        \del_{y} \tilde{\bm{h}} = -4J^2 a \begin{pmatrix}
            c_{x} c_{y} \\
            s_{x} s_{y} \\
            s_{2y}        \end{pmatrix}.
            \label{eq:app:Derivatives-Bloch-Vectors-M-Mtilde}
    \end{align}
Let us now first focus on the single Berry curvature $\Omega_{k_x,k_y}$ component. 
Using~\eqref{eq:app:Derivatives-Bloch-Vectors-M-Mtilde} we first find that 
\begin{align}
    \bm{h}\cdot (\del_{x}\bm{h} \times \del_{y}\bm{h})&=16J^6 a^2[4-(c_{2x} + c_{2y})^2], \nonumber \\ 
    \tilde{\bm{h}}\cdot (\del_{x}\tilde{\bm{h}} \times \del_{y}\tilde{\bm{h}})&=16J^6 a^2[4-(c_{2x} - c_{2y})^2].
\end{align}
Thus, the Berry curvature contributions from the two sectors, $\Omega^{(v_1)}_{xy}$ and $\Omega^{(u_1)}_{xy}$, are given by
    \begin{align}
        \Omega^{(v_1/u_1)}_{xy} &= -\frac{2\sqrt{2}a^2[4-(c_{2x}\pm c_{2y})^2]}{Z^{3/2}},
        \label{eq:app:Berry-Curvature-Contributions-Sectors}
    \end{align}
where have used $h=\tilde{h}=\sqrt{2}J^2 \sqrt{Z}$ from above~\eqref{eq:app:Bloch-Vectors-M-Mtilde}. 
Here, the exponent $(v_1/u_1)$ shall indicate the upper/lower $\pm$ sign in the above expression. 
From~\eqref{eq:app:Berry-Curvature-Two-Level-System} it directly follows that the corresponding Berry curvature contributions for the other sector are given by~$\Omega^{(v_2/u_2)}_{xy} = -\Omega^{(v_1/u_1)}_{xy}$, which will become relevant later for the non-Abelian QGT of the remaining middle two energy bands $E_{\pm,2}$. 
Here, for the lowest band $E_{-,1}$, we finally arrive at the exact analytical for the full Berry curvature $\Omega^1_{xy} =(\Omega^{(u_1)}_{xy} + \Omega^{(v_1)}_{xy})/2$~\eqref{eq:app:Quantum-Metric-Berry-Curvature-Sublattice-Symmetric} 
    \begin{align}
        \Omega^1_{xy} &= -\frac{\sqrt{2}a^2[12-Z]}{Z^{3/2}} \nonumber \\
        &=-\sqrt{2}a^2 \frac{6-\cos(4 k_x a)-\cos(4 k_y a)}{[6+\cos(4 k_x a)+\cos(4 k_y a)]^{3/2}}
        \label{eq:app:Berry-Curvature-Exact-Expression}
    \end{align}
where we have used that $8-2(c_{2x}^2+c_{2y}^2)=12-Z$ with $Z=6+c_{4x}+c_{4y}$ from above~\eqref{eq:app:Bloch-Vectors-M-Mtilde}.
\cref{eq:app:Berry-Curvature-Exact-Expression} is the Berry curvature $\Omega^1_{xy}$ for both $\ket{\Psi_{\pm,1}}$~\eqref{eq:app:Eigenstates-Hofstadter-momentum}, i.e., the lowest and highest energy band of the quarter-flux Harper-Hofstadter model, which is shown in \cref{fig:App:QGT-Band1and4}(d) and (h) as a function of $\bm{k}$ over the FBZ and along a high-symmetry path in quasimomentum space, respectively.
Moreover, note that since $Z\in[4,8]$ for all $\bm{k}$, we have $12-Z>0$ and hence $\Omega^1_{xy}<0, \ \forall \bm{k}$, we find that the Berry curvature does not switch sign over $\bm{k}$-space and is uniformly negative, consistent with the Chern number $C=-1$ of the lowest and highest energy band.
Our result~\eqref{eq:app:Berry-Curvature-Exact-Expression} is consistent with Ref.~\cite{Harper2014}, where it was found, perturbatively, that the Berry curvature of the $N$-band Harper-Hofstadter model, i.e., for flux $\Phi=2\pi/N$ (here $N=4$ bands) admits a $\cos(Nk_x a)+\cos(Nk_y a)$-dependence.

Let us now turn our attention to the quantum metric $g_{\mu\nu}$ components. 
From~\eqref{eq:app:Derivatives-Bloch-Vectors-M-Mtilde} we first find the following inner products between the Bloch vectors themselves and their derivatives: 
\begin{align}
    \del_{x}\bm{h} \cdot \del_{y} \bm{h} &= -8J^4 a^2s_{2x} s_{2y}, \nonumber \\ 
    \del_{x}\tilde{\bm{h}} \cdot \del_{y} \tilde{\bm{h}} &= 8J^4 a^2 s_{2x} s_{2y}, \nonumber \\
    \abs{\del_{\mu} \bm{h}}^2 &= 8 J^4 a^2 (2-c_{2x} c_{2y} -c_{4\mu}), \nonumber \\
    \abs{\del_{\mu} \tilde{\bm{h}}}^2 &= 8 J^4 a^2 (2+c_{2x} c_{2y} -c_{4\mu}), \nonumber \\
    \frac{1}{2}\del_{\mu} h^2 = \bm{h} \cdot \del_{\mu}\bm{h}&= -4J^4 a s_{4 \mu}= \tilde{\bm{h}} \cdot\del_{\mu}\tilde{\bm{h}},
\end{align}
where in the last expression we have used that $h^2=\tilde{h}^2=2 J^4 Z$~\eqref{eq:app:Bloch-Vectors-M-Mtilde}. 
Using these results we arrive at the individual contributions of the two sectors to the quantum metric $g_{\mu\nu}^{(v_1)}$ and $g_{\mu\nu}^{(u_1)}$~\eqref{eq:app:Quantum-Metric-Two-Level-System} given by
    \begin{align}
        g_{\mu\nu}^{(v_1/u_1)} &= \frac{a^2}{Z^2}\left[ Z \mathcal{N}^{(v_1/u_1)}_{\mu\nu} - s_{4\mu} s_{4\nu} \right],
        \label{eq:app:Quantum-Metric-Contributions-Sectors}
    \end{align}
where we introduced the shorthand notation $\mathcal{N}^{(v_1/u_1)}_{\mu\nu} $ defined by
    \begin{align}
        \mathcal{N}^{(v_1)}_{\mu \mu} &= 2-c_{2x} c_{2y} -c_{4\mu}, && \mathcal{N}^{(v_1)}_{xy} = -s_{2x} s_{2y}, \nonumber \\
        \mathcal{N}^{(u_1)}_{\mu \mu} &= 2+c_{2x} c_{2y} -c_{4\mu}, && \mathcal{N}^{(u_1)}_{xy} = s_{2x} s_{2y},
    \end{align}
with $\mu \in \{x,y\}$ and no summation over repeated indices. 
From~\eqref{eq:app:Quantum-Metric-Two-Level-System} it follows that $g_{\mu\nu}^{(v_1)} = g_{\mu\nu}^{(v_2)}$ and $g_{\mu\nu}^{(u_1)} = g_{\mu\nu}^{(u_2)}$ as well, which will become relevant later when we compute QGT of the degenerate middle two energy bands. 

Now, what remains to be computed for the full quantum metric $g_{\mu\nu}$~\eqref{eq:app:Quantum-Metric-Berry-Curvature-Sublattice-Symmetric} is the contribution from the connection difference $\Delta A^1_{\mu} = A^{(u_1)}_{\mu} - A^{(v_1)}_{\mu}$, which we will derive in the next step.
To this end, we first express two eigenstates $\ket{v_{j}}$ of $M$~(\cref{eq:app:Right-Singular-Vectors-D-matrix-Pauli}, $j\in\{1,2\}$) in terms of their Bloch vector $\bm{h}$ components as~\cite{Nakahara2003}
    \begin{align}
        \ket{v_{j}}=\frac{1}{\sqrt{2h(h+\eta_j h_z)}}
    \begin{pmatrix}
        h +\eta_j h_z\\
        \eta_j(h_x+i h_y) 
    \end{pmatrix},
    \label{eq:app:Right-Singular-Vectors-D-matrix-Pauli-Bloch-Vector-Expression}
    \end{align}
where again $\eta_j = (-1)^{j+1}$.
As discussed above, $\ket{u_j}$ must be expressed in the same gauge via the SVD relation $\ket{u_j} = D \ket{v_j}/\sigma_j$.
Using this relation and the eigenvalue equation $D^\dagger D \ket{v_j} = M \ket{v_j} = \sigma_j^2 \ket{v_j}$~\eqref{eq:app:SVD-D-matrix-M-Mtilde},  
the connection differences $\Delta A^j_{\mu}$~\eqref{eq:app:Berry-Connection-Difference} take the form of 
 \begin{align}
        \Delta A^j_{\mu} &= \frac{i}{\sigma_j}\bra{v_j}D^{\dagger}\del_\mu \left( \frac{1}{\sigma_j} D \ket{v_j}\right) - i\braket{v_j}{\del_\mu v_j}  \nonumber \\
        &=i \bigg[\frac{1}{\sigma_j^2}\bra{v_j}D^{\dagger}\del_\mu D\ket{v_j}-\frac{\del_\mu \sigma_j}{\sigma_j} 
        \bigg]
        \label{eq:app:Connection-Differences-Exact-Same-Gauge-Expression}
    \end{align}
where we find that the Berry connection $A^{(v_j)}_{\mu}$ cancels out exactly. 
Using that 
$\del_{\mu} \sigma_j/\sigma_j  = (\eta_j \del_{\mu}\bm{h}\cdot \bm{h})/2h \sigma_j^2$
and $D^{\dagger}\del_x D = -4J^2 \begin{psmallmatrix}
            s_{2x}/2 & i c_{x} s_{y} \\
            c_y s_x & -s_{2x}/2
        \end{psmallmatrix}$
and $D^{\dagger}\del_y D = 4J^2 \begin{psmallmatrix}
            s_{2y}/2 & - c_{x} s_{y} \\
            i s_x c_y & -s_{2y}/2
        \end{psmallmatrix}$, 
we first arrive at the intermediate results of 
    \begin{align}
        \frac{\bra{v_j}D^{\dagger}\del_x D\ket{v_j}}{\sigma_j^2} &= \frac{-2\eta_j J^4 a}{h\sigma_j^2 }\left(s_{4x} + 2i\,s_{2y}c_{2x}\right), \nonumber \\
        \frac{\bra{v_j}D^{\dagger}\del_y D\ket{v_j}}{\sigma_j^2} &= \frac{2\eta_j J^4 a}{h\sigma_j^2 }\left(-s_{4y} + 2i\,s_{2x}c_{2y}\right), \nonumber \\
        \frac{\del_{\mu} \sigma_j}{\sigma_j} &= -2\eta_j J^4 a \frac{s_{4\mu}}{h\sigma_j^2}.
    \end{align}
Thus, the exact expressions for the Berry connection differences $\Delta A^j_{\mu}$~\eqref{eq:app:Connection-Differences-Exact-Same-Gauge-Expression} become
    \begin{align}
        \Delta A^j_{x} = \frac{2\eta_j a}{\mathcal{G}_j} s_{2y}c_{2x}, \quad
        \Delta A^j_{y} = -\frac{2\eta_j a }{\mathcal{G}_j} s_{2x}c_{2y},
        \label{eq:app:Connection-Differences-Exact-Pre-Expression}
    \end{align}
with $\mathcal{G}_j=2\sqrt{2Z} + \eta_j Z$ for brevity.
Finally, using these results and~\eqref{eq:app:Quantum-Metric-Contributions-Sectors}, 
we arrive at exact analytical expressions for the quantum metric components $g^1_{\mu\nu}$~\eqref{eq:app:Quantum-Metric-Berry-Curvature-Sublattice-Symmetric} 
    \begin{align}
        g^1_{xx}/a^2 & 
        =\frac{Z(2-c_{4x}) -s_{4x}^2}{Z^2} + \frac{s^2_{2y}c^2_{2x}}{(\mathcal{G}_1)^2} ,
        \label{eq:app:Quantum-Metric-xx-Exact-Expressions-Simplified} \\
        g^1_{yy}/a^2 & 
        =\frac{Z(2-c_{4y}) -s_{4y}^2}{Z^2} + \frac{s^2_{2x}c^2_{2y}}{(\mathcal{G}_1)^2} , \label{eq:app:Quantum-Metric-yy-Exact-Expressions-Simplified} \\
        g^1_{xy}/a^2 & = -\frac{s_{4x} s_{4y}}{Z^2} - \frac{s_{2x}s_{2y}c_{2x}c_{2y}}{(\mathcal{G}_1)^2}
        \label{eq:app:Quantum-Metric-xy-Exact-Expressions-Simplified} 
    \end{align}
\cref{eq:app:Quantum-Metric-xx-Exact-Expressions-Simplified,eq:app:Quantum-Metric-yy-Exact-Expressions-Simplified,eq:app:Quantum-Metric-xy-Exact-Expressions-Simplified} are the exact analytical expressions for the quantum metric components $g^1_{\mu \nu}$ of the lowest and highest energy band of the quarter-flux Harper-Hofstadter model corresponding to the eigenstates $\ket{\Psi_{\pm,1}}$~\eqref{eq:app:Eigenstates-Hofstadter-momentum}.
These components are depicted in~\cref{fig:App:QGT-Band1and4}(a)-(c) and (e)-(g) as function of $\bm{k}$ over the FBZ and along a high-symmetry path in quasimomentum space, respectively.
Summing the diagonal components gives the metric trace in closed form,
    \begin{align}
        \frac{\mathrm{tr}\,g^1(\bm{k})}{a^2}
        =\frac{Z(10-Z)-2+c_{4x}^2+c_{4y}^2}{Z^2}
        +\frac{s_{2y}^2c_{2x}^2+s_{2x}^2c_{2y}^2}{(2\sqrt{2Z}+Z)^2},
        \label{eq:app:Trace-Metric-Exact}
    \end{align}
where we used $c_{4x}+c_{4y}=Z-6$, and where the second term is the connection-difference contribution
$\tfrac14((\Delta A^1_x)^2+(\Delta A^1_y)^2)$~\eqref{eq:app:Connection-Differences-Exact-Pre-Expression}.

The determinant $\det g^1$ follows in closed form as well.
Writing the decomposition~\eqref{eq:app:Quantum-Metric-Berry-Curvature-Sublattice-Symmetric} as $g^1_{\mu\nu}=\bar{g}^1_{\mu\nu}+\tfrac14 \Delta A^1_{\mu}\Delta A^1_{\nu}$ with the sector average $\bar{g}^1_{\mu\nu} \equiv \tfrac12 (g^{(u_1)}_{\mu\nu} + g^{(v_1)}_{\mu\nu})$~\eqref{eq:app:Quantum-Metric-Contributions-Sectors}, the connection-difference term is an outer product, so its own determinant vanishes. 
Thus, we have
    \begin{align}
        \det g^1 = \det \bar{g}^1 + \tfrac14\big[&\bar{g}^1_{yy}(\Delta A^1_{x})^2 - 2\bar{g}^1_{xy}\Delta A^1_{x}\Delta A^1_{y} \nonumber \\
        &+ \bar{g}^1_{xx}(\Delta A^1_{y})^2\big] ,
        \label{eq:app:Det-Metric-Rank-One-Split}
    \end{align}
with $\Delta A^1_{\mu}$ given by~\eqref{eq:app:Connection-Differences-Exact-Pre-Expression} and $\bar{g}^1_{\mu\nu}$ given by~\eqref{eq:app:Quantum-Metric-Contributions-Sectors}, and 
    \begin{align}
        \frac{\det \bar{g}^1}{a^4}
        =&\frac{(2-c_{4x})(2-c_{4y})}{Z^2} \nonumber \\
        -&\frac{(2-c_{4y})s_{4x}^2 + (2-c_{4x})s_{4y}^2}{Z^3} .
        \label{eq:app:Det-Metric-Sector-Average-Exact}
    \end{align}
Both the determinant $\det g^1(\bm{k})$ and the trace $\mathrm{tr}\,g^1(\bm{k})$~\eqref{eq:app:Trace-Metric-Exact} are depicted in~\cref{fig:App:local-inequalities}(a)-(b) and (e)-(f), respectively. 

\subsection{Non-Abelian Quantum Geometry of sublattice symmetric systems} 
\label{app:Non-Abelian-Quantum-Geometry}

The middle two bands (one super band) of the quarter-flux Harper-Hofstadter model exhibit band touching points (Dirac points) at zero energy at certain high-symmetry points in the Brillouin zone~[see~\cref{fig:Hofstadter-Energy-Spectrum} in the main text], and are hence degenerate and not gapped for all $\bm{k}$. 
Thus, the quantum geometry of these two bands cannot be described by the QGT of an isolated single band~\eqref{eq:app:Quantum-Geometric-Tensor}, but rather requires the notion of quantum geometry for a degenerate subspace. 
This is captured by the so-called non-Abelian quantum geometric tensor~\cite{Ma2010,Rezakhani2010}
    \begin{align}
    \mathcal{Q}_{\mu\nu}^{ab} &= \bra{\partial_\mu \Psi_a} (1 - P) \ket{\partial_\nu \Psi_b} \nonumber \\
                    &=\braket{\del_\mu \Psi_a}{\del_\nu \Psi_b} - \sum_c A_{\mu}^{ac}A_{\nu}^{cb} \nonumber \\
                    &=G^ {ab}_{\mu\nu} - \frac{i}{2}F^{ab}_{\mu\nu},
                    \label{eq:app:non-Abelian-QGT}
    \end{align}
where $a,b,c$ are indices labeling the states within the degenerate subspace, and $P$ the projector onto the degenerate subspace, $A_{\mu}^{ab} = i\braket{\Psi_a}{\del_\mu \Psi_b}=(A_{\mu}^{ba})^{*}$ the non-Abelian (or interband) Berry connection. 
For fixed parameter-indices $\mu,\nu$, the non-Abelian QGT $\mathcal{Q}_{\mu\nu}$ can be understood as $M\times M$ matrix where $M$ labels the dimension the of the degenerate subspace which further satisfies $\mathcal{Q}_{\mu\nu}^\dagger = \mathcal{Q}_{\nu\mu}$ (or $\mathcal{Q}_{\mu\nu}^{ab}=(\mathcal{Q}_{\nu\mu}^{ba})^*$). 
It can be decomposed into a Hermitian symmetric part, $G_{\mu\nu} = ( \mathcal{Q}_{\mu\nu} + \mathcal{Q}_{\mu\nu}^\dagger)/2 =  G_{\nu\mu} = G_{\mu\nu}^\dagger$, the non-Abelian quantum metric, 
and a Hermitian antisymmetric part, $F_{\mu\nu} = i(\mathcal{Q}_{\mu\nu} - \mathcal{Q}_{\mu \nu}^\dagger) = -F_{\nu\mu} = F_{\mu\nu}^\dagger$, the non-Abelian Berry curvature. The latter can be further written as the field strength of the non-Abelian Berry connection $A_{\mu}$~\cite{Wilczek1984} via
    \begin{align}
        F_{\mu\nu} = \del_{\mu} A_{\nu} - \del_{\nu} A_{\mu} - i [A_{\mu},A_{\nu}],
    \end{align}
which is to be understood as a $M\times M$ matrix-valued differential $2$-form. 
Note that under a local $U(N)$ change of basis inside the degenerate subspace, $\ket{\Psi_a}\to\sum_b U^{\ast}_{ab}\ket{\Psi_b}$ with $U\in U(N)$, under which the non-Abelian Berry connection transforms as a gauge field, $A_\mu\to UA_\mu U^\dagger-i(\del_\mu U)U^\dagger$, the non-Abelian QGT~\eqref{eq:app:non-Abelian-QGT} transforms covariantly, $\mathcal{Q}_{\mu\nu}\to U \mathcal{Q}_{\mu\nu} U^{\dagger}$ so that only similarity invariants of the non-Abelian QGT, such as the trace $\mathrm{tr}[\mathcal{Q}_{\mu\nu}]$ and determinant $\det[\mathcal{Q}_{\mu\nu}]$, are truly gauge invariant and can be used to characterize the quantum geometry of the degenerate subspace. 

\subsubsection{General sublattice decomposition}
\label{app:subsubsec:NonAbelian-General}
We now show how the non-Abelian QGT~\eqref{eq:app:non-Abelian-QGT} decomposes for a sublattice-symmetric system with eigenstates of the form~\eqref{eq:app:Sublattice-Symmetric-Eigenstates}. 
Only the sum over intermediate states in the projector term depends on which bands the degenerate group contains.
Readers interested only in the quarter-flux model may skip from~\eqref{eq:app:non-Abelian-QGT-Sublattice-Symmetric-Decomposition-Matrix-Form} directly to~\cref{app:subsubsec:NonAbelian-HH}.

Consider a degenerate group $\mathcal{B}$ of nonzero-energy bands of a sublattice-symmetric system with eigenstates of the form~\eqref{eq:app:Sublattice-Symmetric-Eigenstates}, and let $\mathcal{J}$ be the set of indices $j$ appearing in $\mathcal{B}$. 
With $a\equiv (\epsilon',j')$ and $b\equiv (\epsilon,j)$, the non-Abelian Berry connection splits into the two sectors as
    \begin{align}
        A_{\mu}^{(\epsilon',j'),(\epsilon,j)} &= \frac{1}{2}\left[ A_{\mu}^{(u_{j'},u_{j})} + \epsilon' \epsilon A_{\mu}^{(v_{j'},v_{j})} \right],
        \label{eq:app:non-Abelian-Connection-Sublattice}
    \end{align}
where $A_{\mu}^{(u_{j'},u_{j})} = i\braket{u_{j'}}{\del_\mu u_j}$ and $A_{\mu}^{(v_{j'},v_{j})}= i\braket{v_{j'}}{\del_\mu v_j}$ are the interband Berry connections within the $\mathcal{H}_+$ and $\mathcal{H}_-$ sectors, respectively. 
The derivative overlaps split in the same way, $\braket{\del_\mu \Psi_{\epsilon',j'}}{\del_\nu \Psi_{\epsilon,j}} = \frac{1}{2}\big[ \braket{\del_\mu u_{j'}}{\del_\nu u_j} + \epsilon' \epsilon \braket{\del_\mu v_{j'}}{\del_\nu v_j} \big]$. 

\emph{Sum over intermediate states.} 
Inserting both relations into~\eqref{eq:app:non-Abelian-QGT}, the derivative term decomposes directly, and all of the structure sits in the projector term $\sum_{c\in\mathcal{B}}A^{ac}_\mu A^{cb}_\nu$. 
Its sum runs over the intermediate states $c=(\epsilon'',j'')$ that belong to $\mathcal{B}$, and only over those. 
Multiplying out and using $(\epsilon'')^2=1$, each summand splits into a part independent of $\epsilon''$ and a part linear in $\epsilon''$,
    \begin{align}
        A^{ac}_\mu A^{cb}_\nu
        &=\tfrac{1}{4}\big(A^{(u)}_\mu+\epsilon'\epsilon''A^{(v)}_\mu\big)_{j'j''}\big(A^{(u)}_\nu+\epsilon''\epsilon A^{(v)}_\nu\big)_{j''j} \nonumber \\
        &=\tfrac{1}{4}\big[(A^{(u)}_\mu)_{j'j''}(A^{(u)}_\nu)_{j''j}+\epsilon'\epsilon\,(A^{(v)}_\mu)_{j'j''}(A^{(v)}_\nu)_{j''j}\big] \nonumber \\
        &\quad+\frac{\epsilon''}{4}\big[\epsilon\,(A^{(u)}_\mu)_{j'j''}(A^{(v)}_\nu)_{j''j} \nonumber \\
        &\qquad\qquad+\epsilon'(A^{(v)}_\mu)_{j'j''}(A^{(u)}_\nu)_{j''j}\big],
        \label{eq:app:non-Abelian-Connection-Product-Sublattice}
    \end{align}
where $(A^{(u)}_\mu)_{j'j}\equiv A^{(u_{j'},u_j)}_\mu$ and $(A^{(v)}_\mu)_{j'j}\equiv A^{(v_{j'},v_j)}_\mu$ are matrices over $\mathcal{J}$, and no sum over $j''$ is implied. 
The second bracket mixes the two sectors, and whether it survives the sum over $c\in\mathcal{B}$ 
depends only on which signs $\epsilon''$ are present in $\mathcal{B}$. 

\emph{Groups closed under chiral partners.} 
Let $\bar{\mathcal{B}}=\{(-\epsilon,j):(\epsilon,j)\in\mathcal{B}\}$ denote the chiral mirror of $\mathcal{B}$, i.e., the partner bands at opposite energy. 
We first assume that $\mathcal{B}$ is closed under chiral partners, $\mathcal{B}=\bar{\mathcal{B}}$, so that it contains both $(\pm,j)$ for every $j\in\mathcal{J}$, as is the case for the middle super-band of the quarter-flux Harper-Hofstadter model~[see~\cref{fig:Hofstadter-Energy-Spectrum} in the main text].
The sum over $c\in\mathcal{B}$ then becomes $\sum_{j''\in\mathcal{J}}\sum_{\epsilon''=\pm}$, under which the terms linear in $\epsilon''$ cancel and the $\epsilon''$-independent terms are counted twice, 
    \begin{align}
        \sum_{c\in\mathcal{B}}A^{ac}_\mu A^{cb}_\nu = \tfrac{1}{2}\big[A^{(u)}_\mu A^{(u)}_\nu+\epsilon'\epsilon\,A^{(v)}_\mu A^{(v)}_\nu\big]_{j'j}.
        \label{eq:app:non-Abelian-Projector-Term-Closed}
    \end{align}
The non-Abelian QGT therefore decomposes into the two sectors as
    \begin{align}
        \mathcal{Q}_{\mu\nu}^{(\epsilon',j'),(\epsilon,j)} 
        &=\tfrac{1}{2}\big[\braket{\del_\mu u_{j'}}{\del_\nu u_j} - (A^{(u)}_\mu A^{(u)}_\nu)_{j'j}\big] \nonumber \\
        &\quad+\tfrac{\epsilon'\epsilon}{2}\big[\braket{\del_\mu v_{j'}}{\del_\nu v_j} - (A^{(v)}_\mu A^{(v)}_\nu)_{j'j}\big] \nonumber \\
        &=\tfrac{1}{2}\big[\mathcal{Q}^{(u_{j'},u_j)}_{\mu\nu} + \epsilon' \epsilon\, \mathcal{Q}^{(v_{j'},v_j)}_{\mu\nu}\big],
        \label{eq:app:non-Abelian-QGT-Sublattice-Symmetric-Decomposition}
    \end{align}
where $\mathcal{Q}^{(u_{j'},u_j)}_{\mu\nu}\equiv(\mathcal{Q}^{(u)}_{\mu\nu})_{j'j}$ is the non-Abelian QGT of the sector states $\{\ket{u_j}\}_{j\in\mathcal{J}}$, i.e., the first bracket, and similarly for $\mathcal{Q}^{(v_{j'},v_j)}_{\mu\nu}$. 
The cancellation in~\eqref{eq:app:non-Abelian-Projector-Term-Closed} is the crucial step that uses $\mathcal{B}=\bar{\mathcal{B}}$, and it fails otherwise, as we will show below. 
In the $(\epsilon',\epsilon)\in\{\pm,\pm\}$ basis, with each block a matrix over $\mathcal{J}$, the non-Abelian QGT~\eqref{eq:app:non-Abelian-QGT-Sublattice-Symmetric-Decomposition} takes the form
    \begin{align}
        \mathcal{Q}_{\mu\nu} &= \begin{pmatrix}
            \mathcal{Q}_{\mu\nu}^{++} & \mathcal{Q}_{\mu\nu}^{+-} \\
            \mathcal{Q}_{\mu\nu}^{-+} & \mathcal{Q}_{\mu\nu}^{--} 
        \end{pmatrix} \nonumber \\
        &= \frac{1}{2}\begin{pmatrix}
            \mathcal{Q}_{\mu\nu}^{(u)} + \mathcal{Q}_{\mu\nu}^{(v)} & \mathcal{Q}_{\mu\nu}^{(u)} - \mathcal{Q}_{\mu\nu}^{(v)} \\
            \mathcal{Q}_{\mu\nu}^{(u)} - \mathcal{Q}_{\mu\nu}^{(v)} & \mathcal{Q}_{\mu\nu}^{(u)} + \mathcal{Q}_{\mu\nu}^{(v)} 
        \end{pmatrix}.
        \label{eq:app:non-Abelian-QGT-Sublattice-Symmetric-Decomposition-Matrix-Form}
    \end{align}

\begin{figure*}[tb]
    \begin{center}
        \includegraphics[width=2\columnwidth]{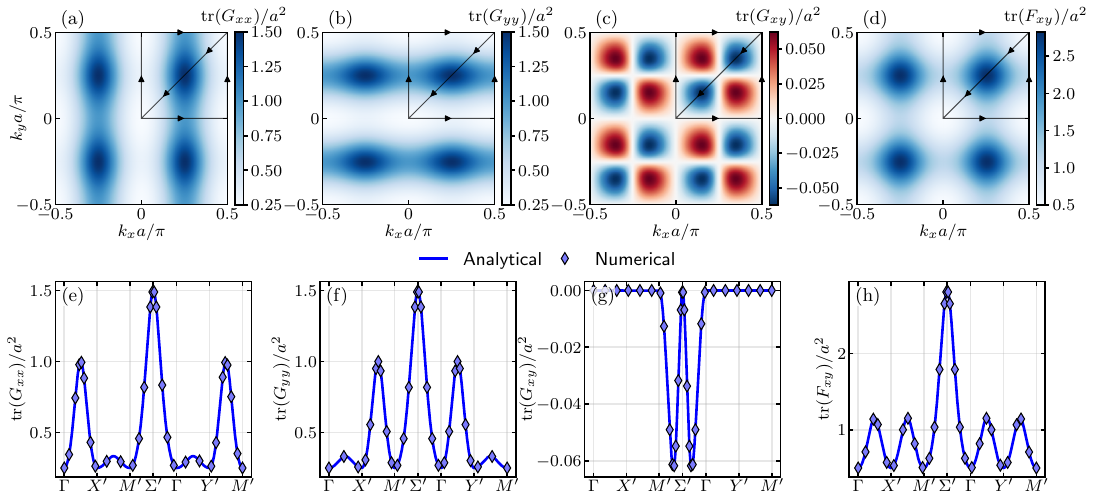}
        \caption{
            Non-Abelian quantum metric and Berry curvature of the degenerate subspace of the middle two energy bands $E_{\pm,2}$ of the quarter-flux Harper-Hofstadter model. 
            (a)-(d) Analytically obtained non-Abelian quantum metric components $\mathrm{tr}(G_{xx})$, $\mathrm{tr}(G_{yy})$, $\mathrm{tr}(G_{xy})$~\eqref{eq:app:non-Abelian-Quantum-Metric-Middle-Two-Bands}, and Berry curvature $\mathrm{tr}(F_{xy})$~\eqref{eq:app:non-Abelian-Berry-Curvature-Middle-Two-Bands} 
            as function of $\bm{k}$ over the FBZ,
            for the degenerate subspace of the middle two energy bands $E_{\pm,2}$. 
            (e)-(h) Same as (a)-(d) but along the high-symmetry path $\Gamma$--$X'$--$M'$--$\Sigma'$--$\Gamma$--$Y'$--$M'$ (as indicated via solid lines with arrows in (a)-(d), see also~\cref{fig:Hofstadter-Energy-Spectrum} in the main text) where solid lines (diamonds) depict the analytical (numerical) results.
        }
        \label{fig:App:QGT-non-Abelian-Band2and3}
    \end{center}
\end{figure*}

\emph{Groups of a single sign.} 
Now let $\mathcal{B}$ contain bands of one sign $\epsilon$ only, e.g., two bands that touch each other on the same side of zero energy while being gapped from all other bands. 
Then $\epsilon'=\epsilon''=\epsilon$ for all states in~\eqref{eq:app:non-Abelian-Connection-Product-Sublattice}, so the terms linear in $\epsilon''$ no longer cancel. 
Summed over $j''\in\mathcal{J}$, the two brackets then combine into a perfect square,
    \begin{align}
        \sum_{c\in\mathcal{B}}A^{ac}_\mu A^{cb}_\nu = \tfrac{1}{4}\big[\big(A^{(u)}_\mu+A^{(v)}_\mu\big)\big(A^{(u)}_\nu+A^{(v)}_\nu\big)\big]_{j'j},
    \end{align}
and the non-Abelian QGT~\eqref{eq:app:non-Abelian-QGT} becomes
    \begin{align}
        \mathcal{Q}_{\mu\nu}^{(\epsilon,j'),(\epsilon,j)} 
        &= \tfrac{1}{2}\big[\braket{\del_\mu u_{j'}}{\del_\nu u_j}+\braket{\del_\mu v_{j'}}{\del_\nu v_j}\big] \nonumber \\
        &\quad- \tfrac{1}{4}\big[\big(A^{(u)}_\mu+A^{(v)}_\mu\big)\big(A^{(u)}_\nu+A^{(v)}_\nu\big)\big]_{j'j}.
        \label{eq:app:non-Abelian-QGT-Same-Sign-Raw}
    \end{align}
To bring this expression into a gauge-covariant form (similar to~\eqref{eq:app:non-Abelian-QGT-Sublattice-Symmetric-Decomposition}), we express the derivative overlaps through the sector QGTs. 
With the sector projector $P^{(u)}_{\mathcal{J}}=\sum_{j\in\mathcal{J}}\projector{u_j}$, the sector QGT reads
    \begin{align}
        \big(\mathcal{Q}^{(u)}_{\mu\nu}\big)_{j'j}
        &=\bra{\del_\mu u_{j'}}\big(1-P^{(u)}_{\mathcal{J}}\big)\ket{\del_\nu u_j} \nonumber \\
        &=\braket{\del_\mu u_{j'}}{\del_\nu u_j}-\big(A^{(u)}_\mu A^{(u)}_\nu\big)_{j'j},
    \end{align}
and similarly for the $v$ sector, so that $\braket{\del_\mu u_{j'}}{\del_\nu u_j}=\big(\mathcal{Q}^{(u)}_{\mu\nu}+A^{(u)}_\mu A^{(u)}_\nu\big)_{j'j}$. 
Inserting this into~\eqref{eq:app:non-Abelian-QGT-Same-Sign-Raw} separates the sector QGTs from a remainder $\mathcal{R}_{\mu\nu}$ that contains only connections,
    \begin{align}
        \mathcal{Q}^{(\epsilon,j'),(\epsilon,j)}_{\mu\nu} &= \tfrac{1}{2}\big[\mathcal{Q}^{(u)}_{\mu\nu}+\mathcal{Q}^{(v)}_{\mu\nu}\big]_{j'j} + \big(\mathcal{R}_{\mu\nu}\big)_{j'j}, \nonumber\\
        \mathcal{R}_{\mu\nu}&=\tfrac{1}{2}\big(A^{(u)}_\mu A^{(u)}_\nu+A^{(v)}_\mu A^{(v)}_\nu\big) \nonumber \\
        &\quad-\tfrac{1}{4}\big(A^{(u)}_\mu+A^{(v)}_\mu\big)\big(A^{(u)}_\nu+A^{(v)}_\nu\big).
    \end{align}
Expanding the product and collecting terms, while keeping the order of the $\mu$ and $\nu$ factors since the connections are matrices, gives
    \begin{align}
        \mathcal{R}_{\mu\nu}
        &=\tfrac{1}{4}\big(A^{(u)}_\mu A^{(u)}_\nu - A^{(u)}_\mu A^{(v)}_\nu - A^{(v)}_\mu A^{(u)}_\nu + A^{(v)}_\mu A^{(v)}_\nu\big) \nonumber\\
        &=\tfrac{1}{4}\big(A^{(u)}_\mu-A^{(v)}_\mu\big)\big(A^{(u)}_\nu-A^{(v)}_\nu\big)\equiv\tfrac{1}{4}\Delta A_\mu\Delta A_\nu,
    \end{align}
with the matrix $(\Delta A_\mu)_{j'j}\equiv A^{(u_{j'},u_j)}_\mu-A^{(v_{j'},v_j)}_\mu$ of connection differences. 
The non-Abelian QGT of a single-sign group therefore reads
    \begin{align}
        \mathcal{Q}_{\mu\nu}^{(\epsilon,j'),(\epsilon,j)} = \tfrac{1}{2}\big[\mathcal{Q}^{(u)}_{\mu\nu} + \mathcal{Q}^{(v)}_{\mu\nu}\big]_{j'j} + \tfrac{1}{4}\big(\Delta A_\mu \Delta A_\nu\big)_{j'j},
        \label{eq:app:non-Abelian-QGT-Sublattice-Symmetric-Same-Sign}
    \end{align}
which is the matrix generalization of the Abelian result~\eqref{eq:app:Quantum-Geometric-Tensor-Sublattice-Symmetric}. 
As in the Abelian case, it does not depend on the sign $\epsilon$ of the group. 

\emph{General groups.} 
In general, $\mathcal{B}$ contains both signs of some indices, $j\in\mathcal{J}_2$, and a single sign $\epsilon_j$ of others, $j\in\mathcal{J}_1$. 
Adding the missing chiral partners $\mathcal{M}=\{(-\epsilon_j,j):j\in\mathcal{J}_1\}$ makes the group closed, so we split the projector term of~\eqref{eq:app:non-Abelian-QGT} as
    \begin{align}
        \sum_{c\in\mathcal{B}}A^{ac}_\mu A^{cb}_\nu=\sum_{c\in\mathcal{B}\cup\mathcal{M}}A^{ac}_\mu A^{cb}_\nu-\sum_{c\in\mathcal{M}}A^{ac}_\mu A^{cb}_\nu.
    \end{align}
By~\eqref{eq:app:non-Abelian-Projector-Term-Closed}, the first sum yields the decomposition~\eqref{eq:app:non-Abelian-QGT-Sublattice-Symmetric-Decomposition} with sector QGTs over all of $\mathcal{J}$, while the second sum enters with a plus sign and equals $\bra{\del_\mu\Psi_a}P_{\mathcal{M}}\ket{\del_\nu\Psi_b}$ with $P_{\mathcal{M}}=\sum_{c\in\mathcal{M}}\projector{\Psi_c}$, the virtual transitions into the missing partners. 
Equivalently, since $\mathcal{B}$ and $\mathcal{M}$ are disjoint sets of orthonormal eigenstates, $P_{\mathcal{B}\cup\mathcal{M}}=P_{\mathcal{B}}+P_{\mathcal{M}}$, so that the complement projector in~\eqref{eq:app:non-Abelian-QGT} splits as $1-P_{\mathcal{B}}=\big(1-P_{\mathcal{B}\cup\mathcal{M}}\big)+P_{\mathcal{M}}$. 
Evaluating~\eqref{eq:app:non-Abelian-Connection-Sublattice} at $\epsilon''=-\epsilon_{j''}$ then gives
    \begin{align}
        \mathcal{Q}_{\mu\nu}^{(\epsilon',j'),(\epsilon,j)} &= \tfrac{1}{2}\big[\mathcal{Q}^{(u)}_{\mu\nu} + \epsilon'\epsilon\,\mathcal{Q}^{(v)}_{\mu\nu}\big]_{j'j} \nonumber \\
        &\quad+ \tfrac{1}{4}\sum_{j''\in \mathcal{J}_1}\big(A^{(u)}_\mu-\epsilon'\epsilon_{j''}A^{(v)}_\mu\big)_{j'j''} \nonumber \\
        &\qquad\times\big(A^{(u)}_\nu-\epsilon_{j''}\epsilon\, A^{(v)}_\nu\big)_{j''j}.
        \label{eq:app:non-Abelian-QGT-Sublattice-Symmetric-General}
    \end{align}
In mixed groups the correction also contributes to $\mathrm{tr}\,F_{\mu\nu}$, but since $\mathrm{tr}\,F_{\mu\nu}=\del_\mu\mathrm{tr}A_\nu-\del_\nu\mathrm{tr}A_\mu$ and $A^{(\epsilon,j),(\epsilon,j)}_\mu=\tfrac{1}{2}\big(A^{(u_j)}_\mu+A^{(v_j)}_\mu\big)$, it always equals $\sum_{(\epsilon,j)\in\mathcal{B}}\tfrac{1}{2}\big(\Omega^{(u_j)}_{\mu\nu}+\Omega^{(v_j)}_{\mu\nu}\big)$, the sum of the sector-averaged curvatures of the bands in~$\mathcal{B}$. 

\emph{Consistency checks.} 
(i)~\emph{Single band.} 
For a group consisting of one band, $\Delta A_\mu$ reduces to the number $\Delta A^j_\mu$ and~\eqref{eq:app:non-Abelian-QGT-Sublattice-Symmetric-Same-Sign} reproduces~\eqref{eq:app:Quantum-Geometric-Tensor-Sublattice-Symmetric}. 
(ii)~\emph{Gauge covariance.} 
For a single-sign group every gauge transformation $\ket{\Psi_{\epsilon,j}}\to\sum_{k\in\mathcal{J}}U^{\ast}_{jk}\ket{\Psi_{\epsilon,k}}$ preserves the paired form~\eqref{eq:app:Sublattice-Symmetric-Eigenstates}, acting with the same $U$ on both sectors, $\ket{u_j}\to\sum_k U^{\ast}_{jk}\ket{u_k}$ and $\ket{v_j}\to\sum_k U^{\ast}_{jk}\ket{v_k}$.
The sector connections $A^{(u)}_\mu$ and $A^{(v)}_\mu$ then pick up the same inhomogeneous term $-i(\del_\mu U)U^\dagger$, which cancels in their difference, so that $\Delta A_\mu\to U\Delta A_\mu U^\dagger$ transforms homogeneously (as an adjoint field) and the extra term in~\eqref{eq:app:non-Abelian-QGT-Sublattice-Symmetric-Same-Sign} transforms as $\mathcal{Q}_{\mu\nu}\to U\mathcal{Q}_{\mu\nu}U^\dagger$.
For groups containing both partners, the transformations preserving the paired form are those acting identically on $(+,j)$ and $(-,j)$, under which the same argument applies, while transformations mixing $(+,j)$ with $(-,j)$ leave the projector $P_{\mathcal{B}}$, and hence all gauge invariants of $\mathcal{Q}_{\mu\nu}$, unchanged.
(iii)~\emph{Curvature and metric.} 
With $\mathcal{Q}_{\mu\nu}=G_{\mu\nu}-\tfrac{i}{2}F_{\mu\nu}$, the term $\tfrac{1}{4}\Delta A_\mu\Delta A_\nu$ adds $\tfrac{1}{8}\{\Delta A_\mu,\Delta A_\nu\}$ to $G_{\mu\nu}$ and $\tfrac{i}{4}[\Delta A_\mu,\Delta A_\nu]$ to $F_{\mu\nu}$. 
The commutator is traceless, so $\mathrm{tr}\,F_{\mu\nu}$ and the Chern number keep the sector-average form, whereas $\mathrm{tr}\,G_{\mu\mu}$ increases by $\tfrac{1}{4}\sum_{j'j}\abs{(\Delta A_\mu)_{j'j}}^2\geq 0$. 

To summarize, the non-Abelian QGT of a sublattice-symmetric degenerate group is always the sector average~\eqref{eq:app:non-Abelian-QGT-Sublattice-Symmetric-Decomposition} plus one correction for every band whose chiral partner lies outside the group~\eqref{eq:app:non-Abelian-QGT-Sublattice-Symmetric-General}.
For groups containing both partners of every band the correction vanishes, for a single band it is the Abelian term $\tfrac{1}{4}\Delta A^j_\mu\Delta A^j_\nu$~\eqref{eq:app:Quantum-Geometric-Tensor-Sublattice-Symmetric}, and for a single-sign group ($\epsilon'=\epsilon=\epsilon_{j''}$, both factors equal to $\tfrac{1}{2}\Delta A$) it is its matrix version~\eqref{eq:app:non-Abelian-QGT-Sublattice-Symmetric-Same-Sign}.
In all cases $\mathrm{tr}\,F_{\mu\nu}$ is the sum of the sector-averaged Berry curvatures of the individual bands, whereas $\mathrm{tr}\,G_{\mu\nu}$ admits no such band-wise form. 
Zero modes $(\ket{u_0},0)^T$ of a rectangular $D$ are not of the paired form~\eqref{eq:app:Sublattice-Symmetric-Eigenstates}, so groups containing them are not covered by these results. 

\subsubsection{Application to the middle super-band of the quarter-flux Harper-Hofstadter model}
\label{app:subsubsec:NonAbelian-HH}
The middle super-band $\mathcal{B}=\{(+,2),(-,2)\}$ contains both chiral partners of its only index $j=2$ ($\mathcal{J}=\{2\}$), so the decomposition~\eqref{eq:app:non-Abelian-QGT-Sublattice-Symmetric-Decomposition} applies without correction. 
Its projector reads $P=\sum_{\epsilon=\pm} \ket{\Psi_{\epsilon,2}}\bra{\Psi_{\epsilon,2}} = \projector{u_2}\oplus \projector{v_2}$,
such that, by the matrix form~\eqref{eq:app:non-Abelian-QGT-Sublattice-Symmetric-Decomposition-Matrix-Form} with $1\times1$ blocks, the non-Abelian QGT becomes a $2\times 2$ matrix for each pair of parameter indices $\mu,\nu$,
\begin{align}
    \mathcal{Q}_{\mu\nu}
    &=\tfrac{1}{2}\big[ (\mathcal{Q}_{\mu\nu}^{(u_2,u_2)} + \mathcal{Q}_{\mu\nu}^{(v_2,v_2)} )\mathbbm{1}_2 \nonumber \\
    &\qquad+(\mathcal{Q}_{\mu\nu}^{(u_2,u_2)} - \mathcal{Q}_{\mu\nu}^{(v_2,v_2)})\sigma_x \big].
\end{align}
Now, we proceed as before noting that $\mathcal{Q}_{\mu\nu}^{(u_2,u_2)}\equiv Q^{(u_2)}_{\mu\nu} = g^{(u_2)}_{\mu\nu}-\frac{i}{2}\Omega^{(u_2)}_{\mu\nu}$ (similarly for $v_2$) can be computed using two-level QGT~\eqref{eq:app:Quantum-Metric-Two-Level-System}-\eqref{eq:app:Berry-Curvature-Two-Level-System} since $\ket{u_2}$ and $\ket{v_2}$, being the left and right singular vectors of $D$, are also eigenstates of 2-by-2 matrices $M$ and $\tilde{M}$~\eqref{eq:app:SVD-D-matrix-M-Mtilde}, respectively. 
The quantum metric components are readily obtained from noting that $g^{(u_2)}_{\mu\nu} =g^{(u_1)}_{\mu\nu}$ and $g^{(v_2)}_{\mu\nu} =g^{(v_1)}_{\mu\nu}$~\eqref{eq:app:Quantum-Metric-Contributions-Sectors}. 
Thus, using $\mathcal{Q}_{\mu\nu}= G_{\mu\nu} - \frac{i}{2}F_{\mu\nu}$~\eqref{eq:app:non-Abelian-QGT}, here we have, first for the quantum metric 
    \begin{align}
        G_{\mu\nu} = \frac{1}{2}\big[ (g^{(u_2)}_{\mu\nu} + g^{(v_2)}_{\mu\nu} )\mathbbm{1}_2
    + (g^{(u_2)}_{\mu\nu} - g^{(v_2)}_{\mu\nu})\sigma_x \big]. 
    \end{align}
The individual components are given by
    \begin{align}
        G_{xx}/a^2&= \frac{1}{Z^2}[(Z(2-c_{4x}) -s_{4x}^2)\mathbbm{1}_2 +(Zc_{2x}c_{2y})\sigma_x], \nonumber \\
        G_{yy}/a^2&= \frac{1}{Z^2}[(Z(2-c_{4y}) -s_{4y}^2)\mathbbm{1}_2 +(Zc_{2x}c_{2y})\sigma_x], \nonumber \\
        G_{xy}/a^2&=\frac{1}{Z^2}[(-s_{4x} s_{4y})\mathbbm{1}_2 + (Z s_{2x}s_{2y})\sigma_x].
        \label{eq:app:non-Abelian-Quantum-Metric-Middle-Two-Bands}
    \end{align}
The total quantum metric of the degenerate subspace of the middle two energy bands is given by the trace of $G_{\mu\nu}$ over the internal degrees of freedom, $\mathrm{tr}(G_{\mu\nu}) = g^{(u_2)}_{\mu\nu} + g^{(v_2)}_{\mu\nu}$, which is given by twice the components of~\eqref{eq:app:non-Abelian-Quantum-Metric-Middle-Two-Bands} proportional to the identity matrix, 
    \begin{align}
        \mathrm{tr}(G_{xx})/a^2&= \frac{2[Z(2-c_{4x}) -s_{4x}^2]}{Z^2}, \nonumber \\
        \mathrm{tr}(G_{yy})/a^2&= \frac{2[Z(2-c_{4y}) -s_{4y}^2]}{Z^2}, \nonumber \\
        \mathrm{tr}(G_{xy})/a^2&=\frac{-2s_{4x} s_{4y}}{Z^2}.
    \end{align}
These components are depicted in~\cref{fig:App:QGT-non-Abelian-Band2and3}(a)-(c) and (e)-(g) as function of $\bm{k}$ over the FBZ and along a high-symmetry path in quasimomentum space, respectively.
This trace shows a commonly overlooked property: the total quantum metric is non-additive~\cite{Mera2022,Peotta2015},
    \begin{align}
        \mathrm{tr}(G_{\mu \nu}) &= g^{(u_2)}_{\mu\nu} + g^{(v_2)}_{\mu\nu} \nonumber \\
        &= g^{(-,2)}_{\mu\nu} +g^{(+,2)}_{\mu\nu} +\mathrm{tr}(\del_\mu P_{-,2} \del_\nu P_{+,2}) \nonumber \\
        &\neq g^{(-,2)}_{\mu\nu} +g^{(+,2)}_{\mu\nu},
    \end{align}
with $g^{(\pm,2)}_{\mu\nu}\equiv g^{2}_{\mu\nu}$~\eqref{eq:app:Quantum-Metric-Berry-Curvature-Sublattice-Symmetric} and $P_{\pm,2} = \projector{\Psi_{\pm,2}}$.
The cross term, $\mathrm{tr}(\del_\mu P_{-,2} \del_\nu P_{+,2}) =2 \mathrm{Re}[\braket{\Psi_{+,2}}{\del_\mu \Psi_{-,2}}\braket{\Psi_{-,2}}{\del_\nu \Psi_{+,2}}]=-\Delta A^2_{\mu}\Delta A^2_{\nu}/2$,
is built from the same connection difference~\eqref{eq:app:Berry-Connection-Difference} that already enters the Abelian metric of each band~\eqref{eq:app:Quantum-Metric-Berry-Curvature-Sublattice-Symmetric}, and describes virtual interband transitions between the two bands of the degenerate subspace~\cite{Mera2022}.

Now, finally for Berry curvature components we have that $\Omega^{(u_2)}_{\mu\nu} = - \Omega^{(u_1)}_{\mu\nu}$ and $\Omega^{(v_2)}_{\mu\nu} = - \Omega^{(v_1)}_{\mu\nu}$~\eqref{eq:app:Berry-Curvature-Contributions-Sectors} such the non-Abelian Berry curvature components become
    \begin{align}
        F_{\mu\nu} = \frac{1}{2}\big[ (\Omega^{(u_2)}_{\mu\nu} + \Omega^{(v_2)}_{\mu\nu} )\mathbbm{1}_2
    + (\Omega^{(u_2)}_{\mu\nu} - \Omega^{(v_2)}_{\mu\nu})\sigma_x \big].
    \end{align}
The individual components are given by 
    \begin{align}
        F_{xy}/a^2&= \frac{\sqrt{2}[12-Z]}{Z^{3/2}}\mathbbm{1}_2 +\frac{4\sqrt{2}}{Z^{3/2}}c_{2x}c_{2y}\sigma_x.
        \label{eq:app:non-Abelian-Berry-Curvature-Middle-Two-Bands}
    \end{align}
We find that the total Berry curvature of the degenerate subspace of the middle two energy bands, given by the trace $\mathrm{tr}(F_{xy}) = \Omega^{(u_2)}_{xy} + \Omega^{(v_2)}_{xy}=-(\Omega^{(u_1)}_{xy} + \Omega^{(v_1)}_{xy})$, is given by
    \begin{align}
        \mathrm{tr}(F_{xy})/a^2&= \frac{2\sqrt{2}[12-Z]}{Z^{3/2}},
        \nonumber \\
        &=2\sqrt{2} \frac{6-\cos(4 k_x a)-\cos(4 k_y a)}{[6+\cos(4 k_x a)+\cos(4 k_y a)]^{3/2}}.
    \end{align}
This non-Abelian Berry curvature is depicted in~\cref{fig:App:QGT-non-Abelian-Band2and3}(d) and (h) as function of $\bm{k}$ over the FBZ and along a high-symmetry path in quasimomentum space, respectively.
Unlike the metric, it is additive, $\mathrm{tr}(F_{xy}) = \Omega^{(-,2)}_{xy}+\Omega^{(+,2)}_{xy}$ with $\Omega^{(\pm,2)}_{xy}\equiv\Omega^{2}_{xy}$~\eqref{eq:app:Quantum-Metric-Berry-Curvature-Sublattice-Symmetric}.
Since $\mathrm{tr}(F_{xy})=-2\Omega^1_{xy}$~\eqref{eq:app:Berry-Curvature-Exact-Expression}, the Berry curvature summed over all four bands vanishes pointwise, and the middle super-band carries $C_{2+3}=(2\pi)^{-1}\smallint_{\text{BZ}} d^2k \,\mathrm{tr}(F_{xy}) =2$, so that $C_1+C_{2+3}+C_4=0$ with $C_1=C_4=-1$~\cite{Xiao2010}.

\section{FCI Stability Conditions}
\label{app:fci-stability-conditions}
A number of criteria have recently been proposed to quantify the stability of a platform to host fractional Chern insulators (FCIs) or fractional quantum Hall (FQH) trial states. 
Interestingly, these criteria can be evaluated from the underlying single-particle wave functions where most of them rest on lowest-Landau-level (LLL) ``mimicry''~\cite{Ledwith2023}. 

Equipped with the exact solutions and quantum geometry of the quarter-flux Harper-Hofstadter model, we can evaluate these criteria explicitly.
We restrict ourselves to the quantum geometric momentum-space conditions, which have been originally formulated in~\Ccite{Parameswaran2012,Parameswaran2013,Roy2014} and further sharpened in~\Ccite{Jackson2015,Wang2021,Mera2021,Mera2021a,Ozawa2021,Mera2022b,Mera2022,Zhang2022b,Varjas2022}.
We note that recently, complementary real-space criteria, in particular vortexability~\cite{Ledwith2023,Fujimoto2025a,Liu2025b}, have also been proposed, but are not the focus of this work.
There are three notions which enter these criteria, which summarize below and go into more detail in the following.

    \emph{Closure.} Whether the band-projected densities $\bar\rho_{\bm q}=\mathcal{P}_n e^{-i\bm{q}\cdot\bm{r}}\mathcal{P}_n$
    with the projector $\mathcal{P}_n$ onto the $n$-th band as a whole~\eqref{eq:app:band-projector} obey \emph{some} closed algebra at all, i.e., whether $[\bar\rho_{\bm q},\bar\rho_{\bm p}]=f(\bm{q},\bm{p})\,\bar\rho_{\bm q+\bm p}$ for a structure constant $f$~\cite{Wang2025a}.
    Evaluating the commutator leaves a coefficient $f(\bm{q},\bm{p})_{\bm{k}}$ that generically still explicitly depends on $\bm{k}$ (varying across the BZ), while the right-hand side carries a single coefficient per momentum transfer. 
    Closure therefore requires this residual $\bm{k}$-dependence to cancel, $f(\bm{q},\bm{p})_{\bm{k}}=f(\bm{q},\bm{p})$~\cite{Wang2025a}, so that the structure constant depends on the transferred momenta alone. 

    \emph{GMP form.} The Girvin-MacDonald-Platzman (GMP) algebra~\cite{Girvin1986} of the LLL is \emph{one} such closed algebra, but it carries a free function, the form factor $F(\bm{q})$, entering as $f\propto [F(\bm{q})F(\bm{p})/F(\bm{q}+\bm{p})]\sin[\tfrac{1}{2}\Omega_{xy}\,(\bm{q}\times\bm{p})_z]$, where the constant $\Omega_{xy}$ appearing here is fixed by the Berry curvature~\cite{Wang2025a}. 
    Different form factors describe genuinely different closed algebras, among them those of the higher Landau levels. These share the uniform Berry curvature of the LLL, but obey $\mathrm{tr}\,g_n=(2n+1)\abs{\Omega_{xy}}$~\cite{Liu2025b} and hence violate the trace condition below for $n\ge1$, so that closure alone does not single out the LLL.

    \emph{Ideal-band condition.} A pointwise saturated trace condition $\mathrm{tr}\,g(\bm{k})=\abs{\Omega_{xy}(\bm{k})}$~\eqref{eq:app:local-trace-inequality}, which is the property we test here and which, following~\Ccite{Ledwith2023}, we call an ideal, or vortexable, band throughout.
    Saturation alone already fixes the metric completely in terms of the curvature: combining it with the determinant inequality $2\sqrt{\det g} \ge\abs{\Omega_{xy}}$~\eqref{eq:app:local-det-inequality} and $2\sqrt{\det g}\le \mathrm{tr}\,g $ forces $g_{\mu\nu}=\tfrac{1}{2}\abs{\Omega_{xy}}\delta_{\mu\nu}$. 
    It does \emph{not}, however, force the curvature itself to be uniform.

    \emph{LLL-mimicry condition.} A uniform Berry curvature $\Omega_{xy}(\bm{k})=\mathrm{const.}$ \emph{in addition} to the saturated trace condition. This is the strictly stronger requirement, and the two must be kept apart: the quarter-flux ground band nearly saturates the trace condition, missing it by $8.3\%$ of $2\pi\abs{C_1}$, while its Berry curvature remains non-uniform. 

There are two results which relate them.
Roy~\cite{Roy2014} showed that the LLL-mimicry condition, i.e., saturation \emph{together with} uniform curvature, is necessary and sufficient for closure \emph{with the form factors of the LLL}.
Recently, Wang and Simon~\cite{Wang2025a} then showed that this exhausts the possibilities, namely, in two and three dimensions, any closed algebra of band-projected density operators must be of GMP form for \emph{some} form factor, and closure in turn forces a $\bm{k}$-independent QGT.
However, the opposite implication is \emph{not} true, i.e., a band with constant Berry curvature and constant quantum metric need not lead to the closure of the projected density algebra~\cite{Wang2025a}. 
The ideal-band condition is therefore the sharp criterion for the LLL member of the GMP family, not for closure in general. 

\subsection{Momentum-space hierarchy: from local positivity to integrated conditions}
\label{app:fci-momentum-hierarchy}

Every momentum-space criterion stems from a single algebraic fact, namely that the quantum geometric tensor (QGT) is positive semi-definite~\cite{Mera2021,Ozawa2021}.
For an isolated Bloch band $\ket{\Psi_n(\bm{k})}$, let $P_{\perp}(\bm{k})=\mathbbm{1}-\projector{\Psi_n(\bm{k})}$ denote the projector onto the orthogonal complement of that band at fixed $\bm{k}$, so that the QGT~\eqref{eq:app:Quantum-Geometric-Tensor} reads $Q^n_{\mu\nu}(\bm{k})=\bra{\del_\mu \Psi_n}P_{\perp}(\bm{k})\ket{\del_\nu\Psi_n}=g^n_{\mu\nu}(\bm{k})-\tfrac{i}{2}\Omega^n_{\mu\nu}(\bm{k})$.
Inserting a resolution of the identity on the orthogonal complement, $P_{\perp}(\bm{k})=\sum_{m\neq n} \projector{\Psi_m(\bm{k})}$ (a sum over all other bands), gives
\begin{align}
    Q^n_{\mu\nu}(\bm{k}) = \sum_{m\neq n} (w^\mu_m)^*\, w^\nu_m, \quad w^\mu_m \equiv \braket{\Psi_m}{\del_\mu \Psi_n},
    \label{eq:app:qgt-gram}
\end{align}
which is a Gram matrix $Q^n=W^\dagger W$ with $(W)_{m \nu}\equiv w^\nu_m$ and is therefore Hermitian and positive semi-definite, denoted $Q^n\succeq 0$, independent of any model details. 
In the remainder of this section we suppress the band index $n$ wherever the statement holds for an arbitrary isolated band.

\paragraph{Local inequalities.} Writing $Q=g-\tfrac{i}{2}\Omega$ with the antisymmetric $\Omega_{\mu\nu}=\Omega_{xy}\varepsilon_{\mu\nu}$ in two dimensions, the semi-definite positivity of the $2\times 2$ Hermitian matrix $Q$ is equivalent to a non-negative determinant, $\det Q = \det g(\bm{k}) - \tfrac{1}{4}\Omega_{xy}(\bm{k})^2 \ge 0$. 
This yields the so-called \emph{local determinant inequality}~\cite{Ozawa2021,Mera2021}
\begin{align}
    2\sqrt{\det g(\bm{k})} \ge \abs{\Omega_{xy}(\bm{k})}, \quad \forall\,\bm{k}. 
    \label{eq:app:local-det-inequality}
\end{align}
Applying the arithmetic-geometric-mean inequality to the non-negative eigenvalues $\lambda_{1,2}$ of the symmetric metric $g$, $\mathrm{tr}\,g=\lambda_1+\lambda_2\ge 2\sqrt{\lambda_1\lambda_2}=2\sqrt{\det g}$, and combining it with~\eqref{eq:app:local-det-inequality} yields the so-called \emph{local trace inequality}~\cite{Mera2021,Ozawa2021}
\begin{align}
    \mathrm{tr}\,g(\bm{k}) \ge \abs{\Omega_{xy}(\bm{k})}, \quad \forall\,\bm{k}.
    \label{eq:app:local-trace-inequality}
\end{align}
In total, we thus have the hierarchy of local inequalities
\begin{align}
    \mathrm{tr}\,g(\bm{k}) \ge 2\sqrt{\det g(\bm{k})} \ge \abs{\Omega_{xy}(\bm{k})} ,
    \label{eq:app:local-hierarchy}
\end{align}
which are pointwise statements, valid for any (isolated) band, with no reference to the global topology of the bands. 
These inequalities are also valid for a filled band with a projector $P(\bm{k})$ of $\mathrm{rank}~>1$~\cite{Mera2021,Ozawa2021}.

\paragraph{Integrated (global) conditions.} 
Integrating~\eqref{eq:app:local-trace-inequality} over the first (magnetic) Brillouin zone (BZ) and using $\int_{\mathrm{BZ}}\abs{\Omega_{xy}}\,d^2k \ge \abs{\int_{\mathrm{BZ}}\Omega_{xy}\,d^2k}=2\pi\abs{C}$ gives the (global) \emph{trace condition}
\begin{align}
    \int_{\mathrm{BZ}} \mathrm{tr}\,g(\bm{k})\,d^2k \ge 2\pi\abs{C},
    \label{eq:app:global-trace-condition}
\end{align}
originally proposed (on LLL-mimicry grounds) by~\Ccite{Parameswaran2012,Parameswaran2013}, with $C$ being the first Chern number.
Note that the defect of~\eqref{eq:app:global-trace-condition} is in general \emph{not} equal to the integrated local defect $\mathcal{T}_{\mathrm{HH}}$~\eqref{eq:trace-defect} we introduced the main text, 
but they agree whenever the Berry curvature does not change sign over the BZ.
Only $\mathcal{T}_{\mathrm{HH}}$ is built from a pointwise non-negative integrand $D(\bm{k})=\mathrm{tr}\,g(\bm{k})-\abs{\Omega_{xy}(\bm{k})}\ge 0$, so that it is zero exactly when the band is locally ideal, whereas the defect of~\eqref{eq:app:global-trace-condition} $\int \mathrm{tr}\,g\,d^2k-2\pi\abs{C}\ge 0$ can be non-zero even for a locally ideal band whose curvature changes sign. 

Integrating~\eqref{eq:app:local-det-inequality} in the same way, $2\int_{\mathrm{BZ}}\sqrt{\det g}\,d^2k \ge \int_{\mathrm{BZ}}\abs{\Omega_{xy}}\,d^2k \ge 2\pi\abs{C}$, gives the global~\emph{determinant condition}~\cite{Ozawa2021,Mera2021}
\begin{align}
    \mathrm{vol}_g \equiv \int_{\mathrm{BZ}}\sqrt{\det g}\,d^2k\ge \pi\abs{C}
    \label{eq:app:global-det-condition}
\end{align}

\begin{table*}[htbp]
    \centering
    \begin{tabular}{@{}llll@{}}
        \toprule
        ID & Criterion & Defect measure & HH value \\
        \midrule
        L1 & local determinant~\eqref{eq:app:local-det-inequality} & $\mathcal{D}_{\det}=2\sqrt{\det g}-\abs{\Omega_{xy}}\ge 0$ & mean $=0.0302$, max $=\tfrac32-\sqrt2\approx 0.0858$ \\
        L1$'$ & anisotropy step of~\eqref{eq:app:local-hierarchy} & $\mathcal{D}_{\mathrm{ani}}=\mathrm{tr}\,g-2\sqrt{\det g}\ge 0$ & mean $=0.0227$, max $\approx 0.0850$ \\
        L2 & local trace~\eqref{eq:app:local-trace-inequality} & $\mathrm{tr}\,g-\abs{\Omega_{xy}}\ge 0$ & mean $=0.0529$, max $=\tfrac54-\tfrac{2}{\sqrt3}\approx 0.0953$ \\
        G1 & curvature flatness & $(\Omega_{\max}-\Omega_{\min})/\abs{\bar\Omega}$ & $\pi(4\sqrt2-1)/8\approx 182.9\%$ ($\Omega\in[-\sqrt{2},-\tfrac{1}{4}]$) \\
        G2 & metric flatness & $(\mathrm{tr}\,g_{\max}-\mathrm{tr}\,g_{\min})/\overline{\mathrm{tr}\,g}$ & $\tfrac54\pi^2/(2\pi+\mathcal{T}_{\mathrm{HH}})\approx 181.3\%$ ($\mathrm{tr}\,g\in[\tfrac{1}{4},\tfrac{3}{2}]$) \\
        G3 & determinant condition~\eqref{eq:app:global-det-condition} & $\int\sqrt{\det g}\,d^2k-\pi\abs{C}$ & $0.1488$ \;($4.74\%$) \\
        G4 & trace condition~\eqref{eq:app:global-trace-condition} & $\mathcal{T}_{\mathrm{HH}}=\int\mathrm{tr}\,g\,d^2k-2\pi\abs{C}$ & $0.5220$ \;($8.31\%$) \\
        \bottomrule
    \end{tabular}
    \caption{Momentum-space FCI-stability criteria evaluated for the quarter-flux Harper-Hofstadter ground band $\ket{\Psi_{-,1}}$ ($C_1=-1$). All values in units of the lattice constant $a=1$, with overbars denoting BZ averages.
    The two flatness ratios of the metric and curvature are closed-form. 
    With the magnetic BZ area $\pi^2/a^2$ one has $\bar\Omega^1_{xy}=2\pi C_1 a^2/\pi^2=-2a^2/\pi$ and $\overline{\mathrm{tr}\,g^1}=(2\pi\abs{C_1}+\mathcal{T}_{\mathrm{HH}})a^2/\pi^2$, so that the ranges $\Omega^1_{xy}\in[-\sqrt2,-\tfrac14]\,a^2$ and $\mathrm{tr}\,g^1\in[\tfrac14,\tfrac32]\,a^2$ giving $\mathrm{(G1)}=\pi(4\sqrt2-1)/8$ and $\mathrm{(G2)}=\tfrac54\pi^2/(2\pi+\mathcal{T}_{\mathrm{HH}})$ for the curvature and metric flatness, respectively.
    The BZ integrals are compared against $2\pi\abs{C_1}$ (G4) and $\pi\abs{C_1}$ (G3). 
    The means of L1 and L1$'$ add up to that of L2 by~\eqref{eq:app:trace-defect-local-split}, whereas their maxima do not, since the two defects peak at different momenta.
    All entries are reproduced from the exact analytic expressions with independent numerical integration over the FBZ.}
    \label{tab:fci-stability-hierarchy}
\end{table*}
A band that in addition carries a $\bm{k}$-independent Berry curvature (\emph{Berry-curvature flatness}) and a $\bm{k}$-independent quantum metric (\emph{metric flatness}), and that saturates~\eqref{eq:app:global-trace-condition}, is band-geometrically a LLL on the lattice~\cite{Ledwith2023}, since the LLL admits $g_{\mu\nu}=\tfrac{1}{2}\ell_B^2\,\delta_{\mu\nu}$ and $\Omega_{xy}=-\ell_B^2$ with $\ell_B=\abs{B}^{-1/2}$ the magnetic length~\cite{Ozawa2021}, saturating~\eqref{eq:app:local-trace-inequality} pointwise.

Curvature flatness together with saturation of~\eqref{eq:app:global-trace-condition} is Roy's ideal-band condition~\cite{Roy2014}, equivalent to closure of the projected density algebra with LLL form factors. 
A $\bm{k}$-independent QGT is a weaker requirement than the ideal-band condition. 
Closure with \emph{any} form factor already enforces $\bm{k}$-independence of the QGT~\cite{Wang2025a}, which is therefore necessary for closure, but sufficient neither for closure nor for ideality.
\subsection{Application to the quarter-flux Harper-Hofstadter ground band}
\label{app:fci-application-ground-band}

We now evaluate every level of the hierarchy on the exact ground band $\ket{\Psi_{-,1}}$, using the closed-form Berry curvature~\eqref{eq:app:Berry-Curvature-Exact-Expression} and quantum metric~\eqref{eq:app:Quantum-Metric-xx-Exact-Expressions-Simplified}--\eqref{eq:app:Quantum-Metric-xy-Exact-Expressions-Simplified}. 
One feature simplifies our following analysis. 
Since $\Omega^1_{xy}=-\sqrt{2}\,a^2(12-Z)/Z^{3/2}<0$ everywhere ($Z\in[4,8]<12$), the curvature never changes sign, so that $\int_{\mathrm{BZ}}\abs{\Omega^1_{xy}}\,d^2k=2\pi \abs{C_1}=2\pi$ exactly with $C_1=-1$. 
Thus, the integrated defect of~\eqref{eq:app:global-trace-condition} is local in origin and not a curvature sign-cancellation effect, and hence equal to the integrated local defect $\mathcal{T}_{\mathrm{HH}}$~\eqref{eq:trace-defect} of the main text. 

We collect the results in~\cref{tab:fci-stability-hierarchy}. 
Although the band is not flat-geometric ideal with both the Berry curvature and the trace of the metric fluctuating, the \emph{integrated} conditions fall short of saturation only modestly.
In particular, the determinant condition~\eqref{eq:app:global-det-condition} by $4.74\%$ and the trace condition by $\mathcal{T}_{\mathrm{HH}}=0.522$~\eqref{eq:app:global-trace-condition}, i.e.\ $8.31\%$.

\begin{figure*}[t]
    \begin{center}
        \includegraphics[width=2\columnwidth]{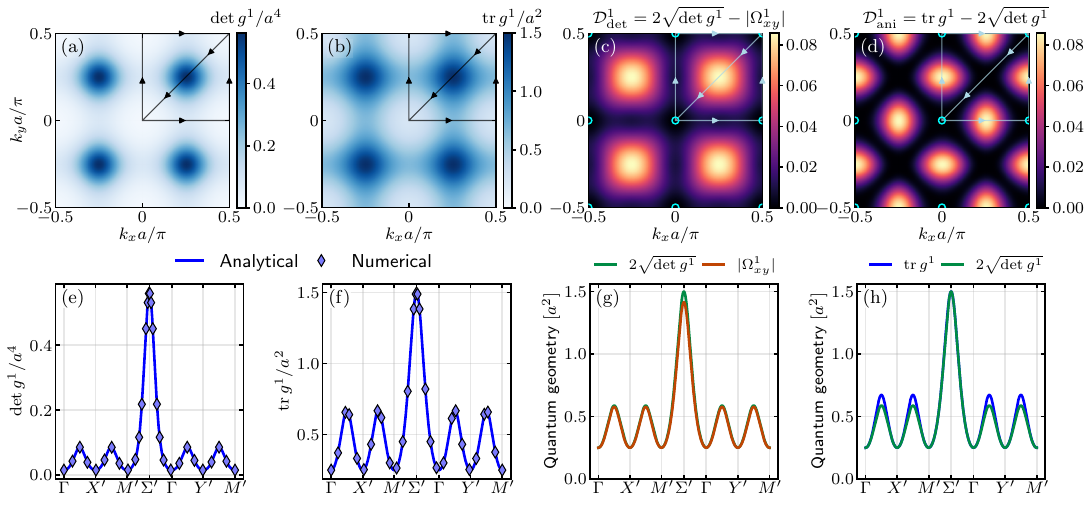}
        \caption{
            Local FCI-stability inequalities~\eqref{eq:app:local-hierarchy} evaluated on the lowest (and highest) energy band of the quarter-flux Harper-Hofstadter model, $\ket{\Psi_{\pm,1}}$~\eqref{eq:app:Eigenstates-Hofstadter-momentum}.
            (a)-(d) Analytically obtained determinant $\det g^1$~\eqref{eq:app:Det-Metric-Rank-One-Split} and trace $\mathrm{tr}\,g^1$~\eqref{eq:app:Trace-Metric-Exact} of the quantum metric, and the two local defects $\mathcal{D}^1_{\det}$~\eqref{eq:app:determinant-defect} and $\mathcal{D}^1_{\mathrm{ani}}$~\eqref{eq:app:anisotropy-defect}, as function of $\bm{k}$ over the FBZ.
            The two contributions to the trace-condition defect $D^1=\mathcal{D}^1_{\mathrm{ani}}+\mathcal{D}^1_{\det}$~\eqref{eq:app:trace-defect-local-split}~[cf.~\cref{fig:trace-defect} in main text] become clear. 
            Both defects vanish at the $Z=8$ momenta $\Gamma$, $X'$, $Y'$, $M'$ (cyan circles), where the band is locally an ideal LLL.
            (e)-(h) Same as (a)-(d) but along the high-symmetry path $\Gamma$--$X'$--$M'$--$\Sigma'$--$\Gamma$--$Y'$--$M'$ (as indicated via solid lines with arrows in (a)-(d), see also~\cref{fig:Hofstadter-Energy-Spectrum} in the main text). 
        }
        \label{fig:App:local-inequalities}
    \end{center}
\end{figure*}

As expected the determinant condition is tighter because the arithmetic-geometric-mean step separating~\eqref{eq:app:local-det-inequality} from~\eqref{eq:app:local-trace-inequality} costs
\begin{align}
    \mathcal{D}^1_{\mathrm{ani}}(\bm{k}) \equiv \mathrm{tr}\,g^1-2\sqrt{\det g^1}=(\sqrt{\lambda_1}-\sqrt{\lambda_2})^2 \ge 0
    \label{eq:app:anisotropy-defect}
\end{align}
with $\lambda_{1,2}$ the eigenvalues of $g^1$, so that it vanishes where the metric is isotropic and grows with its anisotropy. 
Here it vanishes exactly at the $C_4$-equivalent momenta $\Gamma$, $X'$, $Y'$, $M'$ and $\Sigma'$, where $C_4$ enforces $g_{xx}=g_{yy}$ and $g_{xy}=0$, and reaches its maximum $\mathcal{D}^1_{\mathrm{ani}}=[\tfrac{5}{4}-\tfrac{\sqrt3}{3}-\sqrt{\tfrac{3}{2}-\tfrac{2}{\sqrt3}}]a^2\approx0.0850\,a^2$ at $\bm{k}a=(\pi/4,0)$ and its symmetry-related images~[\cref{fig:App:QGT-Band1and4,fig:App:local-inequalities}].

Combining the exact metric trace~\eqref{eq:app:Trace-Metric-Exact} with the curvature~\eqref{eq:app:Berry-Curvature-Exact-Expression}, written as $\abs{\Omega^1_{xy}}/a^2=\sqrt2\,(12-Z)/Z^{3/2}$ (using $6-c_{4x}-c_{4y}=12-Z$, positive since $Z\le8$), the local trace-condition defect $D^1(\bm{k})=\mathrm{tr}\,g^1(\bm{k})-\abs{\Omega^1_{xy}(\bm{k})}$ is obtained in closed form,
    \begin{align}
        \frac{D^1(\bm{k})}{a^2}
        =&\frac{Z(10-Z)-2+c_{4x}^2+c_{4y}^2}{Z^2}\nonumber\\
        +&\frac{s_{2y}^2c_{2x}^2+s_{2x}^2c_{2y}^2}{(2\sqrt{2Z}+Z)^2}
        -\frac{\sqrt2\,(12-Z)}{Z^{3/2}}\, \ge 0,
        \label{eq:app:Trace-Defect-Exact}
    \end{align}
which has been depicted in~\cref{fig:trace-defect} of the main text.

The two steps of the local hierarchy~\eqref{eq:app:local-hierarchy} resolve $D^1$ into two separately non-negative pieces.
Alongside the anisotropy defect~\eqref{eq:app:anisotropy-defect} we define the determinant-condition defect
\begin{align}
    \mathcal{D}^1_{\det}(\bm{k}) \equiv 2\sqrt{\det g^1}-\abs{\Omega^1_{xy}} \ge 0 ,
    \label{eq:app:determinant-defect}
\end{align}
which is non-negative by~\eqref{eq:app:local-det-inequality}, so that the local trace defect splits exactly and pointwise as
\begin{align}
    D^1(\bm{k}) = \mathcal{D}^1_{\mathrm{ani}}(\bm{k}) + \mathcal{D}^1_{\det}(\bm{k}) .
    \label{eq:app:trace-defect-local-split}
\end{align}
Both terms are shown over the FBZ in~\cref{fig:App:local-inequalities}, together with $\det g^1$~\eqref{eq:app:Det-Metric-Rank-One-Split} and $\mathrm{tr}\,g^1$~\eqref{eq:app:Trace-Metric-Exact} themselves.
Since $\Omega^1_{xy}$ does not change sign, we again have $\int_{\mathrm{BZ}}\abs{\Omega^1_{xy}}\,d^2k=2\pi\abs{C_1}$, so that the second term gives twice the defect of the global determinant condition~\eqref{eq:app:global-det-condition},
$\mathcal{T}_{\mathrm{HH}} = \int_{\mathrm{BZ}}\mathcal{D}^1_{\mathrm{ani}}(\bm{k})\,d^2k \;+\; 2\big[\mathrm{vol}_{g^1}-\pi\abs{C_1}\big]$, 
making it not an independent measure, but the sum of the metric anisotropy and twice the determinant-condition defect G3 of~\cref{tab:fci-stability-hierarchy}. 

At the $\Gamma$-point and its $C_4$-symmetric images $X',Y',M'$ ($Z=8$) one finds $\mathrm{tr}\,g^1=2\sqrt{\det g^1}=\abs{\Omega^1_{xy}}=a^2/4$, with $\det g^1=a^4/64$ 
where the entire chain~\eqref{eq:app:local-hierarchy} therefore collapses to equalities, so that both local inequalities saturate exactly, and the band looks locally like an LLL. 
However, at the band-edge point $\Sigma'$ with $\bm{k}a=(\pi/4,\pi/4)$ ($Z=4$) the trace defect is $\mathrm{tr}\,g^1-\abs{\Omega^1_{xy}}=\tfrac32 a^2-\sqrt{2}\,a^2\approx 0.0858\,a^2>0$, and the global maximum of the local defect, $D^1=(\tfrac54-\tfrac{2}{\sqrt3})a^2\approx0.0953\,a^2$, is located on the $\Gamma$--$X'$ line (and its images).
Since the metric is isotropic at $\Sigma'$, the anisotropy defect~\eqref{eq:app:anisotropy-defect} vanishes there, so that there the whole trace defect is of determinant type, $\mathcal{D}^1_{\det}=\tfrac32 a^2-\sqrt2\,a^2$, which at the same time the global maximum of $\mathcal{D}^1_{\det}$ over the FBZ~[\cref{fig:App:local-inequalities}(c)].

Finally, let us note how the sublattice decomposition $g^1=\tfrac12(g^{(u_1)}+g^{(v_1)})+\tfrac14\abs{\Delta A^1}^2$~\eqref{eq:app:Quantum-Metric-Berry-Curvature-Sublattice-Symmetric} resolves the origin of the defect.
Since $\Omega^{(u_1)}_{xy},\Omega^{(v_1)}_{xy}\le 0$ everywhere and never cancel, the trace defect splits cleanly,
\begin{align}
    \mathcal{T}_{\mathrm{HH}} = \tfrac12\,\mathcal{T}^{(u)} + \tfrac12\,\mathcal{T}^{(v)} + \tfrac14\int_{\mathrm{BZ}}\abs{\Delta A^1}^2\,d^2k,
    \label{eq:app:trace-defect-decomposition}
\end{align}
with each sector defect $\mathcal{T}^{(u/v)}=\int_{\mathrm{BZ}}[\mathrm{tr}\,g^{(u/v)}-\abs{\Omega^{(u/v)}_{xy}}]\ge 0$ guaranteed to be non-negative by the PSD applied to the QGT $Q^{u_j/v_j}_{\mu\nu}$~\eqref{eq:app:Quantum-Geometric-Tensor-Sublattice-Symmetric} from the $u/v$ sectors. 
We find that $\mathcal{T}^{(u)}=\mathcal{T}^{(v)}\approx 0.4917$ and $\int\abs{\Delta A^1}^2\,d^2k\approx 0.1211$, reproducing $\mathcal{T}_{\mathrm{HH}}=0.522$ such that the Berry-connection-difference term, 
which is absent for a genuine single-orbital LLL, contributes $\tfrac14\int\abs{\Delta A^1}^2/\mathcal{T}_{\mathrm{HH}}\approx 5.8\%$ of the total defect on its own, identifying the internal $u/v$ structure of the system as an additional intrinsic, gauge-invariant obstruction to ideality.

\subsection{Relation to the projected density algebra}
\label{app:fci-gmp-algebra}

Let us briefly connect the band-geometric criteria above to the language of projected (number) density operators. 
The Fourier transform of the real-space number density operator is $n_{\bm q}=e^{-i\bm q\cdot{\bm{r}}}$, which transfers momentum $\bm{q}$ to the particle.
Below the band gap the particle cannot leave the band, so that every operator of the effective low-energy theory appears in its projected form. 
For instance a density-density interaction becomes $\sim \sum_{\bm q}V(\bm q)\,\bar\rho_{\bm q}\bar\rho_{-\bm q}$, where the algebra of the projected density $\bar\rho_{\bm q}$ controls the low-energy physics~\cite{Parameswaran2012,Roy2014,Wang2025a}.
Since the density operator transfers momentum, the relevant projector here is not the $\bm{k}$-resolved $P_n(\bm{k})=\projector{\Psi_n(\bm{k})}$ used for the QGT above, but the projector onto the band as a whole,
    \begin{align}
        \mathcal{P}_n = \int_{\mathrm{BZ}} d^2k\;\projector{\Psi^{\mathrm{B}}_{n,\bm{k}}},
        \quad \ket{\Psi^{\mathrm{B}}_{n,\bm{k}}}=e^{i\bm{k}\cdot\bm{r}}\ket{\Psi_n(\bm{k})},
        \label{eq:app:band-projector}
    \end{align}
where throughout this work $\ket{\Psi_n(\bm{k})}$ denotes the cell-periodic Bloch function, i.e., the eigenvector of $H(\bm{k})$~\eqref{eq:Bloch-Ham-Case-II}, and $\ket{\Psi^{\mathrm{B}}_{n,\bm{k}}}$ the full Bloch state carrying the plane-wave factor~\footnote{We follow the convention of Ref.~\cite{Wang2025a}, in which the Brillouin-zone measure carries unit total weight, $\int_{\mathrm{BZ}}d^2k=1$, and the Bloch states are normalized accordingly, so that $\mathcal{P}_n^2=\mathcal{P}_n$. Equivalently one may read $\int_{\mathrm{BZ}}d^2k$ as $\sum_{\bm{k}}$ with $\braket{\Psi^{\mathrm{B}}_{n,\bm{k}}}{\Psi^{\mathrm{B}}_{m,\bm{k}'}}=\delta_{\bm{k}\bm{k}'}\delta_{nm}$, as in Refs.~\cite{Parameswaran2012,Roy2014}.}. 
The band-projected number density operator is then $\bar\rho_{\bm q}=\mathcal{P}_n\,n_{\bm q}\,\mathcal{P}_n=\mathcal{P}_n\,e^{-i\bm q\cdot{\bm{r}}}\,\mathcal{P}_n$, which is a single-particle operator. 
Inserting~\eqref{eq:app:band-projector} twice then gives,
    \begin{align}
        \bar\rho_{\bm q} = \int_{\mathrm{BZ}} d^2k\;\braket{\Psi_n(\bm{k})}{\Psi_n(\bm{k}+\bm{q})}\ \ketbra{\Psi^{\mathrm{B}}_{n,\bm{k}}}{\Psi^{\mathrm{B}}_{n,\bm{k}+\bm{q}}}.
        \label{eq:app:projected-density-bloch}
    \end{align}
We see that the plane waves have enforced momentum conservation and the overlap of the cell-periodic Bloch functions $\ketbra{\Psi^{\mathrm{B}}_{n,\bm{k}}}{\Psi^{\mathrm{B}}_{n,\bm{k}+\bm{q}}}$ survives as the weight. 
Expanding that overlap to second order in $\bm{q}$ returns the QGT, which is why the quantum geometry of~\cref{app:fci-momentum-hierarchy} controls the algebra~\footnote{We use the Fourier convention $e^{-i\bm q\cdot\hat{\bm{r}}}$ throughout, which is opposite to that of~\cite{Wang2025a}.}

The projection is what makes this algebra non-trivial. Namely, while unprojected densities multiply as $n_{\bm q}n_{\bm p}=n_{\bm q+\bm p}$ and hence commute, $\bar\rho_{\bm q}\bar\rho_{\bm p}$ do not. 
Acting on a Bloch state at $\bm{k}$, its leading order in the momenta gives the Berry curvature~\cite{Parameswaran2012},
    \begin{align}
        [\bar\rho_{\bm q_1},\bar\rho_{\bm q_2}]\ket{\Psi^{\mathrm{B}}_{n,\bm{k}}}=&-i\,(\bm q_1\times\bm q_2)_z\,\Omega^n_{xy}(\bm k)\,\bar\rho_{\bm q_1+\bm q_2}\ket{\Psi^{\mathrm{B}}_{n,\bm{k}}}\nonumber \\
        &+\mathcal O(q^3).
        \label{eq:app:gmp-leading}
    \end{align}
We immediately see that closure already at this order would demand a constant Berry curvature~\cite{Parameswaran2012} 
while closure at all wavelengths, with LLL form factors, additionally demands pointwise saturation of the trace condition~\cite{Roy2014,Ledwith2023}. 
Both of these requirements are not satisfied here as detailed in the previous section~\cref{app:fci-application-ground-band}.
This is not specific to the quarter-flux model, since a closed band-projected density algebra cannot be realized exactly in any tight-binding model with finitely many orbitals per unit cell~\cite{Varjas2022}, as the four-orbital band studied here. 

\section{Static structure factor and the quantum weight}
\label{app:quantum-weight-sf}

At last, let us now provide some details on the connection between the band geometry and the long-wavelength limits of the static structure factor of a filled band, and its connection to general topological bounds, as recently revealed in~\cite{Onishi2024a,Onishi2024b,Onishi2025a}. 
The static structure factor is a directly measurable quantity which can be obtained in scattering experiments, or via optical conductivity measurements through sum rules~\cite{Onishi2024a,Onishi2024b,Onishi2025a}. 

The static structure factor is defined for any state, interacting or not, as the equal-time correlator of the number density operator $\rho(\bm r)$ of the many-body ground state,
\begin{align}
    S(\bm q) = \frac{1}{V} \ep{\rho_{\bm q}\rho_{-\bm q}},
    \qquad \rho_{\bm q}=\int d\bm r\;e^{-i\bm q\cdot\bm r}\rho(\bm r),
    \label{eq:app:structure-factor-def}
\end{align}
with $V$ the volume, following the normalization of~\Ccite{Onishi2024a}. 
Note that $\rho_{\bm q}$ here is the \emph{many-body} density operator $\int d\bm r\,\hat\Psi^\dagger(\bm r)e^{-i\bm q\cdot\bm r}\hat\Psi(\bm r)$, the second-quantized image of the one-body operator $e^{-i\bm q\cdot\bm r}$. 
Expanding the field operator $\hat\Psi(\bm r)$ in the Bloch basis for non-interacting fermions filling the lowest Hofstadter band, we obtain
    \begin{align}
        \rho_{\bm q}=\sum_{\bm k}\braket{\Psi_{-,1}(\bm k)}{\Psi_{-,1}(\bm k+\bm q)}\ c^\dagger_{\bm k}c_{\bm k+\bm q}.
        \label{eq:app:density-bloch}
    \end{align}
where $c^\dagger_{\bm k}$ denotes the creation of a fermion in mode $\ket{\Psi^{\mathrm{B}}_{1,\bm{k}}}$. 
The emerging overlap factor $\braket{\Psi_{-,1}(\bm k)}{\Psi_{-,1}(\bm k+\bm q)}$
is the entire origin of the band geometry in $S(\bm q)$. 
For the filled Fermi sea of the lowest band, we now insert~\eqref{eq:app:density-bloch} into~\eqref{eq:app:structure-factor-def}, using $\mathrm{Tr}[P_1(\bm k)P_1(\bm k+\bm q)]=\abs{\braket{\Psi_{-,1}(\bm k)}{\Psi_{-,1}(\bm k+\bm q)}}^2$ together with $V^{-1}\sum_{\bm k}\to\int d^2k/(2\pi)^2$, and obtain~\cite{Onishi2024a},
    \begin{align}
        S(\bm q)=\int_{\mathrm{BZ}}\frac{d^2k}{(2\pi)^2}\,\mathrm{Tr}\big[P_1(\bm k)\big(P_1(\bm k)-P_1(\bm k+\bm q)\big)\big].
        \label{eq:app:structure-factor-projector}
    \end{align}
with the ground-band projector $P_1(\bm k)=\projector{\Psi_{-,1}(\bm k)}$ of~\cref{app:fci-momentum-hierarchy}. 
The integrand is the gauge-invariant quantum distance $1-\abs{\braket{\Psi_{-,1}(\bm k)}{\Psi_{-,1}(\bm k+\bm q)}}^2=g_{\mu\nu}^1 q^\mu q^\nu + \mathcal{O}(\abs{q}^3)$, which expanded to second order gives the quantum metric $g^1_{\mu\nu}$ of the lowest band~\cite{Resta2011}. 
Thus, the so-called \emph{quantum weight} $K_{\mu\nu}$~\cite{Onishi2024a,Onishi2024b}, is obtained
\begin{align}
        S(\bm q)=\frac{K_{\mu\nu}}{2\pi}\,q^\mu q^\nu+\mathcal O(q^4),\quad
        K_{\mu\nu}=\frac{1}{2\pi}\int_{\mathrm{BZ}}d^2k\,g^1_{\mu\nu}(\bm k).
        \label{eq:app:obs-structure-metric}
    \end{align}
The $C_4$ symmetry of the model makes the quantum weight isotropic, $K_{xx}=K_{yy}\approx 0.5415$ and $K_{xy}=0$.
Because the Berry curvature does not change sign over the BZ, $K$ is not independent geometric data but actually the integrated trace condition of the main text in disguise,
\begin{align}
    K\equiv K_{xx}+K_{yy}=\abs{C_1}+\frac{\mathcal{T}_{\mathrm{HH}}}{2\pi}\approx1.083\;\ge\;\abs{C_1}=1,
    \label{eq:app:quantum-weight-trace-defect}
\end{align}
i.e., the $8.3\%$ by which $K$ exceeds the topological bound $K\ge\abs{C}$~\cite{Onishi2024a} is exactly $\mathcal{T}_{\mathrm{HH}}/2\pi$, so the structure factor reformulates the same geometric defect as a measurable density correlator.
Note that, up to the factor $2\pi$, $K$ is also the gauge-invariant part of the Marzari-Vanderbilt Wannier spread functional~\cite{Marzari1997,Marzari2012}, equivalently the localization functional of~\Ccite{Souza2000,Onishi2024a}. 
The bound $K\ge\abs{C}$ is thus equivalent to non-zero Chern number obstructing the existence of exponentially localized Wannier functions~\cite{Brouder2007,Hastings2010}, where $\abs{C}$ puts a floor on the Wannier spread. 
For the ground band this floor accounts for $\abs{C_1}/K\approx92.3\%$ of the quantum weight, the remaining $7.7\%$ being the excess spread that the band carries beyond its topological minimum, which is the same $8.3\%$ excess mentioned above measured relative to the bound $\abs{C_1}$ (and not to $K$).
This excess here is a measure of non-ideality (rather than of delocalization), since a band saturating $K=\abs{C}$ would have the most localized Wannier functions its topology permits, while still decaying algebraically due to $C\neq 0$~\cite{Brouder2007,Hastings2010}.

\end{document}